\documentclass[a4paper,fleqn,usenatbib]{mnras}
\usepackage{amsmath}
\usepackage{dsfont}
\usepackage{graphicx,amsmath,bm,natbib,color, multirow}
\usepackage{amssymb}
\usepackage{amsmath} 
\usepackage{appendix}
\usepackage{xspace}
\usepackage{multirow}
\usepackage{mathptmx}
\usepackage{url}
\usepackage{xcolor}
\usepackage{footnote}
\usepackage{natbib}
\usepackage{hyperref}  
\usepackage{graphicx}
\usepackage{wrapfig}
\usepackage{epstopdf}
\usepackage{acronym}
\usepackage{lineno}
\usepackage{eso-pic}

\AddToShipoutPictureBG*{%
  \AtPageUpperLeft{%
    \hspace{1.8\linewidth}%
    \raisebox{-3.5\baselineskip}{%
      \makebox[0pt][l]{\textnormal{DES 2016-0212}}
}}}%

\AddToShipoutPictureBG*{%
  \AtPageUpperLeft{%
    \hspace{1.8\linewidth}%
    \raisebox{-4.5\baselineskip}{%
      \makebox[0pt][l]{\textnormal{FERMILAB-PUB-17-295-AE}}
}}}%

\def\reff@jnl#1{{\rm#1\/}}
\def\aj{\reff@jnl{AJ}}         
\def\araa{\reff@jnl{ARA\&A}}      
\def\apj{\reff@jnl{ApJ}}        
\def\apjl{\reff@jnl{ApJ}}        
\def\apjs{\reff@jnl{ApJS}}       
\def\aap{\reff@jnl{A\&A}}        
\def\aapr{\reff@jnl{A\&A~Rev.}}     
\def\aaps{\reff@jnl{A\&AS}}       
\def\mnras{\reff@jnl{MNRAS}}      
\def\physrep{\reff@jnl{Physics Reports}}
\def\prd{\reff@jnl{Phys.Rev.D}}     
\def\prl{\reff@jnl{Phys.Rev.Lett}}   
\def\pasp{\reff@jnl{PASP}}       
\def\pasj{\reff@jnl{PASJ}}       
\def\nat{\reff@jnl{Nature}}       
\def\jcap{\reff@jnl{JCAP}}   
\def\memsai{\reff@jnl{MemSAI}} 
\def\na{\reff@jnl{New Astronomy}}       
\def\procspie{\reff@jnl{SPIE}}       
\def\pasa{\reff@jnl{PASA}}       

\def\Sref#1{Sec.~\ref{#1}\xspace}

\def\Fref#1{Fig.~\ref{#1}\xspace}

\def\Tref#1{Table~\ref{#1}\xspace}

\def\Eref#1{Eq.~(\ref{#1})\xspace}

\def\Aref#1{Appendix~\ref{#1}\xspace}
\def\Cref#1{Chapter~\ref{#1}\xspace}

\def\imshape{\textsc{Im3shape}\xspace}
\def\metacal{\textsc{MetaCalibration}\xspace}
\def\healpix{\textsc{Healpix}\xspace}
\def\redmapper{\textsc{redMapper}\xspace}
\def\redmagic{\textsc{redMagic}\xspace}

\newcommand{\edth}{\,\eth\,}
\newcommand{\edthbar}{\,\overline{\eth}\,}

\usepackage{todonotes}

\def\southhampton{School of Physics and Astronomy, University of Southampton,  Southampton, SO17 1BJ, UK}
\def\uam{Instituto de Fisica Teorica UAM/CSIC, Universidad Autonoma de Madrid, 28049 Madrid, Spain}
\def\brazil{Departamento de F\'isica Matem\'atica, Instituto de F\'isica, Universidade de S\~ao Paulo, CP 66318, S\~ao Paulo, SP, 05314-970, Brazil}
\def\wataghin{Instituto de F\'isica Gleb Wataghin, Universidade Estadual de Campinas, 13083-859, Campinas, SP, Brazil}
\def\iit{Department of Physics, IIT Hyderabad, Kandi, Telangana 502285, India}
\def\rhodes{Department of Physics and Electronics, Rhodes University, PO Box 94, Grahamstown, 6140, South Africa}
\def\santacruz{Santa Cruz Institute for Particle Physics, Santa Cruz, CA 95064, USA}
\def\uw{Astronomy Department, University of Washington, Box 351580, Seattle, WA 98195, USA}
\def\oakridge{Computer Science and Mathematics Division, Oak Ridge National Laboratory, Oak Ridge, TN 37831}
\def\nyu{New York University, CCPP,  New York, NY 10003, USA}
\def\sorbonne{Sorbonne Universit\'es, UPMC Univ Paris 06, UMR 7095, Institut d'Astrophysique de Paris, F-75014, Paris, France}
\def\cnrs{CNRS, UMR 7095, Institut d'Astrophysique de Paris, F-75014, Paris, France}
\def\lsst{LSST, 933 North Cherry Avenue, Tucson, AZ 85721, USA}
\def\kicp{Kavli Institute for Cosmological Physics, University of Chicago, Chicago, IL 60637, USA}
\def\anl{Argonne National Laboratory, 9700 South Cass Avenue, Lemont, IL 60439, USA}
\def\upenn{Department of Physics and Astronomy, University of Pennsylvania, Philadelphia, PA 19104, USA}
\def\ethz{Department of Physics, ETH Zurich, Wolfgang-Pauli-Strasse 16, CH-8093 Zurich, Switzerland}
\def\ports{Institute of Cosmology \& Gravitation, University of Portsmouth, Portsmouth, PO1 3FX, UK}
\def\ucl{Department of Physics \& Astronomy, University College London, Gower Street, London, WC1E 6BT, UK}
\def\bnl{Brookhaven National Laboratory, Bldg 510, Upton, NY 11973, USA}
\def\fermilab{Fermi National Accelerator Laboratory, P. O. Box 500, Batavia, IL 60510, USA}
\def\stanford{Department of Physics, Stanford University, 382 Via Pueblo Mall, Stanford, CA 94305, USA}
\def\kipac{Kavli Institute for Particle Astrophysics \& Cosmology, P. O. Box 2450, Stanford University, Stanford, CA 94305, USA}
\def\slac{SLAC National Accelerator Laboratory, Menlo Park, CA 94025, USA}
\def\ifae{Institut de F\'{\i}sica d'Altes Energies (IFAE), The Barcelona Institute of Science and Technology, Campus UAB, 08193 Bellaterra (Barcelona) Spain}
\def\ieec{Institut de Ci\`encies de l'Espai, IEEC-CSIC, Campus UAB, Facultat de Ci\`encies, Torre C5 par-2, 08193 Bellaterra, Barcelona, Spain}
\def\ccap{Center for Cosmology and Astro-Particle Physics, The Ohio State University, Columbus, OH 43210, USA}
\def\ohio{Department of Physics, The Ohio State University, Columbus, OH 43210, USA}
\def\manchester{Jodrell Bank Center for Astrophysics, School of Physics and Astronomy, University of Manchester, Oxford Road, Manchester, M13 9PL, UK}
\def\cambridgekavli{Kavli Institute for Cosmology, University of Cambridge, Madingley Road, Cambridge CB3 0HA, UK}
\def\cambridge{Institute of Astronomy, University of Cambridge, Madingley Road, Cambridge CB3 0HA, UK}

\def\lina{Laborat\'orio Interinstitucional de e-Astronomia - LIneA, Rua Gal. Jos\'e Cristino 77, Rio de Janeiro, RJ - 20921-400, Brazil}
\def\on{Observat\'orio Nacional, Rua Gal. Jos\'e Cristino 77, Rio de Janeiro, RJ - 20921-400, Brazil}

\def\michigan{Department of Physics, University of Michigan, Ann Arbor, MI 48109, USA}

\def\maxplanck{Max Planck Institute for Extraterrestrial Physics, Giessenbachstrasse, 85748 Garching, Germany}

\def\lmu{Faculty of Physics, Ludwig-Maximilians-Universit\"at, Scheinerstr. 1, 81679 Munich, Germany}
\def\ctio{Cerro Tololo Inter-American Observatory, National Optical Astronomy Observatory, Casilla 603, La Serena, Chile}
\def\aao{Australian Astronomical Observatory, North Ryde, NSW 2113, Australia}

\def\jpl{Jet Propulsion Laboratory, California Institute of Technology, 4800 Oak Grove Dr., Pasadena, CA 91109, USA}
\def\ciemat{Centro de Investigaciones Energ\'eticas, Medioambientales y Tecnol\'ogicas (CIEMAT), Madrid, Spain}
\def\uiuc{Department of Astronomy, University of Illinois, 1002 W. Green Street, Urbana, IL 61801, USA}
\def\ncsa{National Center for Supercomputing Applications, 1205 West Clark St., Urbana, IL 61801, USA}
\def\ua{University of Arizona, Department of Physics, 1118 E. Fourth St., Tucson, AZ 85721, USA}
\def\jpl{Jet Propulsion Laboratory, California Institute of Technology, 4800 Oak Grove Dr., Pasadena, CA 91109, USA}
\def\sussex{Department of Physics and Astronomy, Pevensey Building, University of Sussex, Brighton, BN1 9QH, UK}
\def\cluster{Excellence Cluster Universe, Boltzmannstr.\ 2, 85748 Garching, Germany}
\def\barcelona{Instituci\'o Catalana de Recerca i Estudis Avan\c{c}ats, E-08010 Barcelona, Spain}

\def\edinburgh{Institute for Astronomy, University of Edinburgh, Edinburgh EH9 3HJ, UK}
\def\princeton{Department of Astrophysical Sciences, Princeton University, Peyton Hall, Princeton, NJ 08544, USA}
\def\ceasaclay{DEDIP/DAP, IRFU, CEA, Universit\'e Paris-Saclay, F-91191 Gif-sur-Yvette, France}
\def\cea{Universit\'e Paris Diderot, AIM, Sorbonne Paris Cit\'e, CEA, CNRS, F-91191 Gif-sur-Yvette, France}

\begin{document}

\title[Dark Energy Survey Year 1 Results: Curved-Sky Weak Lensing Mass Map]{Dark Energy Survey Year 1 Results: \\ Curved-Sky Weak Lensing Mass Map}
\author[C.~Chang et al.]{
\parbox{1.03\textwidth}{
\Large{C.~Chang$^{1*}$,
A.~Pujol$^{2,3,4}$, 
B.~Mawdsley$^{5}$,
D.~Bacon$^{5}$,
J.~Elvin-Poole$^{6}$,
P.~Melchior$^{7}$,
A.~Kov\'acs$^{8}$,
B.~Jain$^{9}$,
B.~Leistedt$^{10,11}$
% ...
   \begin{center} (DES Collaboration) \end{center}
}
\vspace{0.2cm}
\parbox{\textwidth}{ \small
\textit{The authors' affiliations are shown in \Aref{sec:affiliations}. \\
$^{*}$e-mail address: chihway@kicp.uchicago.edu}}
}}

\maketitle

\begin{abstract}
We construct the largest curved-sky galaxy weak lensing mass map to date from the DES first-year 
(DES Y1) data. The map, about 10 times larger than previous work, is constructed over a contiguous 
$\approx1,500$~deg$^2$, covering a comoving volume of $\approx10$~Gpc$^3$. 
The effects of masking, sampling, and noise are tested using simulations. We 
generate weak lensing maps from two DES Y1 shear catalogs, \metacal\ and \imshape, with sources at 
redshift $0.2<z<1.3,$ and in each of four bins in this range. In the highest signal-to-noise map, 
the ratio between the mean signal-to-noise in the E-mode and the B-mode map is $\sim$1.5 ($\sim$2) when 
smoothed with a Gaussian filter of $\sigma_{G}=30$ (80) arcminutes. 
The second and third moments of the 
convergence $\kappa$ in the maps are in agreement with simulations. We also find no significant 
correlation of $\kappa$ with maps of potential systematic contaminants. 
Finally, we demonstrate two applications of the mass 
maps: (1) cross-correlation with different foreground tracers of mass and (2) exploration of the largest 
peaks and voids in the maps. 
\end{abstract}

\begin{keywords}
gravitational lensing: weak, cosmology: dark matter, surveys
\end{keywords}

\maketitle

\section{Introduction}
\label{sec:intro}

One way to map the mass distribution of the Universe is by using the technique of weak gravitational lensing.
\citep{Kaiser1993,Massey2007,VanWaerbeke2013,Vikram2015,Chang2015,Oguri2017}. 
The motivations for generating these mass maps using weak lensing are twofold. First, it is easy 
to pick out distinct features and understand the qualitative characteristics of the mass distribution from 
maps. Second, as the maps ideally preserve the full, uncompressed information for the field, they enable 
the extraction of non-Gaussian information beyond the standard two-point statistics used in cosmology 
\citep[e.g.][]{Abbott2016, Kwan2017,Hildebrandt2017}. These non-Gaussian 
statistics are being explored using 3-point statistics \citep{Cooray2001,Dodelson2005}, 
peak counts \citep{Dietrich2010, Kratochvil2010,Kacprzak2016}, and the full Probability Density 
Function (PDF) of the map \citep{Clerkin2015,Patton2016}. As the statistical 
uncertainties in 
the current and future data sets decrease, we expect these higher-order statistics to offer new 
constraints that are complementary to the more traditional two-point approaches.  

Physically, a weak lensing mass map, or convergence map, represents the integrated total matter density 
along the line-of-sight, weighted by a broad lensing kernel that peaks roughly half-way between the 
observer and the source galaxies from which the measurement is made.
Since lensing does not distinguish between the species and dynamical state of the mass, or the 
``lens'', one can directly probe mass with weak lensing, which is a key difference from maps 
constructed from biased tracers of mass such as galaxies. The theoretical framework of constructing 
weak lensing convergence maps from the weak lensing observable, shear, has 
been developed since \citet[][hereafter KS]{Kaiser1993} and \citet{Schneider1996}. 
Shear and convergence are second derivatives of the same lensing potential 
field, which makes it possible to convert between them up to a constant. 

Small-field weak lensing mass 
maps have been commonly used in galaxy cluster fields to study the detailed structure of the cluster 
mass distribution and compare with the distribution of baryonic matter \citep{Clowe2006,
vonderLinden2014,Melchior2015}. These maps have relatively high signal-to-noise because
the cluster lensing signal is $\sim10$ times larger than the lensing signal from the large-scale structure 
\citep{Bartelmann2001}, and the information about the fields was obtained using deep imaging to 
achieve a high number density of 
source galaxies for weak lensing measurements. A number of algorithms beyond KS 
were developed to specifically tackle the mass reconstruction with clusters and have been 
successfully implemented on data \citep{Seitz1998,Marshall2002,Leonard2014}. 

Wide-field convergence maps, on the other hand, have only been constructed recently, thanks to the 
development of dedicated weak lensing surveys that cover patches of sky on the 
order of hundreds of square degrees or larger. This includes the Canada-France-Hawaii Telescope 
Lensing Survey \citep[CFHTLenS,][]{Erben2013}, the KIlo-Degree Survey \citep[KIDS,][]{deJong2015}, 
the Hyper SuprimeCam Survey \citep[HSC,][]{Aihara2017} and the Dark Energy Survey \citep[DES,][]{Flaugher2005}.  
\citet{VanWaerbeke2013} was the first to study in detail these wide-field weak lensing mass maps in four 
fields (adding up to a total of 154 deg$^{2}$) of the CFHTLenS data, including the noise properties, 
high-order moments, and the cross-correlation with galaxies. In \citet{Vikram2015} and \citet{Chang2015}, 
we carried out a similar analysis with early DES Science Verification (SV) data using a 139 deg$^{2}$ 
contiguous region of the sky. 
Recent work from HSC \citep{Mandelbaum2017, Oguri2017} also carried out an analysis of mass map 
reconstruction using the HSC data in both 2D and 3D. Although the area of these maps are not as large 
(the total area of 
the data set is 136.9 deg$^{2}$, split into six separate fields), the number density of the sources is 
several times larger than in the other data sets (25 galaxies per arcmin$^{2}$), which allows for 
reconstruction on much smaller scales. \citet{Oguri2017} looked at cross-correlation of the mass maps 
with galaxy distributions and several systematics tests.  
All three studies described above use the KS method under flat-sky approximation, and show that the 
mass maps contain significant extractable cosmological information.

Continuing from the SV work described above to the first year of DES data (DES Y1), we present in this paper 
a weak lensing mass map of $\sim1,500$ deg$^{2}$, more than ten times larger than the SV map. A few 
advances over the SV studies were made: First, given the large area of the mass 
map on the sky, it was necessary to go beyond the flat-sky approximation and employ curved-sky estimators. 
The implementation of the curved-sky reconstruction borrows from tools developed for CMB polarisation 
analyses and has been studied in detail in the context of weak lensing mass mapping and cosmic shear 
\citep{Heavens2003,Castro2005,Heavens2006,Kitching2014,Leistedt2017,Wallis2017}. 
The first all sky curved weak lensing maps constructed from simulations were presented in \citet{Fosalba2008}, 
which was an extension from the work on constructing mock galaxy catalogs in \cite{Gaztanaga1998}. Second, 
we move from a single redshift bin to multiple redshift bins, a first step towards constructing a 3D weak lensing map. 
These tomographic bins match 
those used in the DES Y1 cosmology analysis, thus making our maps very complementary to the series of 
DES Y1 papers that focus on two-point statistics  \citep{DES2017, Troxel2017, Prat2017, MacCrann2017}. 
Specifically, this paper presents the spatial configuration and phase information of the data that goes into 
these cosmological analyses. Finally, 
we explore for the first time the possibility of constructing the lensing potential and deflection fields. These 
fields are commonly studied in the CMB lensing community, but seldom constructed and visualised using 
measurements of galaxy lensing except in some theoretical studies \citep{Vallinotto2007,Dodelson2008,
Chang2014}. 
The primary reason that potential and deflection fields are seldom used in galaxy lensing is that the 
information of the potential and deflection fields are on scales much larger (or lower $\ell$ modes) 
than the convergence field. This means that in previous smaller data sets, there is not enough low $\ell$ 
information in the data to reconstruct the potential and deflection fields. However, with the wide-field data 
used in this work, we are just beginning to enter the era where the reconstruction is not dominated 
by noise and interesting applications can be explored. For example, with an accurate deflection field, one 
can ``delens'' the galaxy fields and move the observed galaxy positions back to their unlensed position, which 
would improve measurements such as galaxy-galaxy lensing \citep{Chang2014}. Similarly, having a good 
estimate of the lensing potential could in principle provide foreground information for delensing the CMB 
\citep{Marian2007, Manzotti2017, Yu2017}.     

This paper is organised as follows. In \Sref{sec:formalism} we introduce the formalism used for constructing 
the curved-sky convergence map from shear maps. In \Sref{sec:data}, the data and simulations used in 
this paper are described. We then outline in \Sref{sec:method} the practical procedure of constructing the 
maps from the DES Y1 shear catalogs. In \Sref{sec:sim_test} we present a series of tests using simulated data to 
quantitatively understand the performance of the map-making method as well as how that method interacts with the 
different sources of noise in the data. We then present our final DES Y1 mass maps in \Sref{sec:results} for 
different redshift bins and test for residual systematic effects by cross-correlating the maps with observational 
quantities. We follow up with two 
applications of the mass maps in \Sref{sec:application}: (1) cross-correlation of the mass maps with 
different foreground galaxy samples, and (2) examination of the largest peaks and voids in the maps. We 
conclude in \Sref{sec:conclusion}. 
In \Aref{sec:inpainting} we investigate the different approaches of masking and their effect on the reconstruction. 
In \Aref{sec:deflection_test_sim} we demonstrate the possibility of 
reconstructing the weak lensing potential and deflection maps in addition to the convergence map, which 
will become more interesting in future datasets as the sky coverage increases. Finally in \Aref{sec:im3shape} we present 
convergence maps from the \imshape shear catalog (in addition to the maps from the \metacal shear catalog 
presented in the main text) to show the consistency between the catalogs.

\section{Formalism}
\label{sec:formalism}

As mentioned in \Sref{sec:intro}, the construction of convergence ($\kappa$) maps from shear ($\bm{\gamma}$) 
maps in data has been done assuming the flat-sky approximation in most previous work \citep{VanWaerbeke2013, Vikram2015, 
Chang2015} due to the relatively small sky coverage involved. In fact, as shown in \citet{Wallis2017}, the gain in moving 
from flat-sky to curved-sky is very marginal in the case where the data is on the order of ~100 deg$^{2}$. In this 
paper, our data set is sufficiently large to warrant a curved-sky treatment, which also prepares us for future, even 
larger, data sets. The fundamental mathematical operation that we are interested in is to decompose a spin-2 field 
($\bm{\gamma}$) into a curl-free component and a divergence-free component. The curl-free component corresponds 
to the convergence signal, which is also referred to as the E-mode convergence field 
$\kappa_{E}$. The divergence-free component, which we refer to as $\kappa_{B}$, is expected to be negligible 
compared to $\kappa_{E}$ for gravitational lensing, but can arise from noise and systematics in the shear estimates. 
Mathematically, this 
operation is the same as the classical Helmholtz decomposition, but generalised onto the spherical coordinates. 
We sketch below the formalism of converting between the $\kappa$ and $\bm{\gamma}$ maps as well as the 
deflection field $\bm{\eta}$ and the potential field $\psi$. For detailed derivations, we refer the reader to  
\citet{Bartelmann2010,Castro2005,Wallis2017}.

Consider the 3D Newtonian potential $\Psi$ defined at every given comoving distance $\chi$ and angular 
position $(\theta, \phi)$ on the sky. The effective lensing potential $\psi$ is defined by projecting $\Psi$ along 
the line-of-sight. That is \citep{Bartelmann2001}, 
\begin{equation}
\psi( \chi_{s}, \theta, \phi) = 2 \int d\chi' \frac{f_{K}(\chi_{s}-\chi')}{f_{K}(\chi') f_{K}(\chi_{s})} \Psi(\chi',\theta, \phi),
\end{equation}
where 
$f_{K}$ depends on the curvature $k$ of the 
Universe: $f_{K}(\chi) = \sin \chi$, $\chi$, $\sinh \chi$ for closed ($k=1$), flat ($k=0$) and open ($k=-1$) universe 
respectively. The 3D potential is related to the distribution of the matter overdensity $\delta( \chi, \theta, \phi)$ 
via the Poisson equation 
\begin{equation}
\nabla^{2}_{\chi} \Psi ( \chi, \theta, \phi) = \frac{3 \Omega_{m}H_{0}^{2}}{2a} \delta( \chi, \theta, \phi),
\end{equation} 
where $\Omega_{m}$ is the total matter density today, $H_{0}$ is the Hubble constant today, and $a=1/(1+z)$ 
is the scale factor. Note that the gradient $\nabla_{\chi}$ is taken in the comoving radial direction.
 
Expanding the lensing potential at a given comoving distance $\chi$ in spherical harmonics, we have
\begin{align}
\psi(\chi) &= \sum_{\ell m} \psi_{\ell m}(\chi) \: _{0}Y_{\ell m}(\theta, \phi), \notag \\  
\psi_{\ell m}(\chi) &= \int d\Omega \psi(\chi) \: _{0}Y^{*}_{\ell m}(\theta, \phi),
\label{eq:psi}
\end{align} 
where $_{0}Y_{\ell m}$ are the spin-0 spherical harmonic basis set and $\psi_{\ell m}(\chi)$ is the coefficient associated 
with $_{0}Y_{\ell m}$ at $\chi$. Below we will omit the $\chi$ reference in our notation for simplicity, but note that 
these equations apply to the fields on a given redshift shell. 

To derive the spherical harmonic representation of shear and convergence, we have
\begin{equation}
\kappa = \frac{1}{4} (\edth \edthbar +\edthbar \edth) \psi, 
\label{eq:kappa_deriv}
\end{equation}
\begin{equation}
\bm{\gamma} = \gamma^{1}+i\gamma^{2} =\frac{1}{2}\edth \edth \psi,
\label{eq:gamma_deriv}
\end{equation}
where $\edth$ and $\edthbar$ 
are the raising and lowering operators that act on spin-weighted spherical harmonics,
$_{s}Y_{\ell m}$ and follow a certain set of rules \citep[see e.g.,][for details]{Castro2005}.
We can now define the spherical representation of the convergence field and the shear field to be
\begin{equation}
\kappa = \kappa_{E} + i \kappa_{B} = \sum_{\ell m} ( \hat{\kappa}_{E, \ell m} +i \hat{\kappa}_{B,\ell m} )\: _{0}Y_{\ell m},
\label{eq:kappa_alm}
\end{equation} 
and
\begin{equation}
\bm{\gamma} = \gamma^{1} + i \gamma^{2} = \sum_{\ell m} \hat{\gamma}_{\ell m} \; {}_{2}Y_{\ell m}.
\label{eq:gamma}
\end{equation}
Here $_{2}Y_{\ell m}$ are spin-2 spherical harmonics. From \Eref{eq:kappa_deriv} and \Eref{eq:gamma_deriv} 
it follows that
\begin{equation}
\hat{\kappa}_{E, \ell m}  + i\hat{\kappa}_{B, \ell m} = -\frac{1}{2}\ell (\ell +1)\psi_{\ell m},
\label{eq:kaprel}
\end{equation}
 \begin{align}
\hat{\gamma}_{\ell m} &= \hat{\gamma}_{E, \ell m} + i \hat{\gamma}_{B, \ell m} \notag \\
&= \frac{1}{2}[\ell (\ell +1)(\ell -1)(\ell +2)]^{\frac{1}{2}}\psi_{\ell m}  \notag \\
&= - \sqrt{\frac{(\ell +2)(\ell -1)}{\ell (\ell +1)}} (\hat{\kappa}_{E,\ell m} +i \hat{\kappa}_{B,\ell m}).
\label{eq:gamphi}
\end{align} 
That is, one can convert between the three fields: $\kappa$, $\bm{\gamma}$ and $\psi$ by manipulating 
their spherical harmonics decompositions. 
The mathematical operation described above is entirely analogous to a description of linear polarisation such as that 
in the CMB polarisation measurements. In this analogy, the $Q$ and $U$ Stokes parameters correspond 
to the $\gamma^{1}$ and $\gamma^{2}$. 
In the flat-sky limit, we have $\edth \rightarrow \partial$ and the decomposition into spherical harmonics is 
replaced by the Fourier transform, 
$\Sigma \psi_{\ell m} Y_{\ell m} \rightarrow \int \frac{d^2 \ell}{(2\pi)^{2}} \psi(\bm{\ell})e^{i \bm{\ell}  \cdot \bm{\theta}}$. 
The above equations then reduce to the usual KS formalism.

One can derive the lensing deflection field $\bm{\eta}$ in a similar fashion. The lensing deflection 
field is defined as the first derivative of the lensing potential 
\begin{equation}
\bm{\eta} = \eta^{1} + i \eta^{2} = \edth \psi,
\label{eq:alpha_def}
\end{equation}
so the deflection field is a spin-1 field and can be expressed as 
\begin{equation}
\bm{\eta} = \eta^{1} + i \eta^{2} = \sum_{\ell m} \hat{\eta}_{\ell m}\; {}_{1}Y_{\ell m}.
\label{eq:alpha}
\end{equation}
Carrying through the derivation, we get 
\begin{equation}
\hat{\eta}_{\ell m} = [\ell (\ell +1)]^{\frac{1}{2}} \psi_{\ell m}, 
\label{eq:alpha_phi}
\end{equation}
which is again related to the other lensing quantities via a simple linear operation in the spin-harmonic space.
That is, once $\bm{\gamma}$ is measured, the other fields ($\kappa$, $\bm{\eta}$ and $\psi$) can be 
constructed using the formalism described above. 
 
From \Eref{eq:kaprel} and \Eref{eq:gamphi} we observe from which $\ell$ modes $\kappa$, $\bm{\eta}$ and $\psi$ 
receive their dominant contributions: $\psi$ receives most contribution 
from the lowest $\ell$ modes, $\bm{\eta}$ receives contribution 
from slightly higher $\ell$ modes, and $\kappa$ receives contribution from even higher $\ell$ 
modes. Therefore, $\kappa$ is more strongly influenced by the smaller scale effects (e.g. noise) 
and $\psi$ is affected by large scale effects (e.g. masking). This can also be seen from the fact that 
the $\kappa$ ($\bm{\eta}$) field is derived from applying a Laplacian (derivative) operator on the 
$\psi$ field, which means that the power spectrum of $\kappa$ ($\bm{\eta}$) scales like $\ell^{4}$ 
($\ell^{2}$) times the power spectrum of $\psi$.
The main focus of this paper is to construct the $\kappa$ map. 
However, we also explore the construction of the $\bm{\eta}$ and $\psi$ in \Aref{sec:deflection_test_sim} 
to show that the quality of the reconstruction for these fields is indeed sensitive to the mask on large-scales 
and less sensitive to shape noise on small scales. 
We also show that with the 1,500 deg$^{2}$ sky coverage of DES Y1, reconstructing the $\bm{\eta}$ and 
$\psi$ maps are just starting to be interesting.  
 
In practice, the main observable for weak lensing is the galaxy shape $\bm{\varepsilon}$, which in the weak lensing 
regime, is a noisy estimate of $\bm{\gamma}$. When averaged over a large number of galaxies, 
$\langle \bm{\varepsilon} \rangle \approx  g  = \frac{\bm{\gamma}}{1-\kappa}$, where $g$ 
is the reduced shear. As $\kappa \ll 1$ in the weak lensing regime, $\bm{\varepsilon}\approx \bm{\gamma}$. 
The noise in $\varepsilon$ is dominated by the 
intrinsic shape of the galaxies, or ``shape noise'', but also includes measurement noise. That is,
\begin{equation}
\bm{\varepsilon} = \bm{\gamma} + \bm{\varepsilon}_{\rm int} +  \bm{\varepsilon}_{\rm m}, 
\label{eq:shapenoise}
\end{equation} 
where $\bm{\varepsilon}_{\rm int}$ is the intrinsic shape of the galaxy and $\bm{\varepsilon}_{\rm m}$ 
is the error on the measured shape due to the measurement. One often quantifies the combined 
effect of $\bm{\varepsilon}_{\rm int}$ and $\bm{\varepsilon}_{\rm m}$ using $\sigma_{\varepsilon}$, 
the standard deviation of the distribution of $\bm{\varepsilon}_{\rm int}+\bm{\varepsilon}_{\rm m}$. 
As we will see in \Sref{sec:method}, one needs to average $\bm{\varepsilon}$ over a large number of 
galaxies to suppress this noise. Note that here we have not considered the effect of intrinsic alignment 
\citep[IA,][]{Troxel2015, Blazek2015}, where $\langle \bm{\varepsilon} \rangle \approx  g $ no longer holds. 

\section{Data and simulations}
\label{sec:data}

DES  is an ongoing wide-field galaxy and supernova survey that began in August 2013 
and aims to cover a total of 5000 deg$^{2}$ in five filter bands ($grizY$) to a final median depth 
of $g\sim$24.45, $r\sim$24.3, $i\sim$23.5, $z\sim$22.9, $Y\sim$21.7 \citep[10-$\sigma$ PSF limiting magnitude, see][]{DES2016}
at the end of the survey. The survey instrument is the Dark Energy Camera \citep{Flaugher2015} 
installed on the 4m Blanco telescope at the Cerro Tololo Inter-American Observatory (CTIO) in Chile. 
This work is based on the DES first-year cosmology data set (Y1A1 GOLD) including photometrically 
calibrated object catalogs and associate ancillary coverage and depth maps \citep{DrlicaWagner2017}. 
We focus on
the Southern footprint of the DES Y1 data, which overlaps 
with the South Pole Telescope survey \citep{Carlstrom2011}. 
This is the largest contiguous area in the 
Y1 data set and ideal for constructing weak lensing mass maps. We briefly describe below the data 
products and simulations used in this work. 

\subsection{Photometric redshift (photo-$z$) catalog}

We use the photometric redshifts (photo-$z$'s) derived using a code closely following 
the Bayesian Photometric Redshifts (\textsc{BPZ}) algorithm developed in 
\citet{Benitez2000} and \citet{Coe2006}. \textsc{BPZ} is a template-fitting code using templates from 
\citet{Coleman1980,Kinney1996,Bruzual2003}. 
The catalog generation in Y1 is similar to the SV analysis \citep{Bonnett2016}, but with several 
improvements described in \citet{Hoyle2017}.

BPZ calculates a redshift PDF for each galaxy in that sample. The mean of this PDF, 
$z_{\rm mean}$, is used to place source galaxies into redshift bins, 
while the $n(z)$ for each of the samples is estimated by randomly drawing a redshift from the PDF of each galaxy. 
These $n(z)$'s are validated in \citet{Hoyle2017} 
using two orthogonal methodologies: comparison with precise redshifts 
and clustering-based inference, see \cite{Hoyle2017, Cawthon2017, Gatti2017, Davis2017}. 

\subsection{Weak lensing shape catalogs}
\label{sec:shear}

Two DES Y1 weak lensing shape catalogs are used in this paper ---  the \metacal catalog based on 
\cite{Huff2017} and \cite{Sheldon2017}, and the \imshape catalog based on \citet{Zuntz2013}. Both catalogs have been 
tested thoroughly in \citet[][hereafter Z17]{Zuntz2017}. 
Given that the two algorithms are fundamentally different and that the pipelines were developed independently, 
obtaining consistent results from the two catalogs is a non-trivial test of the catalogs themselves.       

Briefly, the \metacal algorithm relies on a self-calibration framework using the data itself, instead of a large number 
of image simulations as is used in many other algorithms \citep[e.g. \imshape,][]{Bruderer2016, FenechConti2016}. 
The basic idea is to apply a small, known shear on the \textit{deconvolved} galaxy images in different directions 
and re-measure the 
post-shear reconvolved galaxy shapes. Since the input shear is known, the change in the measured galaxy shapes due to the 
artificial shearing gives a direct measure of how the shear estimators responds to shear. This quantity is referred 
to as the \textit{response}. In addition, selection effects\footnote{Here we refer to the fact that the response is 
different when one selects a subsample of the galaxies based on signal-to-noise, sizes, redshift etc..} can be 
easily corrected in this framework by measuring the response for the full sample and for the subsample. 
The final signal-to-noise and size selection for the catalog is S/N$>$10 and $T/T_{PSF}>0.5$ ($T$ and $T_{PSF}$ 
are the sizes of the galaxy and the PSF, respectively). 
Following Z17, the residual systematic errors are quoted in terms of $m$ (the multiplicative bias), $\bm{\alpha}$ 
(the additive bias associated with the PSF model ellipticity $\bm{\varepsilon}_{\rm PSF}$) and $\bm{\beta}$ (the 
additive bias associated with the errors on the PSF model ellipticity $\delta \bm{\varepsilon}_{\rm PSF}$). 
For \metacal, Z17 estimated $m=0.0\pm1.2\%$, 
$\bm{\alpha}\sim 0$, and $\bm{\beta} \sim -1$. In \citet{Troxel2017}, however, it is 
found that the $\bm{\beta}$ correction has very little effect on the final measurements. We therefore do not 
correct for $\bm{\beta}$ when making the mass maps. We have also checked that setting $\bm{\beta}=-1$ leads 
to negligible changes in the second moments of the map. This selection gives a total of 
$\sim34,800,000$ galaxies in the full Y1 catalog. The shear measurement method in \metacal is based on the 
\textsc{ngmix} method \citep{Sheldon2014}. The full implementation of \metacal is publicly available and hosted 
under the \textsc{ngmix} code repository\footnote{\url{https://github.com/esheldon/ngmix}}.

The \imshape algorithm is one of the algorithms also used in the DES SV analyses \citep{Jarvis2015}. It is a 
maximum likelihood fitting code using the Levenberg-Marquardt minimisation that models the galaxies either as 
an exponential disk or a de Vaucouleurs profile --- fitting is done with both models and the one with a better likelihood 
goes into the final catalog. 
Calibration of bias in the shear estimate associated with noise \citep{Kacprzak2012, Refregier2012} is 
based on the image simulation package \textsc{GalSim}\footnote{\url{https://github.com/GalSim-developers/GalSim}}, but 
is significantly more complex and incorporates many effects seen in the DES Y1 data as described in 
Z17 and Samuroff et al. (2017). 
The final signal-to-noise and size selections are $12<S/N<200$ and $1.13<R_{gp}/R_{p}<3$, where $R_{gp}$ is the 
size of the galaxy and $R_{p}$ is the size of the PSF. 
The catalog has an estimated $m\sim 0.0 \pm 2.5\%$, $\bm{\alpha}\sim0$, and $\bm{\beta}\sim-1$. Similar to \metacal, 
we do not correct for $\bm{\beta}$ as \citet{Troxel2017} showed that the correction has a negligible effect on the 
measurements.
The final catalog contains $\sim21,900,000$ galaxies. The lower number relative to \metacal is due to the 
fact that \imshape operates on $r$-band images while \metacal use all images from the bands $r$,$i$ and $z$. 
The \imshape code is publicly available\footnote{\url{https://bitbucket.org/joezuntz/im3shape/}}.

Details for both shape catalogs and the tests performed on these catalogs can be found in Z17. 
We mainly show results for \metacal as it has the higher S/N, but also constructed \imshape maps and performed 
several systematics tests with these. Also, as noted above, we only use the SPT wide-field region with Dec$<-35$ as 
it has been the region where most 
testing was done for both the shear and the photo-$z$ catalogs. We generate 5 maps for each catalog with different 
source $z_{\rm mean}$ selections: $0.2<z<1.3$, $0.2<z<0.43$, $0.43<z<0.63$, $0.63<z<0.9$, $0.9<z<1.3$. The first redshift 
bin combines galaxies in a broad redshift range to allow for a large source number density and therefore higher 
signal-to-noise for the mass maps. This is the map with which most quantitative studies are done in this paper. The other 
four redshift bins match those defined by \citet{Troxel2017}, which are well-tested samples that meet the criterion for 
cosmic shear measurements. These maps are noisier, but allow us to explore the 3D tomographic aspect of the maps. 
Basic characteristics of the samples associated with the five maps are listed in \Tref{tab:data_sample} and Table 1 of 
\citet{Troxel2017}.

Finally, both shear catalogs were blinded with a multiplicative factor during the entire analysis and only unblinded 
after all tests were finalised. See Z17 for the detailed blinding procedure.

\subsection{Flux-limited galaxy catalog}
\label{sec:lss}

In \Sref{sec:crosscorr} we use a flux-limited galaxy sample as a tracer of the foreground mass of the mass maps. 
This sample is constructed to be a simple, 
clean flux-limited sample from the DES Y1 catalog \citep{DrlicaWagner2017}, 
which is easier to compare with simulations as it puts less pressure on having other 
galaxy properties (colour, galaxy type) in the catalogs being matched to the data.   

The catalog is built by applying the following selections to the DES Y1 catalog: 
$17.5<i<22.0$; $-1<g-r<3$, $-1<r-i<2.5$ and $-1<i-z<2$ 
to remove galaxies that potentially have very incorrect photo-$z$'s; \texttt{flags\_gold$=$0} to remove any blended, 
saturated, incomplete or problematic galaxies; \texttt{flags\_badregion $\leq$3} to remove problematic regions with 
e.g. high stellar contamination; \texttt{modest\_class$=$2} to select objects as galaxies. The full catalog 
contains $\sim$34,800,000 objects, to which we further impose photo-$z$ cuts to construct two samples, $0.2<z<0.4$ 
and $0.4<z<0.6$, together with a cut in Dec$<-35$ to select the SPT region. The two samples are then pixelated into 
\healpix maps of ${\rm nside}=2048$ using the associated masks, which is then used for computing the cross-correlation.

\subsection{Simulations}
\label{sec:sims}

Two types of simulations are used in this work to investigate the performance of the convergence map 
reconstruction and the effects of noise and masking. First, we generate fully sampled, Gaussian maps 
with a given power spectrum using the \texttt{synfast} routine in \textsc{Healpix} \citep{Gorski2005}. 
We use the software package \textsc{Cosmosis} \citep{Zuntz2015}, which wraps around the \textsc{CAMB} software 
\citep{Lewis2002}, to generate the input power 
spectrum with the cosmological parameters: $\Omega_{m}=0.295$, $\Omega_{b}=0.047$, $\sigma_{8}=0.8$, 
$h=0.69$, $n_{s}=0.97$, and $w=-1$, although the particular details of the power spectrum are not very 
important for the tests we perform with these Gaussian simulations. 

Second, we use the ``Buzzard v1.3'' mock galaxy catalogs based on N-body simulations as described in 
DeRose et al. (in prep). Briefly, three flat $\Lambda$CDM dark-matter-only N-body simulations were used, 
with $1050^3$, $2600^3$ and $4000^3$ $Mpc^3 h^{-3}$ boxes and $1400^3$, $2048^3$ and $2048^3$ particles, 
respectively. These boxes were run with \textsc{LGadget-2} \citep{Springel2005} with \textsc{2LPTic} initial conditions 
from \citep{Crocce2006} and \textsc{CAMB}. The cosmology assumed was $\Omega_{m}=0.286$, 
$\Omega_{b}=0.047$, $\sigma_{8}=0.82$, $h=0.7$, $n_{s}=0.96$, and
$w=-1$ (consistent with the best-fit cosmological parameters from the DES
Y1 3$\times$2-pt anaylsis \citep{DES2017}. Particle lightcones were created from 
these boxes on the fly. 
Galaxies were then placed into the simulations 
and $grizY$ magnitudes and shapes are assigned to each galaxy using the algorithm Adding Density Determined 
Galaxies to Lightcone Simulations 
(ADDGALS, Wechsler et al. in prep., DeRose et al. in prep.). Galaxies are assigned to dark matter particles and 
given r-band absolute magnitudes based on the distribution $p(\delta|M_{r})$ measured from a high resolution 
simulation populated with galaxies using subhalo abundance matching (SHAM) \citep{Conroy2006, Reddick2013}, 
where $\delta$ is a large scale density proxy. 
Each galaxy is assigned an SED from SDSS DR6 \citep{Cooper2006} by finding neighbors in the space of 
$M_{r}-\Sigma_{5}$, where $\Sigma_{5}$ is the projected distance to the fifth nearest neighbor in redshift slices of 
width $\delta z = 0.02$. 
These SEDs are k-corrected and integrated over the appropriate bandpasses to generate $grizY$ magnitudes.

Finally, the weak lensing parameters ($\kappa$ and $\bm{\gamma}$) in the simulations are based on the ray-tracing 
algorithm Curved-sky grAvitational Lensing for Cosmological Light conE simulatioNS \citep[CALCLENS;][]{Becker2013} 
which builds on previous work by \citet{Gaztanaga1998} and \citet{Fosalba2008} to make all sky weak lensing maps 
from projected density fields in simulations. 
The ray-tracing resolution is accurate to $\simeq 6.4$ arcseconds.
The catalogs were then post-processed and trimmed to match the quality of our data sample. This 
includes adding photometric noise using the DES Y1 depth map, running the same photo-$z$ pipeline 
(\textsc{BPZ}) on the photometry, adding shape noise\footnote{The Buzzard catalogs include shape noise that 
are modeled from external Subaru data sets, which are not fully representative of our data. In order to have a 
better matching between simulation and data, we instead randomly draw the galaxy shapes from the \metacal 
catalog and add the simulated shear to the galaxy shape.}, imposing redshift, size, signal-to-noise cuts 
to match the shear catalog described in \Sref{sec:shear} (here the cuts are tailored to the \metacal catalog) and 
the flux-limited galaxy catalog described in \Aref{sec:lss}. 
We note, however, that due to the setup of the simulation box, the footprint of the simulations 
is 26\% smaller than the data, with the area of RA$>100^{\circ}$ removed. For the purpose 
of testing in this work, this does not impose a significant problem. We also note that the galaxy number density 
is 20\% lower than our data set. To account for that, we scale the shape noise by a factor of 
$\sqrt{n_{\rm gal, Buzzard}/n_{\rm gal,DES}}$, where $n_{\rm gal, Buzzard}$ and $n_{\rm gal, DES}$ are the number density 
of source galaxies in the simulations and data respectively.  

\section{Methodology}
\label{sec:method}

We describe here the steps taken to construct the convergence map for the two shear catalogs. The only difference 
between the two catalogs is that different calibration schemes are applied to the shear estimates prior to making the 
maps.

All the maps are constructed using \healpix pixelisation, which is a natural choice for map making on the sphere and 
includes the necessary tools to manipulate the data on a sphere. This includes the decomposition of the spin fields into 
spin harmonics, 
which is essential for the transformation between shear $\bm{\gamma}$, convergence $\kappa$, the lensing potential 
$\psi$ and the deflection angle $\bm{\eta}$, as we outlined in \Sref{sec:formalism}. We use a \healpix map of 
${\rm nside}=1024$, which approximately corresponds to a mean pixel spacing of 3.44 arcminutes. This resolution is chosen 
based on the 
density of our source galaxies, and provides a good balance between the resolution of the map and the simplicity of the 
mask geometry. 
As the completeness of the source galaxies is not a concern here, no additional cuts on e.g. depth are needed beyond 
the selection from the shear catalog. 
The mask is defined to be 1 where there are galaxies within the pixel and 0 where there are no galaxies. This yields a 
total map area of $\sim 1,500$ deg$^{2}$, which appears larger than the naive footprint of our data in the SPT region 
\citep[$\sim 1,300$ deg$^{2}$,][]{Troxel2017}. This is because we are using a coarser pixel resolution than what is 
used to estimate the footprint (${\rm nside}=4096$).  
The final mask used in this paper still contains small ``holes'' in the otherwise contiguous footprint. 
We have considered interpolating over the holes to prevent edge effects, but found that these 
procedures make little difference in the reconstruction in terms of our metric defined in \Sref{sec:sim_test} 
(we give more details about these tests on \Aref{sec:inpainting}).

The first step in the reconstruction of the mass map involves making pixelised ellipticity (or shear estimate) 
maps $\varepsilon^{1}$ 
and $\varepsilon^{2}$ from the galaxy shape catalogs. To do this, we follow the procedure outlined in Section 7 
of Z17 for calculating the mean shear per pixel. Note that both the response $R$ for the \metacal catalog and the 
multiplicative noise-bias calibration (NBC) factor $m$ for the \imshape catalog are noisy within each pixel of 
our maps. We therefore use the mean $R$ and $m$ values for each sample instead of calculating them in each 
pixel when constructing the maps. That is, for \metacal, we have
\begin{equation}
\varepsilon^{\nu}_{i} = \frac{\sum_{j}^{n_{i}}\varepsilon^{\nu}_{ij}}{n_{i} \bar{R}^{\nu}}, \; \nu=1,2,
\end{equation} 
where $n_{i}$ is the number of source galaxies in pixel $i$ and $\varepsilon^{\nu}_{ij}$ is the shape estimate 
of each individual galaxy $j$ in that pixel. $\bar{R}^{\nu}$ is the mean response of the full sample. 
The $\bar{R}^{\nu}$ values vary from $\sim0.7$ to $\sim0.5$ going from low to high redshift. 
Typically 1--2\% of $\bar{R}^{\nu}$ comes from the correction of the selection effects.
For the \imshape, 
we have
\begin{equation}
\varepsilon^{\nu}_{i} = \frac{1}{n_{i}(1+\bar{m})}\frac{\sum_{j=1}^{n_{i}}(\varepsilon^{\nu}_{ij} - c^{\nu}_{ij}) w_{ij} }{\sum_{j=1}^{n_{i}} w_{ij}}, \; \nu=1,2, 
\end{equation}
where $c^{\nu}_{ij}$ and $w_{ij}$ are the additive NBC factor and weight for galaxy $j$ in pixel $i$,
and $\bar{m}$ is the average multiplicative NBC factor for each sample. 
Typical $m$ values range from -0.08 to -0.18 going from low to high redshift. 
We then subtract the mean shear for each sample from the maps as suggested by Z17 Section 7.1. 
 
\begin{table}
\begin{center}
\caption{Characteristics of the source galaxy samples and the maps. The number preceding the semicolon 
is for the \metacal catalog while the number after the semicolon is for the \imshape catalog. $\bar{z}$ is the 
mean redshift estimate from \textsc{BPZ} for each sample, while $\sigma_{\varepsilon}$ is the mean 
of $\sigma_{\varepsilon^{1}}$ and $\sigma_{\varepsilon^{2}}$, the standard 
deviation of the weighted galaxy shapes reported from the catalogs \citep[see last column in Table 1 of][]{Troxel2017}. 
The area of the map is $\sim$1,500 deg$^{2}$ for both \metacal and \imshape, where the exact size 
changes slightly from different photo-z bins and shear catalogs. 
The \healpix maps have a resolution of nside$=1024$.}
\begin{tabular}{ccc}
\hline
Redshift range & $\bar{z}$ &$\sigma_{\varepsilon}$ \\ \hline 
$0.2<z<1.3$  & 0.60; 0.56   &  0.28; 0.27 \\ 
$0.2<z<0.43$ & 0.38; 0.36  & 0.26; 0.26  \\ 
$0.43<z<0.63$  & 0.51; 0.52  & 0.30; 0.28  \\ 
$0.63<z<0.9$  & 0.74; 0.75  & 0.27; 0.24  \\ 
$0.9<z<1.3$  & 0.96; 1.03   &  0.28; 0.26 \\ 
\hline
\end{tabular}
\label{tab:data_sample}
\end{center}
\end{table}

Next, we perform the spin transformation which converts the ellipticity maps (which combine to form a spin-2 field 
$\varepsilon^{1}+i \varepsilon^{2}$) into a curl-free E-mode convergence map $\kappa_{E}$ and a divergence-free B-mode 
convergence map $\kappa_{B}$. The \healpix functions \texttt{map2alm} performs this decomposition in spherical 
harmonic space and returns the E- and B-mode coefficients, which are equivalent to the 
$\hat{\gamma}_{E, \ell m}$ and $\hat{\gamma}_{B, \ell m}$ in \Eref{eq:gamphi}. We calculate $\hat{\kappa}_{E, \ell m}$ 
and $\hat{\kappa}_{B, \ell m}$, then use the \healpix function 
\texttt{alm2map} to convert these coefficients back to the real space $\kappa_{E}$ and $\kappa_{B}$ maps. Similarly, $\psi$ 
and $\bm{\eta}$ maps can be constructed using \Eref{eq:gamphi} and \Eref{eq:alpha_phi}.

For all the convergence map visualisation in this paper, we further smooth the maps with a Gaussian kernel. The noise 
associated with each pixel after smoothing can be calculated through \citep{VanWaerbeke2013}
\begin{equation}
\sigma^{2}_{\kappa} = \frac{\sigma_{\varepsilon^{1}}^{2}+\sigma_{\varepsilon^{2}}^{2}}{4 \pi \sigma_{G}^{2} n_{\rm gal}},
\label{eq:noise}
\end{equation}
where $\sigma_{\varepsilon^{1}}$ and $\sigma_{\varepsilon^{2}}$ are the standard deviation of the two components 
for the measured galaxy shapes, 
$\sigma_{G}$ is the width of the Gaussian filter used to smooth the maps, and $n_{\rm gal}$ is the number 
density of the source galaxies.  

Finally, for all measurements in this work, we estimate the error bars and the covariance matrix using a 
standard Jackknife approach. We divide the footprint into $N_{JK}$ Jackknife regions using a kmeans 
clustering code\footnote{\url{https://github.com/esheldon/kmeans_radec}} and divide the mask into 
$N_{JK}$ approximately equal-area regions. Throughout this paper, we use $N_{JK}=100$.
For angular correlation measurements, a fast 
tree-based code \textsc{treecorr}\footnote{\url{https://github.com/rmjarvis/TreeCorr/wiki}} is used.

\begin{figure*}
\centering
\includegraphics[width=0.99\linewidth]{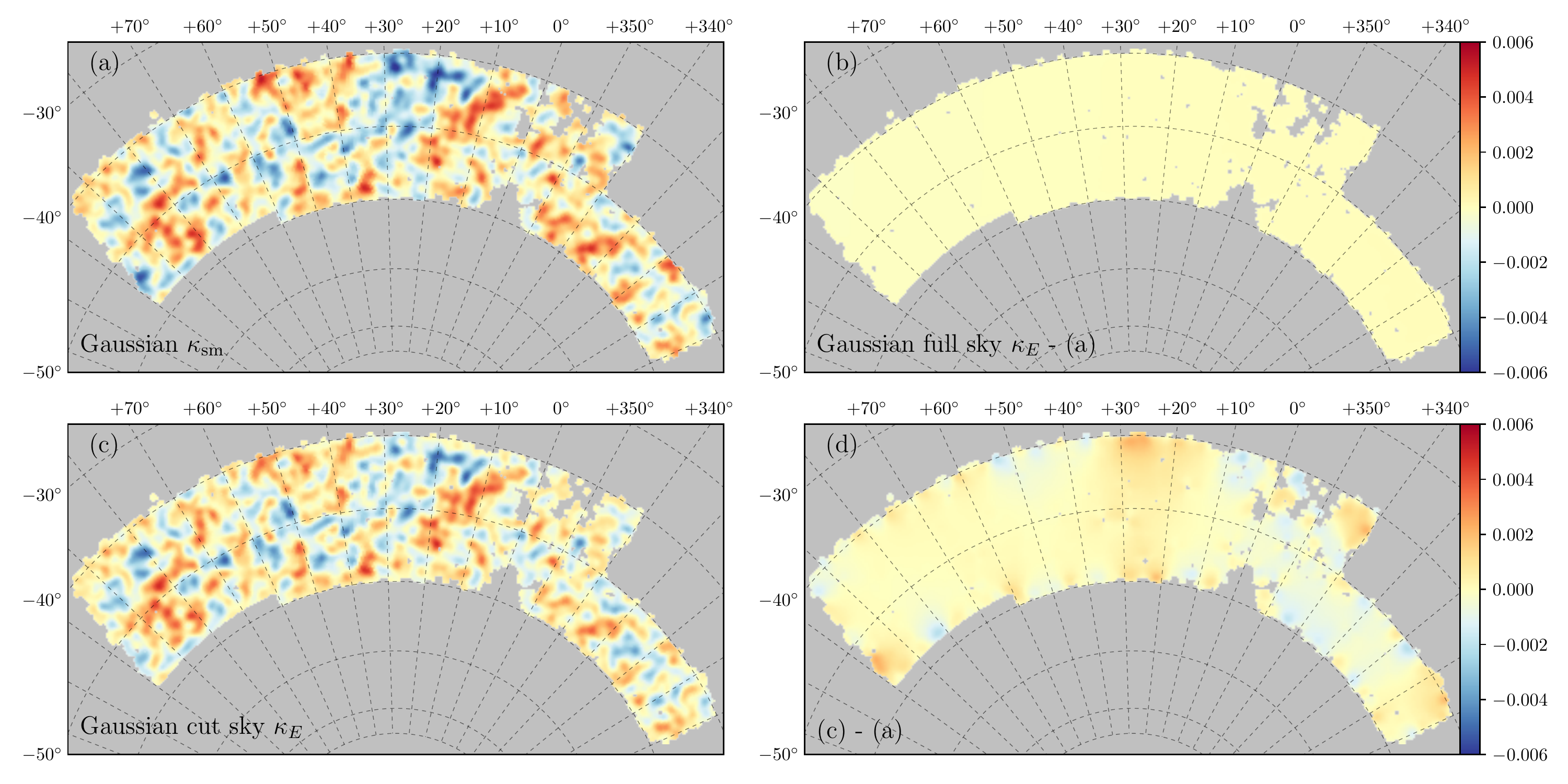}
\caption{Tests from maps of simulated Gaussian fields. All maps are 
smoothed with a Gaussian filter of $\sigma_{G}=$30 arcminutes, mean-subtracted and projected onto a conic 
projection. Panel (a) shows the original Gaussian $\kappa_{\rm sm}$ map; panel (b) is the difference map between 
the full-sky reconstructed $\kappa_{E}$ map and panel (a); panel (c) shows the cut-sky reconstructed $\kappa_{E}$ 
map, and panel (d) shows the difference map between panel (c) and panel (a). The series of maps shows that 
the reconstruction on the edges is degraded when introducing the mask.}
\label{fig:sim_map_synfast}
\end{figure*}

\section{Simulation tests}
\label{sec:sim_test}

In this section we present a series of simulation tests to validate our procedure for map generation and quantify 
the uncertainties associated with the various source of systematics and noise. 
We start with an idealised setup of a Gaussian, fully-sampled, full-sky map in \Sref{sec:synfast} to quantify 
the errors associated with the shear-to-convergence conversion alone, then we impose a DES Y1-like mask 
to investigate the degradation introduced by the mask. Next in \Sref{sec:bcc}, we repeat the tests in 
\Sref{sec:synfast} with a mock galaxy catalog based on an N-body simulation. We test the effect of shot noise 
(finite sampling) and shape noise. 

For both \Sref{sec:synfast} and \Sref{sec:bcc}, we quantify the quality of the 
reconstruction using the following statistics: 
\begin{equation}
F_{1} = \sqrt{ \frac{\langle \kappa_E^2 \rangle}{\langle \kappa_{\rm sm}^2 \rangle}}; \; 
F_{2} = \frac{\langle \kappa_E \kappa_{\rm sm} \rangle}{\langle \kappa_{\rm sm}^2 \rangle},
\label{eq:f1f2}
\end{equation}
where $\kappa_E$ is the reconstructed map, $\kappa_{\rm sm}$ is the true convergence map degraded 
to the same resolution as $\kappa_E$ (see \Sref{sec:synfast} for details), $\langle X Y \rangle$ is the 
zero-lag cross-correlation between two maps $X$ and $Y$, or 
\begin{equation}
\langle X Y \rangle = \frac{1}{N}\sum_{i=1}^{N} X_{i}Y_{i}.
\end{equation}
The index $i$ runs over all pixels in the map where the pixels are not masked. $F_{1} $ is the square-root of the ratio 
of the second moments of the map. $F_{2}$ on the other hand, tests in addition that the phases (in addition 
to the amplitudes) of the map are reconstructed correctly, or in other words, that the patterns in the maps are correctly 
reconstructed. $F_{1}$ and $F_{2}$ are designed to have the same units as $\kappa_E /\kappa_{\rm sm}$. 
We require that for our final reconstruction (including all noise and systematics effects) of both $F_{1}$ and $F_{2}$ 
be consistent with 1 within the 2$\sigma$ measurement errors. In \Aref{sec:deflection_test_sim}, we perform a subset 
of the tests above on reconstructing the lensing potential and deflection field described in \Sref{sec:formalism}. 

In \Sref{sec:pdf}, we take the maps in \Sref{sec:bcc} one step further and examine the PSF of the maps and the 
second and third moments as a function of smoothing. We require the reconstructed map to have second and 
third moments consistent with expectation from simulations within 2$\sigma$ of the measurement errors, which 
then assures that the reconstruction preserves the distribution of structures on different scales.

We note that the requirements on the reconstruction performance depends on the specific application. 
Passing the requirements on $F_{1}$, $F_{2}$ and the moments means that the mean variance, phase, 
and distribution of power on different scales (on the scales we tested) in the maps are robust. Extending 
to further applications would require additional tests. 

\subsection{Gaussian lensing convergence maps}
\label{sec:synfast}

We consider a set of full-sky, noiseless, Gaussian lensing maps ($\bm{\gamma}$ and $\kappa$) generated 
using the \healpix routine \texttt{synfast}. These maps are constructed using an input lensing power spectrum for 
a flat $\Lambda$CDM model with 
cosmological parameters: $\Omega_{m}=0.3$, $h=0.7$, $\Omega_{b}=0.047$, $\sigma_{8}=0.82$, $w=-1$. The 
source redshift distribution $n_{s}(z)$ is approximately matched to the redshift estimate of BPZ for redshift bin 
$0.2<z<1.3$ in 
\Tref{tab:data_sample}. We have chosen to demonstrate all the tests on this redshift bin since it contains the 
highest signal-to-noise. We generate the maps 
with ${\rm nside}=1024$ and $\ell_{\rm max}=2 \times {\rm nside}$. Note that the $\ell_{\rm max}$ cut is 
necessary for further \healpix manipulations, since the modes close to the pixel scale can introduce undesired noise. 
This means that these maps do not contain information on scales beyond $\ell_{\rm max}$. 
The \texttt{synfast} routine outputs three maps that are consistent with the input power spectrum: a 
spin-0 map and two maps for the two components of the spin-2 field. We can then identify the spin-0 map as the 
convergence map $\kappa_{\rm sm}$ and 
the spin-2 maps as the shear maps $\bm{\gamma}_{\rm sm}$. Since all the lensing maps are effectively smoothed, 
we use the `sm' subscript to distinguish these maps (which do not contain information on scales beyond $\ell_{\rm max}$) 
from the true underlying field with infinite resolution. We denote $\kappa_{E}$ and $\kappa_{B}$ to be the E- and B-mode 
convergence generated from the smoothed shear maps $\bm{\gamma}_{\rm sm}$.

\begin{figure*}
\centering
\includegraphics[width=0.8\linewidth]{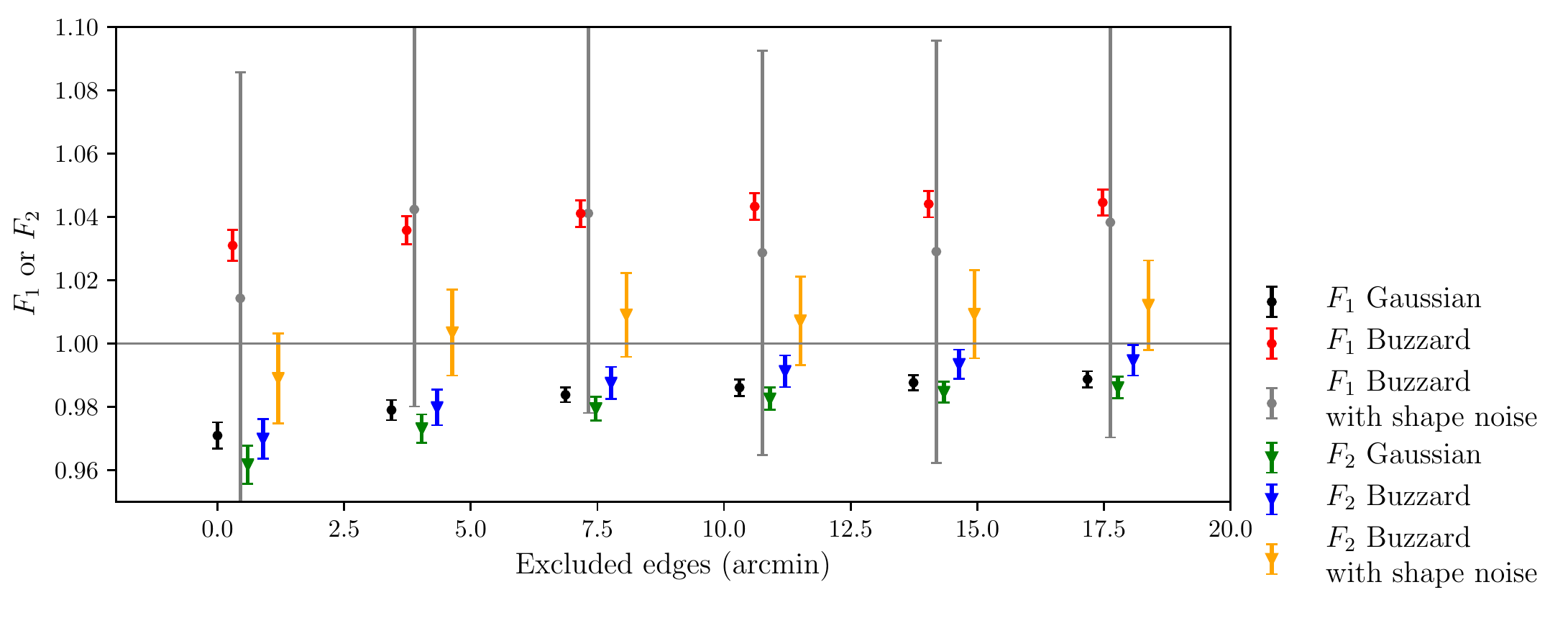}
\caption{$F_1 = \sqrt{ \langle \kappa_E^2 \rangle / \langle \kappa_{\rm sm}^2 \rangle}$ (circle markers) and 
$F_2 = \langle \kappa_E \kappa_{\rm sm} \rangle/\langle \kappa_{\rm sm}^2 \rangle$ (triangle markers) for the 
reconstructed convergence map from the Gaussian simulations and the Buzzard mock galaxy catalogs. 
$F_1$ measures how well the variance of the map is reconstructed, while $F_2$ measures in addition 
how well the phase information is reconstructed. $F_{1}=F_{2}=1$ means perfect reconstruction. 
The plot shows how $F_{1}$ and $F_{2}$ 
changes when we exclude pixels within a certain distance from the edge of the mask. The larger the exclusion, 
the less effected the reconstruction is from the edge effects.}
\label{fig:edges}
\end{figure*}

\begin{figure*}
\centering
\includegraphics[width=0.99\linewidth]{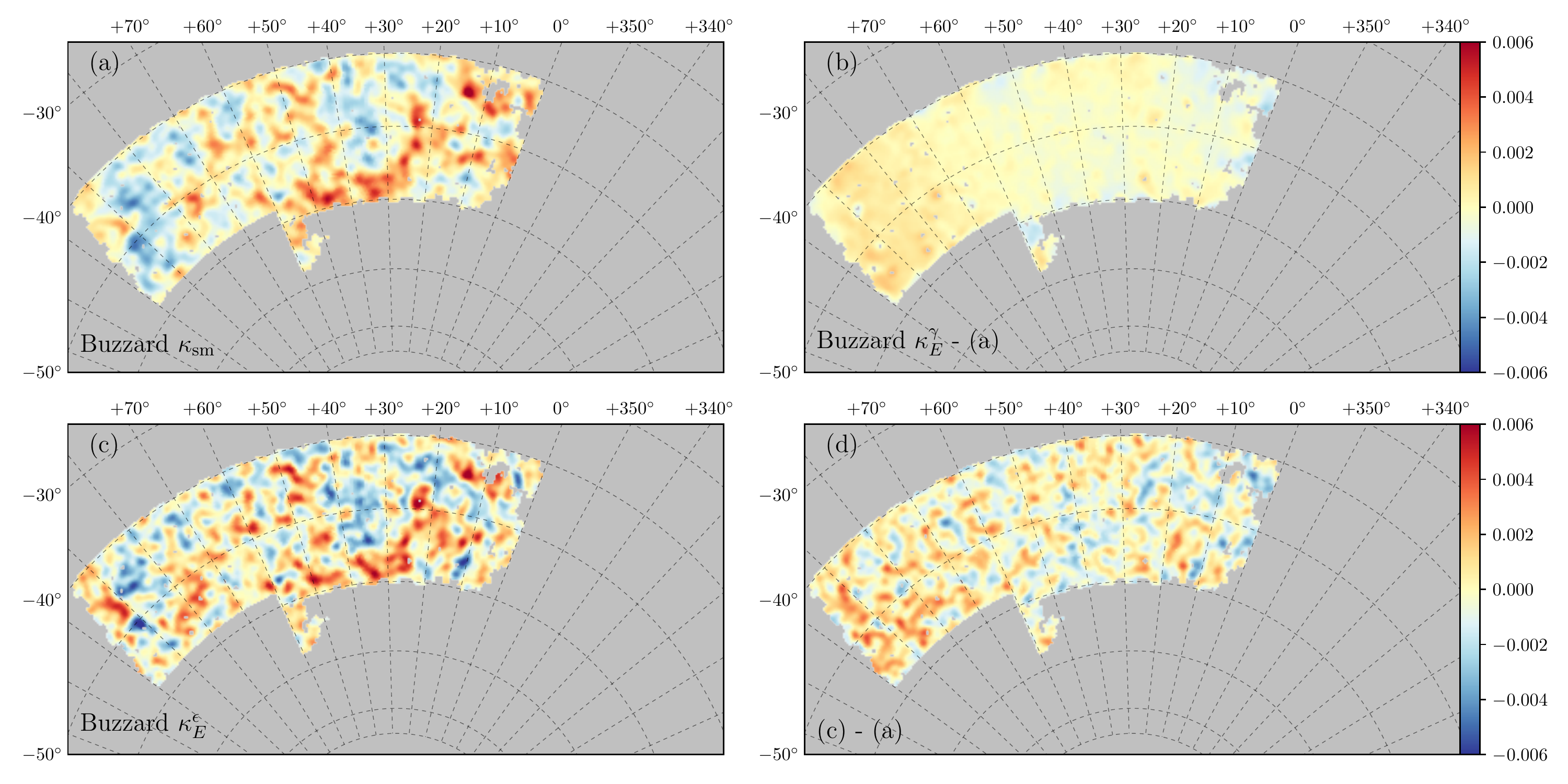}
\caption{This figure is similar to \Fref{fig:sim_map_synfast} but using the Buzzard mock galaxy simulations. Panel (a) 
shows the original Buzzard $\kappa_{\rm sm}$ map; panel (b) shows the difference between the reconstructed 
$\kappa_{E}^{\gamma}$ map (without shape noise) and panel (a); panel (c) shows the reconstructed 
$\kappa_{E}^{\varepsilon}$ map (with shape noise), and panel (d) shows the difference map between panel (c) 
and panel (a).}
\label{fig:sim_map_buzzard}
\end{figure*}

For visualisation purpose, all maps presented in this paper are first smoothed with a Gaussian filter of $\sigma_{G}=$30 
arcminutes, then mean-subtracted\footnote{Since lensing reconstruction is only valid up to a constant offset, we subtract the mean to 
avoid this constant additive bias.}, and finally projected onto a plane with Albers equal-area conic using the code 
\textsc{SkyMapper}\footnote{\url{https://github.com/pmelchior/skymapper}} (for quantitative analyses later we use the raw map themselves). 
The smoothing scale is chosen so that the highest peaks in the E-mode S/N maps have S/N values greater than $\sim3$ 
and so that one can clearly see the difference between the E and B mode maps. Each of the panels in \Fref{fig:sim_map_buzzard} are 
described below:
\begin{itemize}
\item Panel (a): noiseless $\kappa_{\rm sm}$ map directly from \texttt{synfast}, cutout in the Y1 footprint. 
\item Panel (b): subtracting panel (a) from a full-sky, fully sampled, noiseless $\kappa_{E}$ reconstruction. This shows that 
in this ideal situation, the reconstructed $\kappa_{E}$ is able to recover $\kappa_{\rm sm}$ very well with negligible 
residuals, validating our basic implementation of the shear-to-convergence transformation. 
\item Panel (c): $\kappa_{E}$ reconstruction when applying the Y1 mask to the shear maps. This illustrates overall 
good reconstruction of the spatial pattern of the maps compared to panel (a). As we have set the mask to zero, the 
amplitude of the $\kappa_{E}$ map is slightly lower than panel (a) at this relatively large smoothing scale.
\item Panel (d): subtracting panel (a) from panel (c). We can see 
edge effects resulting from the Y1 mask, as the pixels on the edge have less information to infer the 
convergence than the pixels in the centre of the field. In addition, the residuals are small but anti-correlated 
with the real structure, since the overall amplitude of panel (c) is lower than panel (a).

\end{itemize}

In \Fref{fig:edges} we show in black and green how $F_{1}$ and $F_{2}$ (\Eref{eq:f1f2}) change when we exclude regions up to 30 
arcminutes away from the mask edge. For $F_{1}$, we find a value $\sim0.97$ when no pixels are excluded and this 
improves up to about 0.99 
when areas 15 arcminutes around the edges are excluded. The fact that $F_{1}<1$ is because we have set the 
empty pixels to be zero, which dilutes the signal during the reconstruction. We see that $F_{2}$ behaves very similar to 
$F_1$, which confirms that the reconstruction is good to $\sim1\%$ in these ideal scenarios with only small effects 
coming from the dilution due to the edges. 
We note that the above analysis was evaluated for the map at $0^{\circ}<$RA$<100^{\circ}$ in 
order to compare to the Buzzard simulations. 

Alternative approaches to dealing with the mask and edge effects include filling in the empty pixels via a smooth 
interpolation from neighboring pixels and more sophisticated inpainting techniques \citep{Pires2009}. 
We investigate the former in \Aref{sec:inpainting} and find that it does not improve the performance of the map 
reconstruction significantly given the noise level and mask geometry of our data, while the latter is 
beyond the scope of this paper.

\subsection{Convergence maps from simulated galaxy catalogs}
\label{sec:bcc}

Next, we turn to using mock galaxy catalogs generated from N-body simulations. The main differences between 
these and the Gaussian simulations are that (1) they only sparsely sample the lensing 
fields at a given thin redshift slice, effectively introducing shot noise, (2) they are derived from a ray-traced 
lensing field which contains non-Gaussian information, and (3) as discussed in the previous section, the maps naturally 
contain a small amount of information on scales beyond $\ell_{\rm max} = 2 \times{\rm nside}$ that we cannot 
reconstruct when we enforce a $\ell_{\rm max}$ smoothing during the reconstruction. We would like to 
understand how these factors affect the reconstruction of the convergence maps. 
In this section, we mainly use the Buzzard mock 
galaxy catalogs described in \Sref{sec:sims} for testing, but we have also tested on an independent set of simulations 
\citep[the Marenostrum Institut de Ciencias de l'Espai Simulations, or the MICE simulations,]
[]{Fosalba2015, Fosalba2015b, Crocce2015} and found consistent results.

We carry out a series of tests using the convergence map generated for redshift bins that are matched to that used for 
the data (see \Sref{sec:results}). That is, we bin the galaxies using the mean redshift reported by the photo-$z$ code and 
check that the resulting $n(z)$ reported by \textsc{BPZ} is close to that of our data. Next, we make three maps using directly 
the quantities provided by the simulation:
\begin{itemize}
\item $\kappa_{\rm pix}$: convergence
\item $\bm{\gamma}_{\rm pix}$: shear
\item $\bm{\varepsilon}_{\rm pix}$: galaxy shapes.
\end{itemize}
These maps are constructed with the same resolution (${\rm nside}=1024$) as before. The subscript `pix' denotes 
pixelised quantities. 
Next, we generate several other versions of convergence maps. 
\begin{itemize}
\item $\kappa_{\rm sm}$: to ensure that all maps we compare later have the same resolution, we smooth 
the $\kappa_{\rm pix}$ map by removing all $\ell$ modes beyond $\ell_{\rm max}=2 \times {\rm nside}$;
\item $\kappa_{E}^{\gamma}$, $\kappa_{B}^{\gamma}$: E- and B-mode convergence constructed using 
shear $\gamma_{\rm pix}$;
\item $\kappa_{E}^{\varepsilon}$, $\kappa_{B}^{\varepsilon}$: E- and B-mode convergence constructed using 
galaxy shapes $\bm{\varepsilon}_{\rm pix}$.
\end{itemize}
In \Fref{fig:sim_map_buzzard} we compare visually several of these reconstructed maps: 
\begin{itemize}
\item Panel (a): $\kappa_{\rm sm}$ map from the Buzzard simulation. 
Comparing with 
panel (a) of \Fref{fig:sim_map_synfast}, one can see that the convergence map from the galaxy catalog 
has similar amplitudes and characteristic spatial patterns as the Gaussian  map. The Buzzard maps appear 
slightly more clustered, which comes from the non-Gaussian nature of these maps compared to 
the pure Gaussian simulations. 
\item Panel (b): subtracting panel (a) from the reconstructed $\kappa_{E}^{\gamma}$ map from Buzzard, which 
includes shot noise from the finite sampling from the galaxies and the Y1 mask but no shape noise. 
Similar to panel (d) of \Fref{fig:sim_map_synfast}, there is an anti-correlation of the low-level residuals 
with the true structures.  
\item Panel (c): reconstructed $\kappa_{E}^{\varepsilon}$ map from the Buzzard simulation, which includes shot noise 
from the finite sampling from the galaxies, the Y1 mask and shape noise. We 
find the amplitude of the map to be higher than the $\kappa_{\rm sm}$ map in panel (a) and that there are spurious structures 
that arise from noise which do not correspond to real structures in the $\kappa_{\rm sm}$ map. However, the resemblance of the 
$\kappa_{E}^{\varepsilon}$ map to the $\kappa_{\rm sm}$ map is still very obvious, especially the large-scale patterns in the maps. 
This suggests that despite of noise, the majority of the structures in the $\kappa_{E}^{\varepsilon}$ map are associated with real 
structures on this smoothing scale.
\item Panel (d): subtracting panel (a) from panel (c). We see more clearly the shape noise-induced small-scale noise peaks 
as well as a large scale pattern that is very similar to that in panel (b). The edge effect, in comparison, 
becomes less visible in the presence of shape noise.
\end{itemize}

\begin{figure*}
\centering
\includegraphics[width=0.95\linewidth]{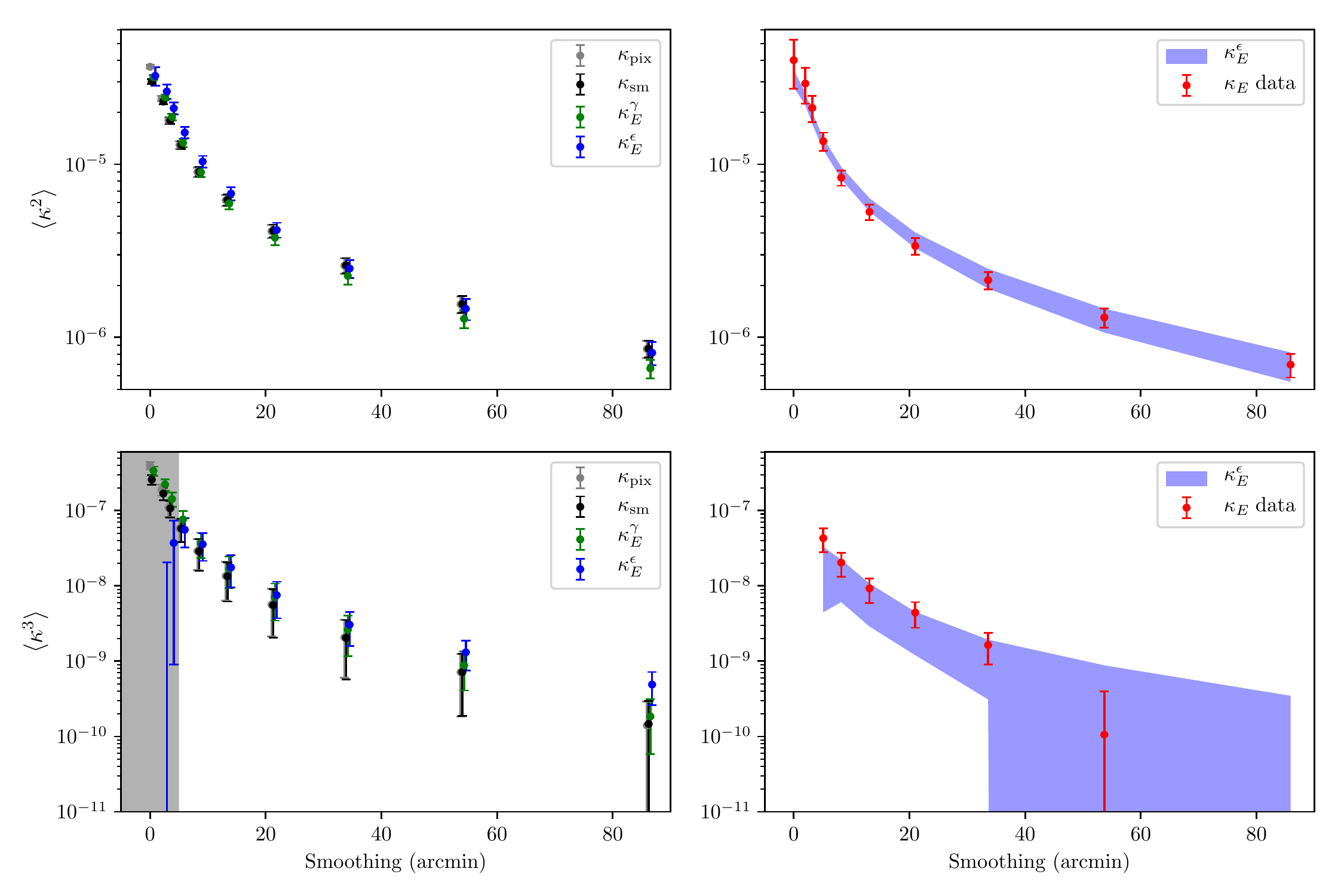}
\caption{The upper left panel shows the second moments of the maps as a function of smoothing scale for different 
$\kappa$ maps in one Buzzard simulation, from the most idealised noiseless case (grey), to two intermediate 
stages (black and green), and to the $\kappa^{\varepsilon}_{E}$ map that includes observational noise that match to 
the data (blue). The shaded 
blue band in the upper right panel shows the mean and standard deviation of the $\kappa^{\varepsilon}_{E}$ 
measurement for 12 independent Buzzard realisations. The measurement from the data is shown in red. 
The lower panels show the same as the upper panels, except for the third moments. The grey band in the lower left 
panel marks the scales that we remove for third moments analyses due to noise on small scales. 
All maps are generated for the redshift bin $0.2<z<1.3$.}
\label{fig:second_moment_sims}
\end{figure*}

In \Fref{fig:edges} we again show the $F_{1}$ (red) and $F_{2}$ (blue) statistics as a function of the area excluded 
around the mask. We find that $F_{2}$ behaves very similar to the Gaussian version shown in green, while $F_{1}$ appears 
systematically higher than the Gaussian simulations. This indicates that the reconstruction with the mock galaxy catalogs 
introduces un-correlated noise in $\kappa_{E}$, causing the overall variance in the map to be larger, while the phase 
remains the same. This additional noise comes from the finite sampling of the shear field inside each pixel --- the mean 
shear over all galaxies inside each pixel is different from the true mean shear in that pixel. This noise can be suppressed 
by smoothing the maps at a scale slightly larger than the pixel scale, as can be seen in \Fref{fig:second_moment_sims}.
Both $F_{1}$ and $F_{2}$ increase by a few percent when excluding the edges. 
When introducing shape noise, the error bars on $F_{2}$ increase, but the amplitude stays roughly unchanged, 
suggesting that on average, shape noise does not change the phase information. The raw $F_{1}$ with shape noise is 
will be dominated by shape noise in the denominator, therefore we show instead the ``de-noised'' version $F_{1}$ 
defined as $\sqrt{ (\langle \kappa_E^2 \rangle - \langle \kappa_{E, {\rm{ran}}} ^2 \rangle) / \langle \kappa_{\rm sm}^2 \rangle}$, 
where $\kappa_{E, {\rm{ran}}}$ is a convergence map constructed by randomizing the ellipticities. In the remaining of the 
paper, ``$F_{1}$ with shape noise'' refers to this de-noised quantity.

Overall, we find that at the number density and pixel resolution of this particular map ($0.2<z<1.3$), the performance 
of the reconstruction from the galaxy catalog is similar to that from the Gaussian map in terms of the effect of masking, 
though the reconstruction is noisier for the galaxy catalogs which results in a higher $F_{1}$. After including 
shape noise, both $F_{1}$ and $F_{2}$ are consistent with 1 even without exclusion of the edge pixels.
We also note that if we 
perform the same tests on a different redshift bin where the number density of galaxies is lower, the 
performance of the reconstruction using both the Gaussian and the Buzzard simulations becomes worse 
with the same pixel resolution. That is, the three factors --- resolution of the map, effect of the edges, and number density of 
the source galaxies --- are tightly coupled. If the chosen pixel resolution is sub-optimal for the data set, the reconstruction could 
be significantly biased. For example, if the pixel size is much smaller than the typical separation of source galaxies, there
will be a large number of empty pixels, which would result in a lower amplitude in the reconstructed maps. 
For our sample of the DES Y1 shear catalog, we perform quantitative studies only on the highest S/N 
map at $0.2<z<1.3$ with the pixel scale of 3.44 arcminutes. 
We test the $F_1$ and $F_2$ statistics for this map in different resolutions and find that increasing or 
decreasing the resolution by a factor of 2 in the noiseless Buzzard simulation changes $F_1$ and $F_2$ by at most 3\%.

\begin{figure*}
\centering
\includegraphics[width=0.95\linewidth]{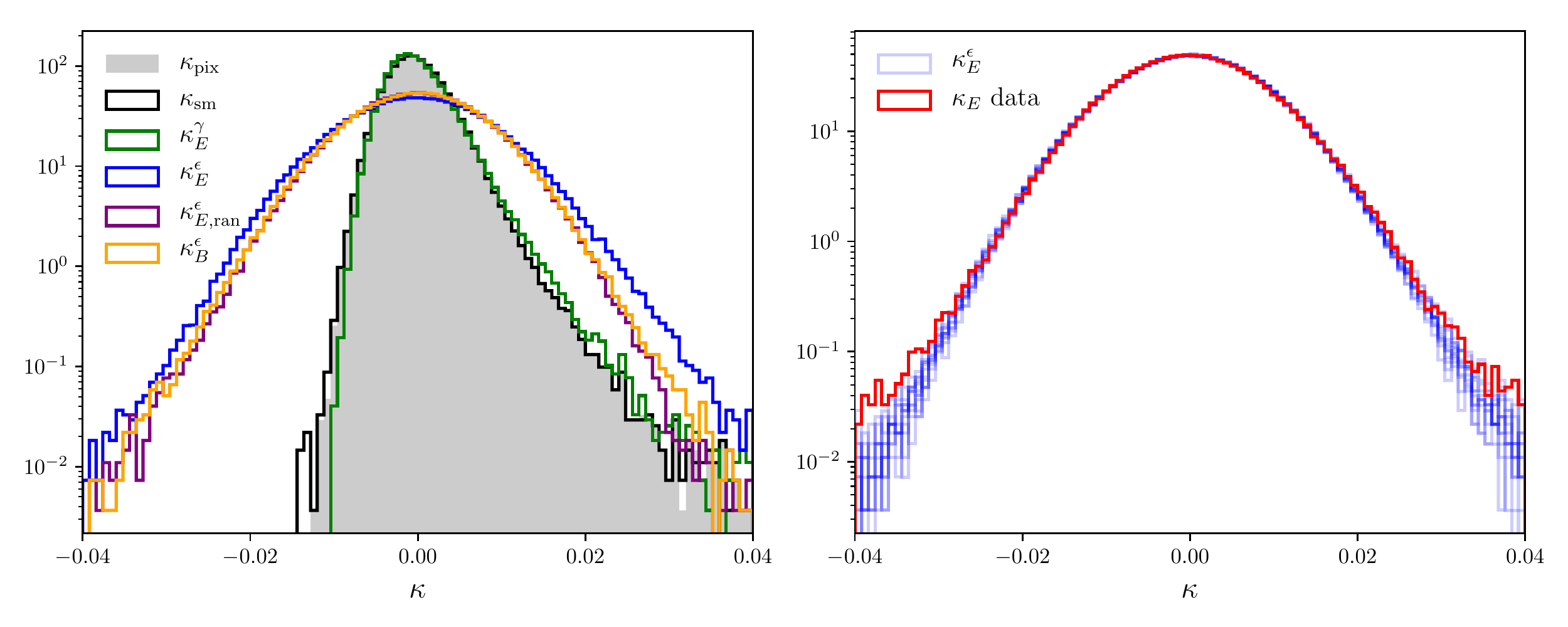}
\caption{Pixel histograms for various maps in simulation and data when smoothed with a Gaussian filter of 
$\sigma_{G}=5.1$ arcminutes. The left panel shows the pixel histograms for maps generated from one Buzzard 
simulation going from the most ideal noiseless scenario (grey) to two intermediate stages (black and green), to 
the final simulation with noise properties matched to the data (blue). Also shown are the histogram for the 
random map (purple), which is consistent with the B-mode 
map (orange). The faint blue lines in the right panel shows the histograms for $\kappa_{E}^{\varepsilon}$ in 12 
independent Buzzard realisations, while the red line shows the pixel histogram for the data $\kappa_{E}$ 
map (See Section 6.1 for discussion). All maps are generated for the redshift bin $0.2<z<1.3$.}
\label{fig:histogram}
\end{figure*}

\subsection{Moments and PDF}
\label{sec:pdf}

One final powerful test of the reconstruction is to look at the moments and the PDF of the maps. 
In this section, we examine the second and third moments of the various maps used in \Sref{sec:bcc} as we 
progressively smooth the maps on increasingly larger scales. Since these moments of the convergence 
maps as a function of smoothing scale are sensitive to cosmology \citep{Bernardeau1997, Jain1997, Jain2000}, 
it is important to verify how well the reconstructed maps preserve these characteristics. A similar test was 
performed in \citet{VanWaerbeke2013}, where they checked up to the 5th moment of the maps. We only 
consider the second and the third moments as the galaxy number density in our maps is lower compared to 
that used in \citet{VanWaerbeke2013}, and the higher moments are more sensitive to the noise in the maps. 
We begin with the set of Buzzard maps described in the previous section: $\kappa_{\rm pix}$, 
$\kappa_{\rm sm}$, $\kappa_{E}^{\gamma}$ and $\kappa_{E}^{\varepsilon}$. For each map, we smooth with a 
Gaussian filter with $\sigma_{G}=[0.0, 2., 3.2, 5.1, 8.2, 13.1, 21.0, 33.6, 53.7, 85.9]$ arcminutes, where the first case is 
equivalent to the unsmoothed map examined previously. To correct for the effect of smoothing on the edge pixels, 
we smooth the mask with the same filter and dividing the map by the smoothed mask. 
We then calculate the second moment $\langle \kappa^{2} \rangle$ and third moments $\langle \kappa^{3} \rangle$ of 
these maps for the different smoothing scales. For $\kappa_{E}^{\varepsilon}$, we follow the de-noising prescription 
described in \citet{VanWaerbeke2013}. That is
\begin{equation}
\langle(\kappa_{E, {\rm denoise}}^{\varepsilon})^2 \rangle = \langle (\kappa_{E}^{\varepsilon})^2 \rangle - \langle (\kappa_{E, {\rm ran}}^{\varepsilon})^2 \rangle, 
\end{equation}
where $\kappa_{E, \rm ran}^{\varepsilon}$ is obtained from shuffling the positions of the galaxies while keeping 
their ellipticities fixed. $\kappa_{E, \rm ran}^{\varepsilon}$ is a measure of the contribution from shape noise to 
the second moments and thus needs to be subtracted from the raw measured second moments. 

The second and third moments of the various $\kappa$ maps as a function of smoothing scale are shown in 
the left panels of \Fref{fig:second_moment_sims}. The error bars are estimated via the standard Jackknife 
approach. We find that $\kappa_{\rm pix}$ and $\kappa_{\rm sm}$ disagree slightly with 
no smoothing, 
but once a small amount of smoothing is applied, which removes the very small scale information in the 
$\kappa_{\rm pix}$ map, they agree vey well. $\kappa_{\rm sm}$ and $\kappa_{E}^{\gamma}$ are also consistent  
within the error bars, suggesting that the reconstruction does not distort the information about how the 
structures of different scales are distributed in the maps. Finally, $\kappa_{E}^{\gamma}$ and 
$\kappa_{E}^{\varepsilon}$ agree with each other within 1$\sigma$ for the second moments on all scales and for the third moments 
on scales $>5$ arcminutes. The error bars for $\kappa_{E}^{\varepsilon}$ are larger due to shape noise. We note 
that the third moment measurements on small scales are not recovered due to the noise on small scales (for a 
shear signal of 1\%, a smoothing scale of 5 arcminutes would result in an effective S/N of $\sim$0.5). We therefore 
remove scales smaller than 5 arcminutes in further analyses on the third moments. We also find 
that on scales $>40$ arcminutes, noise can cause the third moments to be negative. We repeat the measurement for 12 
independent realisations of the Buzzard simulations. The mean and standard deviation of the 12 measurements for 
$\kappa_{E}^{\varepsilon}$ are shown in the right panels of \Fref{fig:second_moment_sims}. This provides a 
measure of the contribution from cosmic variance. We find that, within the uncertainties from the measurement 
and cosmic variance, we can indeed recover the second and third moments as a function of smoothing scales 
with our reconstruction method for scale larger than 5 arcminutes in the map corresponding to $0.2<z<1.3$. The 
data point for the third moment on the largest scale is (-2.9$\pm$1.6)$\times 10^{-10}$, which is not shown on the log plot, 
but is consistent within 2$\sigma$ with the simulation value of  (0.97$\pm$2.5)$\times 10^{-10}$.

\begin{figure*}
\centering
\includegraphics[width=0.9\linewidth]{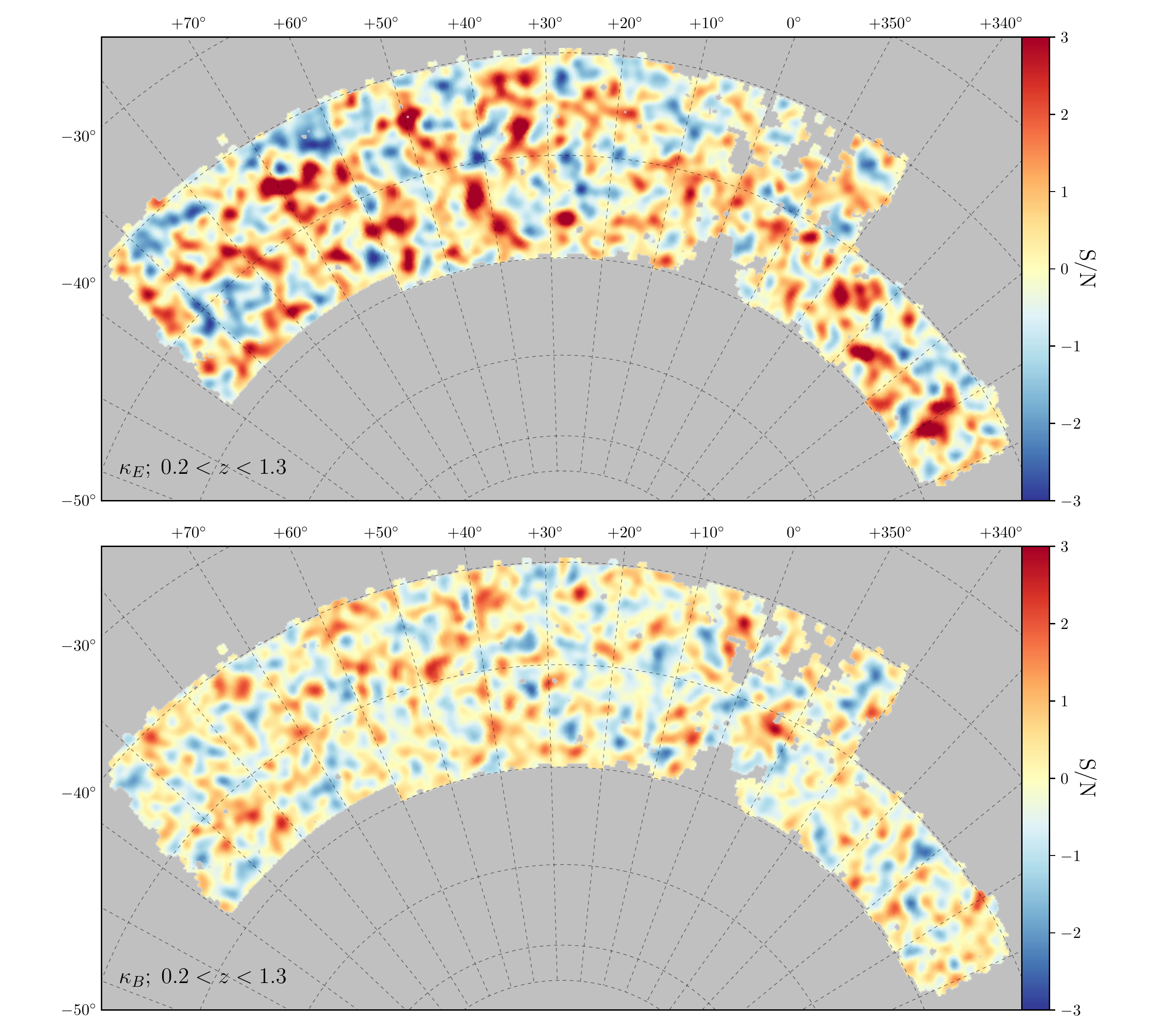}
\caption{Pixel signal-to-noise (S/N) $\kappa_{E}/\sigma(\kappa_{E})$ maps (top) and $\kappa_{B}/\sigma(\kappa_{B})$ maps 
(bottom) constructed from the \metacal catalog for galaxies in the redshift range of $0.2<z<1.3$, smoothed by a 
Gaussian filter of $\sigma_{G}=30$ arcminutes. $\sigma(\kappa_{E})$ and $\sigma(\kappa_{B})$ are estimated by \Eref{eq:noise}. 
}
\label{fig:map_full}
\end{figure*}

It is also instructive to look at the PDF of the different maps for one smoothing scale in \Fref{fig:second_moment_sims}. 
The left panel of \Fref{fig:histogram} shows the distribution of $\kappa_{\rm pix}$ (grey shaded), $\kappa_{\rm sm}$ (black) 
and $\kappa_{E}^\gamma$ (green) when smoothed by a Gaussian filter of 5.1 arcminutes. We find that the three histograms agree 
very well, and the non-Gaussian nature of the PDF is apparent. These distributions closely resemble the log-normal distribution 
and is consistent with the results shown in \citet{Clerkin2015}.
The distribution of $\kappa_{E}^{\varepsilon}$ (blue), $\kappa_{B}^{\varepsilon}$ (orange) and $\kappa_{E, \rm ran}^{\varepsilon}$ 
(purple) are also shown. Due to the added shape noise, these three fields appear much more Gaussian and the shape 
of the PDF is much broader. The fact that the distribution of $\kappa_{E, \rm ran}^{\varepsilon}$ is consistent with 
$\kappa_{B}^{\varepsilon}$ suggests that shape noise is the main contributor of the B-mode map on these smoothing scales, 
rather than B-mode leakage due to imperfect reconstruction. We also check by looking at the B-mode signal in the noiseless 
reconstruction scenario, and find it to be negligible compared to the B-mode from shape noise. 
The shape of the of $\kappa_{E}^{\varepsilon}$ PDF is qualitatively 
different from $\kappa_{E, \rm ran}^{\varepsilon}$ and $\kappa_{B}^{\varepsilon}$ --- the $\kappa_{E}^{\varepsilon}$ map contains 
more extreme high and low values, which correspond to real peaks and voids in the mass distribution. The $\kappa_{E}^{\varepsilon}$ 
PDF is also slightly skewed towards positive values, which is the imprint of the skewed true $\kappa$ distribution seem in 
$\kappa_{\rm pix}$.

\begin{figure*}
\centering
\includegraphics[width=0.98\linewidth]{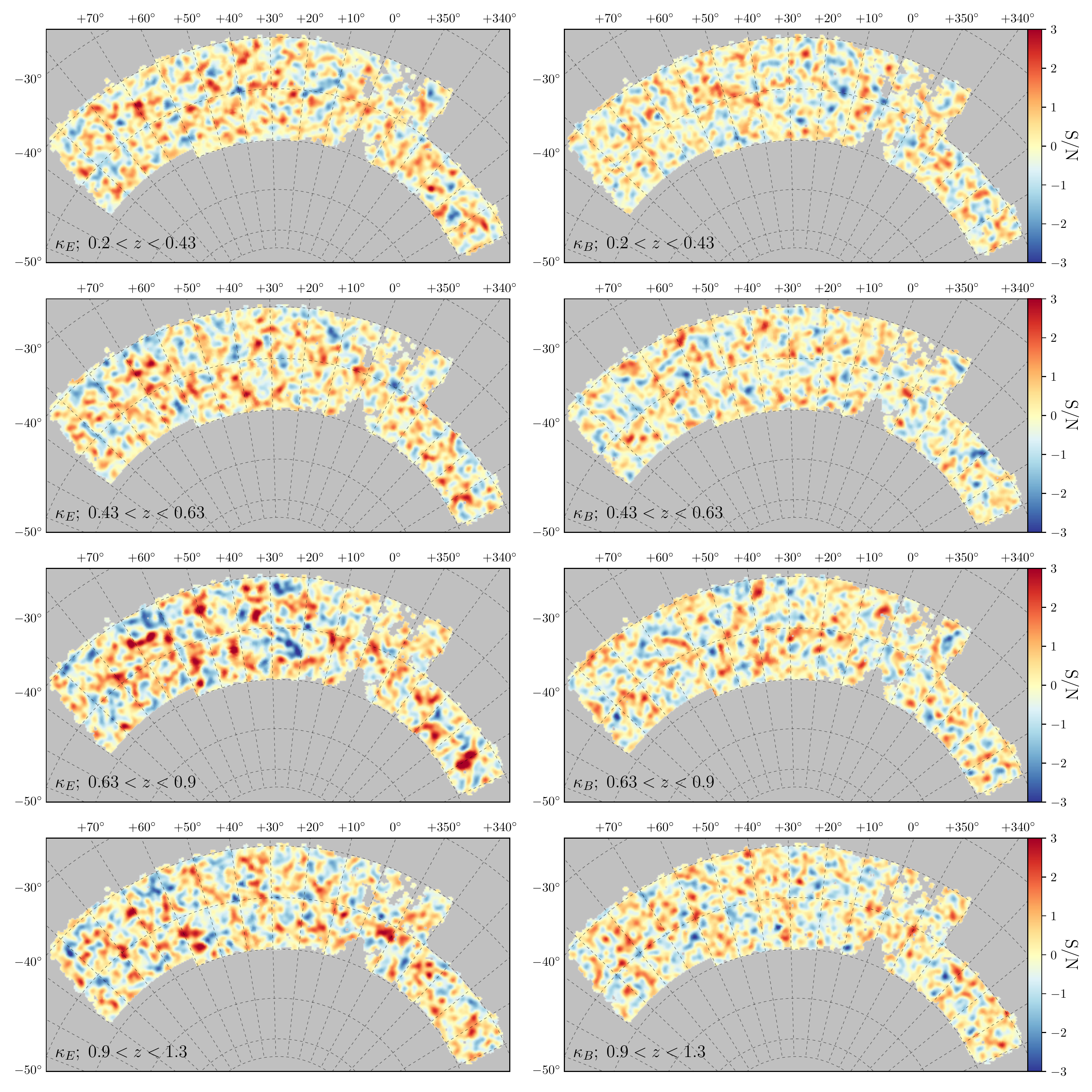}
\caption{Same as \Fref{fig:map_full} but for the four tomographic maps. The $\kappa_{E}/\sigma(\kappa_{E})$ maps 
are shown on the left and the $\kappa_{B}/\sigma(\kappa_{B})$ maps are shown on the right.}
\label{fig:map}
\end{figure*}

\section{DES Y1 weak lensing maps}
\label{sec:results}

\subsection{Convergence maps}
\label{sec:kappa_maps}

Now we present the main goal of the paper. 
In \Fref{fig:map_full} we show the signal-to-noise (S/N) maps associated with the E-mode and B-mode convergence 
generated from the \metacal catalog for galaxies in the redshift range $0.2<z<1.3$ and smoothed with $\sigma_{G}=$30 
arcminutes. The S/N in these maps apply both to the positive (peaks) and negative (voids) values --- extreme positive and 
negative values are significant, while values close to zero are more likely to be consistent with noise. 
In \Fref{fig:map}, maps for the four tomographic bins are shown. The \imshape convergence maps in all the redshift bins 
are shown in \Aref{sec:im3shape} for comparison, together with maps generated using the Science Verification data 
\citep{Vikram2015, Chang2015}.  

\begin{figure*}
\centering
\includegraphics[width=0.9\linewidth]{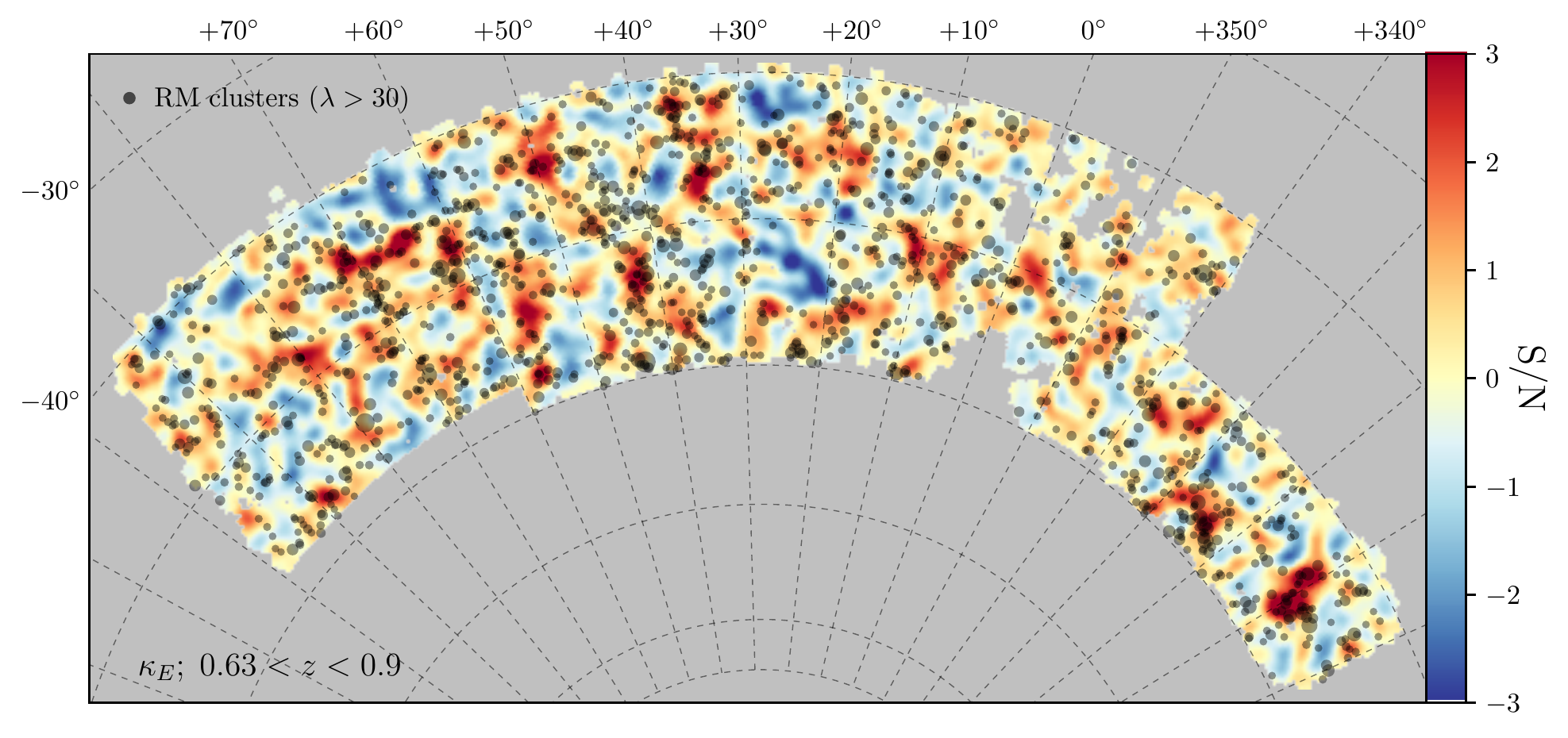}
\caption{Top panel shows the $\kappa_{E}$ map at $0.63<z<0.9$, overlaid with \redmapper (RM) clusters at $\lambda>30$ and 
$0.2<z<0.5$ (black solid circles). The size of the circles scale linearly with $\lambda$, or the cluster mass. }
\label{fig:rm_overlay}
\end{figure*}

We first look at the E-mode maps. \Fref{fig:map_full} includes the full redshift range ($0.2<z<1.3$) and thus has much 
higher signal-to-noise compared to the tomographic maps in \Fref{fig:map}, as expected from the higher number density 
of source galaxies. The visual impression of the map is very similar to the maps generated from the mock galaxy catalogs 
shown in \Fref{fig:sim_map_buzzard}, where there is an imprint of large-scale structure stretched over tens of degrees. 
The area close to RA$\sim0^{\circ}$ suffers from a more complicated mask structure as well as shallower depth, which 
results in a lower S/N in the map in that region. In \Fref{fig:map}, we find that the redshift bin $0.63<z<0.9$ has the highest S/N, 
which is due to both the higher signal at higher redshift and the lower noise coming from the higher number density of 
source galaxies. Structures that show up in a given map are likely to also show up in the neighbouring redshift bins, 
since the mass that is contributing to the lensing in one map is likely to also lens galaxies in neighbouring redshift bins. 
This is apparent in e.g. the structures at (RA, Dec)=(35$^{\circ}$, -48$^{\circ}$) and (58$^{\circ}$, -55$^{\circ}$). 
Next, we compare the E-mode maps with their B-mode counterpart in \Fref{fig:map_full} and \Fref{fig:map}. In 
general, the B-mode maps have lower overall amplitudes. The mean absolute S/N of the E-mode map is $\sim$1.5 times 
larger than the B-mode map at this smoothing scale. For a smoothing scale of $\sigma_{G}=$80 arcminutes, this ratio 
increases to $\sim2$. There are no significant correlations between the E- and the B-mode maps in \Fref{fig:map_full} and \Fref{fig:map}: 
we find that the Pearson correlation coefficients\footnote{The Pearson correlation coefficient two maps $X$ and $Y$ 
is defined as $\langle (X-\bar{X}) (Y-\bar{Y}) \rangle/ (\sigma_{X}\sigma_{Y})$, where $\bar{X}$ and $\bar{Y}$ are the mean 
pixel values for the two maps, the $\langle \rangle$ averages over all pixels in the map, and $\sigma$ indicates the standard 
deviation of the pixel values in each map.} are all consistent with zero, as expected for maps where systematic effects are 
not dominant. Comparing the four tomographic B-mode maps in 
\Fref{fig:map}, there is no obvious correlation between the structures in one map with maps of neighboring redshift 
bins. We find that the Pearson correlation coefficient between the second and third (third and fourth) redshift bins 
for the B-mode maps is 8 (5.5) times lower than that for the E-mode maps. 
The E and B-mode maps for the lowest redshift bin $0.2<z<0.43$ have similar levels of S/N, which is expected since 
the lensing signal at low redshift is weak and the noise level is high.  

We now examine the second and third moments of the $\kappa_{E}$ maps similar to the tests in \Sref{sec:bcc}. 
For direct comparison with simulations, the measurements are done using the map with the full redshift range 
$0.2<z<1.3$ and in the region of $0^{\circ}<$RA$<100^{\circ}$. Our results are shown in the 
right panels of \Fref{fig:second_moment_sims}, where the mean and standard deviation of the 12 noisy 
simulation results are also overlaid. 

We note that we do not expect perfect agreement between the simulation and data for several reasons: first, 
the detailed shape noise incorporated in the simulations is only an approximation to the \metacal shape noise. In 
particular, there is no correlation of the shape noise with other galaxy properties in our simulations. 
This, however, should be a second-order effect, since we do not expect the galaxy properties to correlate with the 
true convergence. Second, the number density and $n(z)$ in the simulations only approximately match the data as we 
discussed in \Sref{sec:sims}. This is also a second-order effect since lensing is mainly sensitive to the mean redshift of the 
lensing kernel. The detailed shape of the $n(z)$ will not significantly alter the convergence maps. Finally, the simulations 
assume a certain cosmology that may not be the true one. 
As $\langle \kappa^2 \rangle \propto \sigma_{8}^{2}\Omega_{m}^{1.5}$ and 
$\langle \kappa^3 \rangle /  \langle \kappa^2 \rangle^2 \propto \sigma_{8}^{-0.8}$ 
\citep{Bernardeau1997,Jain2000}, these measurements are directly sensitive to the cosmological parameters. Given the 
current constraints in $\sigma_{8}$ and $\Omega_{m}$ from \citet{Planck2016b}, changing the cosmological parameters 
by 2$\sigma$ does not affect the comparison carried out here. 

From \Fref{fig:second_moment_sims}, we find very good agreement between the measurements from data and 
simulations in the overall amplitude and trend of the second and third moments as a function of smoothing scale. 
The fact that our measurements are in agreement with the simulations suggests that they are also in agreement 
with the cosmology assumed in the simulations (see \Sref{sec:sims}), though the error bars are fairly large compared 
to e.g. \citet{Troxel2017, DES2017}.
The histograms of the $\kappa_{E}$ and $\kappa_{B}$ maps smoothed with a 5.1 arcminute Gaussian filter are 
shown in the right panel of \Fref{fig:histogram}, together with the simulation counterparts generated from the 12 
Buzzard simulations. Again, we find good agreement in the shape and width of the $\kappa_{E}$ PDF between 
the simulation and the data. The slightly narrower width of the simulation PDF at the extreme $\kappa_{E}$ values 
is likely due to the lack of spatial variation of shape noise, which is not properly incorporated in the simulations.

\begin{figure}
\centering
\includegraphics[width=0.92\linewidth]{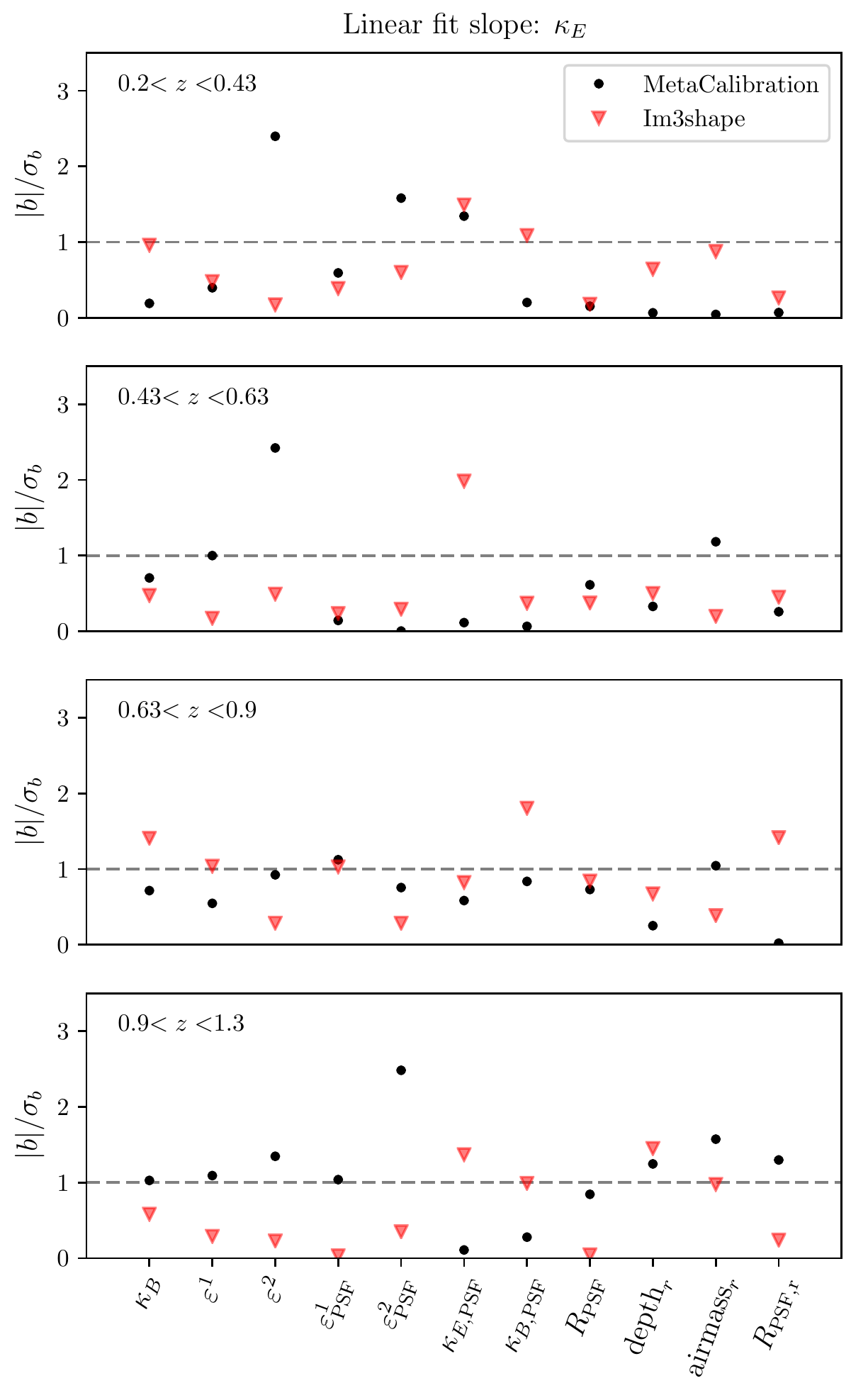}
\includegraphics[width=0.92\linewidth]{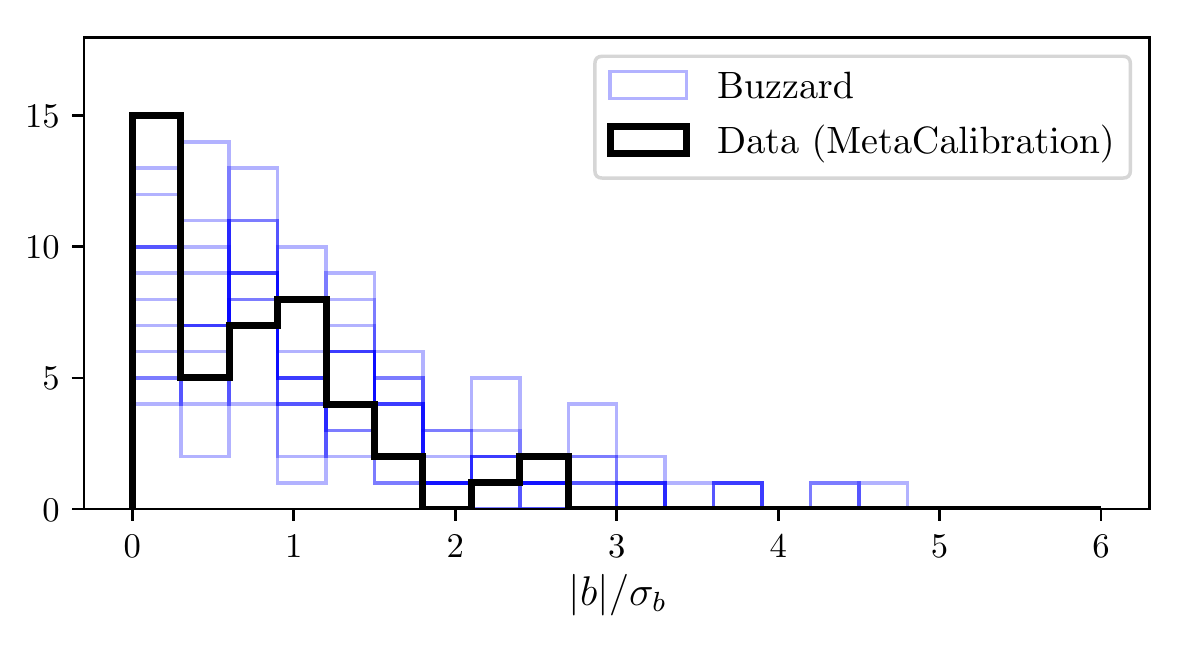}
\caption{The upper four panels show the absolute value of the best-fit slope $b$ divided by the uncertainty of $b$ 
for the linear fit of $\kappa_{E}$ vs. various systematics templates. $|b|/\sigma_{b}$ measures the significance of 
a trend between the convergence and the systematics templates. The four panels correspond to the four 
tomographic bins which we construct the $\kappa_{E}$ maps, and the two set of points correspond to the two 
shear catalogs. The list of systematics templates are labeled for the last redshift bin. The bottom panel shows a 
histogram of $|b|/\sigma(b)$ measured from the 12 Buzzard simulations (thin blue lines) compared to \metacal 
data points (thick black line).}
\label{fig:sys_1d}
\end{figure}

\begin{figure}
\centering
\includegraphics[width=0.92\linewidth]{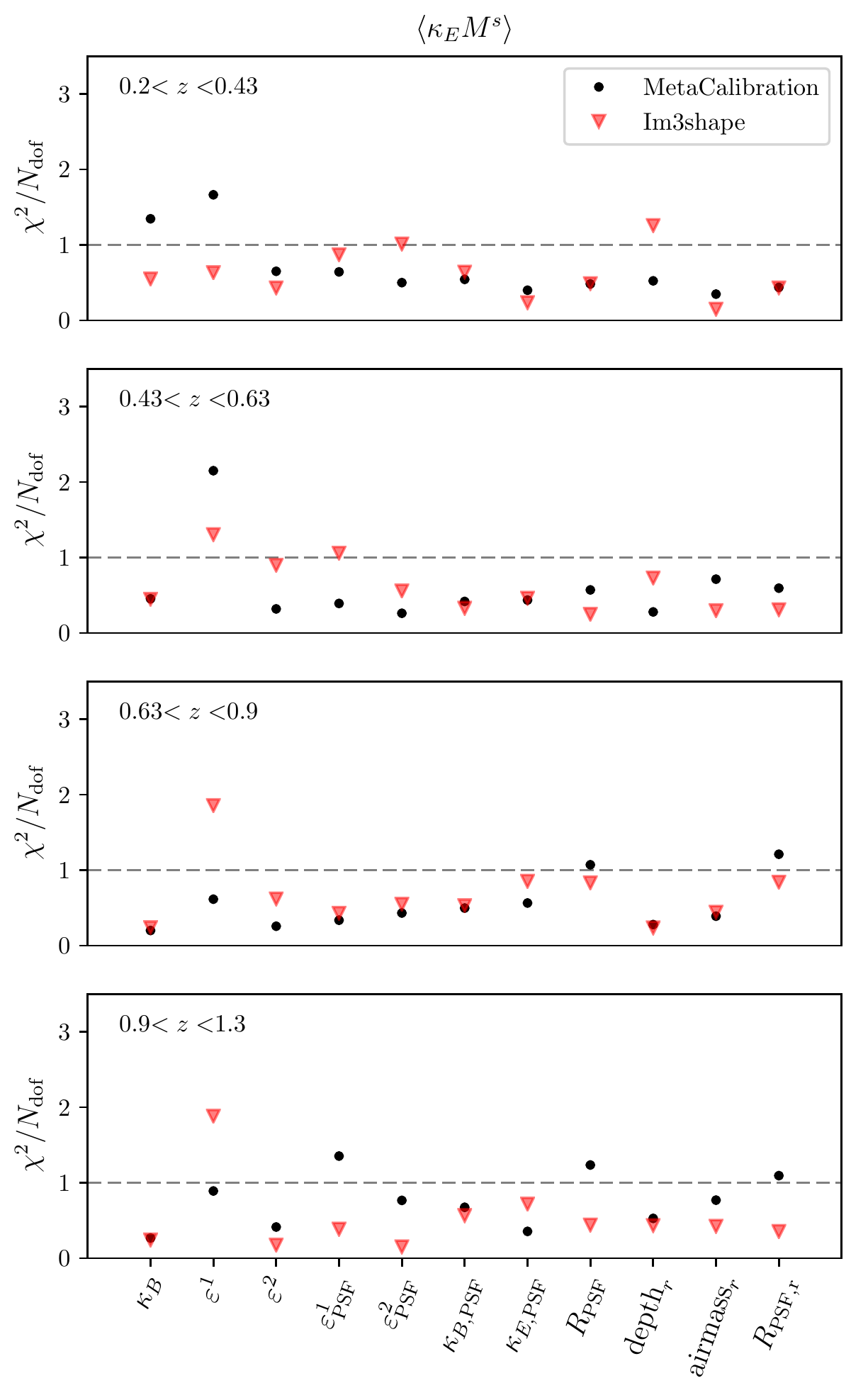}
\includegraphics[width=0.92\linewidth]{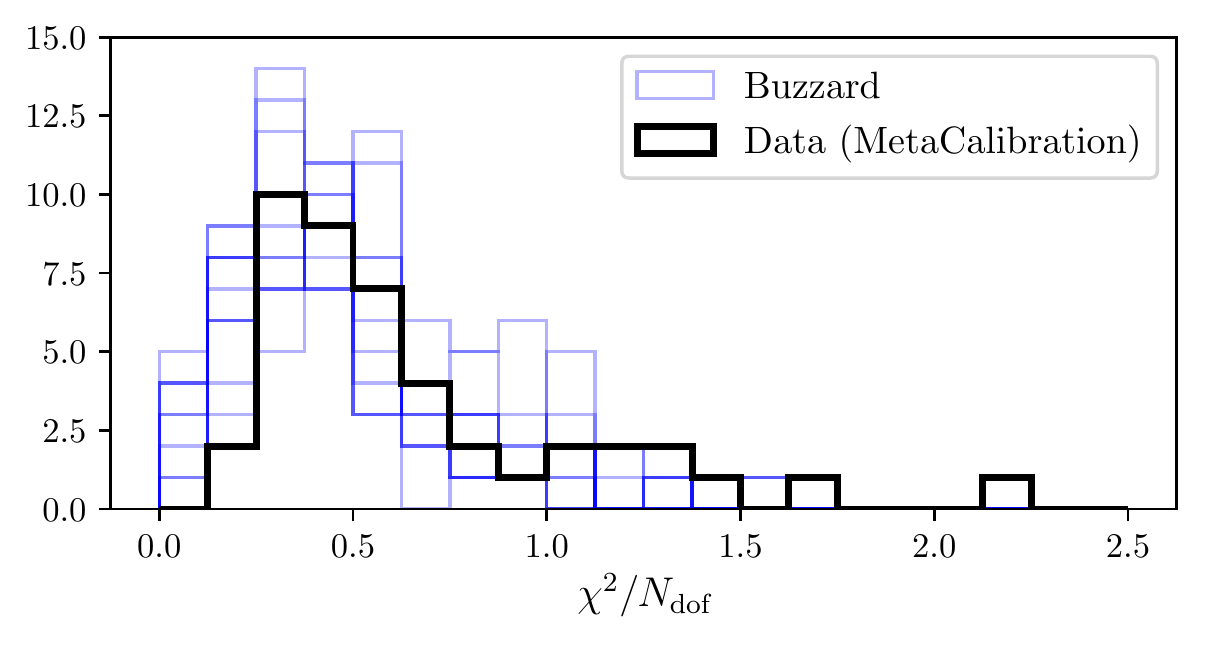}
\caption{Similar to \Fref{fig:sys_1d}, the upper four panels show the reduced $\chi^{2}$ for the cross-correlation 
of the $\kappa_{E}$ maps with various systematics templates to be consistent with zero. The cross-correlation is 
measured for a range of scales, with a total of 8 data points, thus the number of degrees-of-freedom ($N_{\rm dof}$) 
for the $\chi^{2}$ is 8. The bottom panel shows a histogram of the reduced $\chi^{2}$ measured from the 12 
Buzzard simulations (thin blue lines) compared to \metacal data points (thick black line).}
\label{fig:sys}
\end{figure}

Finally, as an additional visual inspection, we overlay a sample of \redmapper galaxy clusters \citep{Rykoff2016} 
onto the E-mode map at $0.63<z<0.9$ as shown in \Fref{fig:rm_overlay}. The galaxy clusters are selected to be 
in the redshift range $0.2<z<0.5$ (roughly the peak of the lensing efficiency of the map) and the richness range 
$\lambda>30$ \citep[corresponding to roughly a mass greater than $2\times10^{14}$ M$_{\odot}$;][]{Melchior2017}. 
Each circle indicates a cluster, with the size of the circle proportion to 
the richness (mass) of the cluster. Visually we can see the correlation between the cluster positions and 
the region of the map with high $\kappa$ values. It is noticeable that the high $\kappa$ regions in the map are often 
associated with an ensemble of smaller clusters rather than one large cluster, while there is a clear lack of clusters 
inside most of the ``void'' regions in the map. There are exceptions, though, where very high S/N peaks do not line up 
with the cluster distributions. For example, the peaks at (RA, Dec)=(55.9$^{\circ}$, -53.8$^{\circ}$) and (34.3$^{\circ}$, -47.5$^{\circ}$) 
do not correspond to any clusters at the centre of the peak, and the void area around (RA, Dec)=(60.3$^{\circ}$,-43.3$^{\circ}$) 
overlaps with several clusters. 
This could be in part due to the shape noise moving the locations of peaks and voids, as we have seen in 
\Fref{fig:sim_map_buzzard}. Nevertheless, further investigation of these structures would be interesting in identifying 
e.g. massive structures with relatively low luminosity. Overall, in \Fref{fig:rm_overlay} we find that there are $\sim$30\% 
of clusters in pixels above S/N$>$1, and $\sim$6.5\% in pixels S/N$<$-1; $\sim$13\% of clusters in pixels above S/N$>$2, 
and none in pixels S/N$<$-2.

\subsection{Systematic tests}
\label{sec:sys}

We have explored, in \Sref{sec:sim_test}, the systematic effects associated with the reconstruction algorithm, masking, 
shot noise, and shape noise using simulations. We also examined the zeroth order 
systematic effects in the data by looking at the B mode convergence maps in \Sref{sec:kappa_maps}. In this 
section we concentrate on examining other potential sources of systematic effects that could contaminate 
our maps. Specifically, we look at whether there exists any spurious correlation between our maps and 
quantities that are not expected to correlate with the convergence maps. This technique is similar to that 
used in \citet{ElvinPoole2017}.

We first identify a number of potential systematics that could contaminate the $\kappa_{E}$ 
maps. The potential systematics presented here are listed below: 
 
\begin{itemize} 
\item $\kappa_{B}$: B-mode convergence map
\item $\varepsilon^{1}$, $\varepsilon^{2}$: the mean galaxy ellipticity 
\item $\varepsilon^{1}_{\rm PSF}$, $\varepsilon^{2}_{PSF}$: the mean PSF ellipticity
\item $\kappa_{E,{\rm PSF}}$, $\kappa_{B,{\rm PSF}}$: $\kappa_{E}$ and $\kappa_{B}$ maps generated from 
$\varepsilon^{1}_{\rm PSF}$ and $\varepsilon^{2}_{\rm PSF}$
\item $R_{\rm PSF}$: the mean PSF size used for galaxy shape measurement\footnote{We use the 
quantity \texttt{mean\_psf\_fwhm} in the \imshape catalog and \texttt{psfrec\_T} in the \metacal catalog.}
\item $R_{{\rm PSF},r}$: the mean $r$-band PSF FWHM size  
\item ${\rm depth}_{r}$: the mean $r$-band magnitude limit\footnote{These are 10$\sigma$ detection limits 
for galaxies.} 
\item ${\rm airmass}_{r}$: the mean $r$-band airmass.
\end{itemize}  
Note that we have checked the PSF size, depth and airmass quantities for other filter bands but only 
present here the $r$-band quantities. 
For potential systematics $s$, we construct map $M^{s}$.
For quantities where we expect the mean to be close to zero ($\kappa_{B}, \varepsilon^{1}, 
\varepsilon^{2}, \varepsilon^{1}_{\rm PSF}, \varepsilon^{2}_{\rm PSF}, \kappa_{E, {\rm PSF}}, 
\kappa_{B, {\rm PSF}}$), $M^{s}$ is constructed using the mean-subtracted values; whereas for the 
rest of the quantities where the mean is non-zero, we use the fractional contrast of the map 
$M^{s} = \delta_{s} = \frac{s-\bar{s}}{\bar{s}}$.

We first look for correlation at the pixel level between the four tomographic $\kappa_{E}$ 
maps with each of the above potential systematics $s$. That is, we are interested in whether the high 
$\kappa_{E}$ values are associated with a certain systematic quantity being high or low. 
To do this, we bin the pixels in the systematics templates into 10 bins depending on the value of the 
pixels, and measure the average convergence in the pixels assigned to each of the 10 bins. The  
error bars are evaluated using a Jackknife approach. We then perform a linear fit with intercept $a$ 
and slope $b$ to the measurements. 
In order to see whether there is a significant correlation between the value of the convergence 
and the value of the systematics template, we plot $|b|/\sigma_{b}$ in \Fref{fig:sys_1d}. There is one 
data point that has a $|b|/\sigma_{b}$ value larger than 3 ($\kappa_{E, {\rm PSF}}$ for \metacal in highest 
redshift bin), which we show in the histogram in the bottom panel.
To understand whether these $|b|/\sigma_{b}$ values are a cause of concern, we perform the same analysis for the 12 
Buzzard simulation maps by cross-correlating them with the systematics templates. Since these simulations 
cannot be possibly correlated with the data, this measurement provides a quantitative way to interpret the 
results. The distribution of all $|b|/\sigma_{b}$ values are shown in the bottom panel of \Fref{fig:sys_1d}, together 
with the \metacal results (as the simulations are matched to the characteristics of the \metacal catalog). 
We find that 97\% (88\%) of the points in the simulations are below 3$\sigma$ (2$\sigma$), which is in reasonable  
agreement with that from the data (98\% of the points below 3$\sigma$ and 91\% of the points below 2$\sigma$). 
The overall distribution of $|b|/\sigma_{b}$ values in the simulations also agrees well with the data.

Next, we compute the two-point angular cross correlation between the convergence maps and the systematics 
templates. This measurement tests the potential contamination of cross-correlating the $\kappa$ maps with other 
maps, such as that investigated in \Sref{sec:crosscorr}. We measure
\begin{equation}
\langle \kappa M^{s} \rangle (\theta_{i}) = \frac{1}{N_{i}} \sum_{j}^{N_{i}} (\kappa M^{s})_{j},
\end{equation}
where $M^{s}$ is the systematics template of interest and the sum is over all pairs of pixels $j$ in the maps separated 
by angular distance within the bin $\theta_{i}$. The correlation function is evaluated in 8 logarithmically separated 
angular bins $\theta_{i}$ between 10 and 200 arcminutes. The covariance matrix is derived from the Jackknife 
approach. 
We then calculate the reduced $\chi^{2}$ of each correlation for it to be consistent with null signal, ie. 
$\langle \kappa M^{s} \rangle (\theta_{i}) = 0$, at all $\theta_{i}$, where the $\chi^{2}$ is defined through
\begin{equation}
\chi^2 = D Cov_D^{-1} D^{T}
\end{equation}
where $D=\langle \kappa M^{s} \rangle (\theta_{i})$ is the angular correlation function and $Cov_D$ is the 
covariance matrix between the 8 angular bins.

The results of the two-point cross-correlation are shown in \Fref{fig:sys}. We also perform similar measurements 
using the 12 Buzzard simulations and show the total distribution of the reduced $\chi^{2}$ in the bottom panel. 
We find that reduced $\chi^{2}$ for all combinations of maps, shear catalogs, and redshift bins, all fall below 3, 
indicating no significant contamination in the maps directly introduced by these potential systematics quantities 
on the two-point level. 
Comparing with simulations also shows that the overall distribution of these reduced $\chi^{2}$ values are consistent 
with no correlation between the $\kappa_{E}$ maps and the systematics templates. 
We find that 100\% (92\%) of the points in the simulations are below 2$\sigma$ (1$\sigma$), which is in reasonable  
agreement with that from the data (98\% of the points below 2$\sigma$ and 80\% of the points below 1$\sigma$). 

\begin{figure*}
\centering
\includegraphics[width=0.95\linewidth]{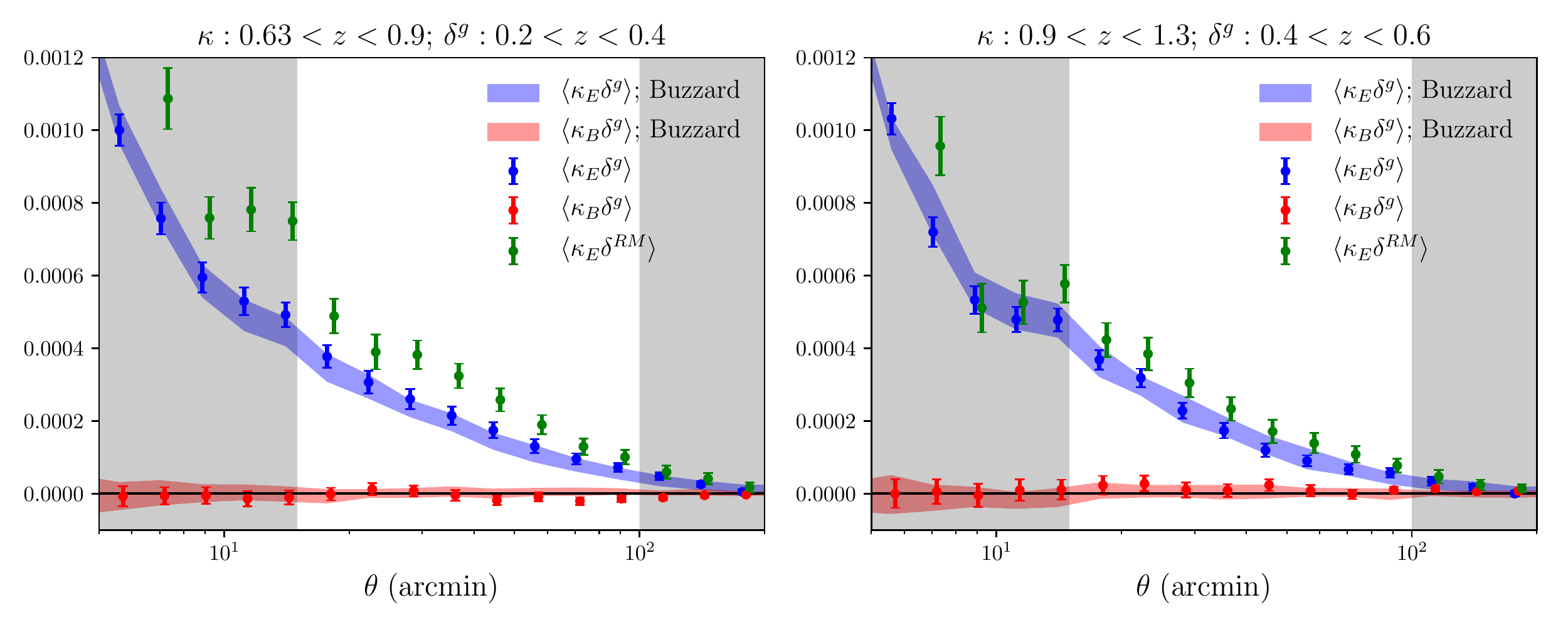}
\caption{Cross-correlation of the $\kappa$ maps with foreground galaxy samples. The blue (red) data points show 
the cross-correlation between the $\kappa_{E}$ ($\kappa_{B}$) map with the foreground flux-limited sample for two 
redshift bins. The shaded band show the mean and standard deviation of the 12 Buzzard simulations, 
while the data points show the DES Y1 data. 
The green data points show $\langle \kappa_{E} \delta^{RM} \rangle$, the cross-correlation of the same 
$\kappa_{E}$ maps with the foreground \redmagic sample which have similar redshift distributions as the two 
flux-limited samples but higher galaxy bias (therefore higher amplitudes). 
The grey shaded region is not used for the calculation of galaxy bias.}
\label{fig:mass_light_corr}
\end{figure*}

\section{Applications of DES Y1 mass map}
\label{sec:application}

In this section we present two applications of the convergence map constructed from this work. 
In \Sref{sec:crosscorr} we cross-correlate the convergence maps with foreground mass tracers 
to demonstrate that our maps do indeed contain significant signal and is consistent with 
expectation. In \Sref{sec:superstructure} we take a closer look at some of the high signal-to-noise 
structure in the maps and discuss the physical interpretation for the largest peaks and voids 
respectively. We defer some of the more involved applications (e.g. cross-correlation of the 
convergence maps to CMB lensing maps, peak statistics) to future work.

\subsection{Cross correlation of mass and light}
\label{sec:crosscorr}

One of the motivations for generating a convergence map instead of using the weak lensing shear 
directly is that in many cases a scalar field is easier to manipulate and cross-correlate with other data 
sets compared to a spin-2 field. Here we demonstrate some of the usages by cross-correlating the 
convergence maps in \Fref{fig:map} with other tracers of mass. Specifically, we look at a flux-limited 
galaxy sample (described in \Sref{sec:lss}) and the \redmagic 
Luminous Red Galaxies (LRG) sample \citep{Rykoff2016}. The amplitudes of these 
cross-correlations will be a direct measure of the galaxy bias for the different samples 
\citep[see e.g.][]{Pujol2016, Chang2016}. Note that the 
cross-correlation can naturally extend to include maps of other wavelengths such as X-ray, Gamma ray 
\citep{Shirasaki2014}, H$_{\rm I}$ neutral hydrogen \citep{Kirk2015}, the CMB, CMB lensing 
\citep{Liu2015b, Hand2015, Kirk2016} and others.

In this analysis, we opt for calculating the real-space 2-point correlation function similar to that used in 
\Sref{sec:sys},
\begin{equation}
\langle \kappa \delta^{X} \rangle (\theta_{i}) = \frac{1}{N_{i}} \sum_{j}^{N_{i}} (\kappa \delta^{X})_{j},
\end{equation}
where $X$ denotes the specific sample of interest (flux-limited galaxy sample or \redmagic galaxies in 
different redhshift ranges). $\delta = \frac{n-\bar{n}}{\bar{n}}$ is the density contrast of the sample, where 
$n$ is the number of counts per pixel and $\bar{n}$ is the mean number count over the full map. The 
average is calculated for all pairs of points $j$ whose angular separation $\theta$ fall in the angular bin 
$\theta_{i}$. 
The cross-correlation is calculated for scales 2.5 to 250 arcminutes. In later analyses where 
we compare the cross-correlation between the convergence map and the two foreground samples, 
we exclude scales larger than 100 arcminutes and smaller than 15 arcminutes. The small-scale cutoff corresponds 
to about 3 times the scale corresponding to $\ell_{\rm max}$, while the large-scale cutoff corresponds to the 
size of the Jackknife region.

We begin with testing whether the cross-correlation between the $\kappa_{E}$ and $\kappa_{B}$ map 
with a foreground flux-limited galaxy sample is consistent with expectation from the simulations. 
We use the same set of mock galaxy catalogs used in \Sref{sec:bcc}, with the addition of a simulated 
foreground sample that matches with the flux-limited sample. 
We perform the cross-correlation for various redshift combinations of the $\kappa$ map and the galaxy map, 
as well as the two shear catalogs. We find very good agreement between the two shear catalogs and between 
the simulation and data.   

In \Fref{fig:mass_light_corr}, we show two examples of the measurements: 
cross-correlation of the \metacal $\kappa$ maps at $0.63<z<0.9$ and $0.9<z<1.3$ with the flux-limited galaxy 
sample $\delta^{g}$ maps at $0.2<z<0.4$ and $0.4<z<0.6$, respectively. We show the data measurements 
together with the mean and standard deviation of the 12 measurements from the Buzzard simulations. Both the 
E-mode and the B-mode cross-correlation show excellent agreement between the data and the simulations.  
As the amplitude of the cross-correlation is sensitive to the cosmological model, galaxy bias, and the photo-$z$, 
the agreement between simulations and data suggests that there is no outstanding differences between the 
simulations and the data that could be potentially a sign of systematic effects. 

\begin{figure*}
\centering
\includegraphics[width=0.35\linewidth]{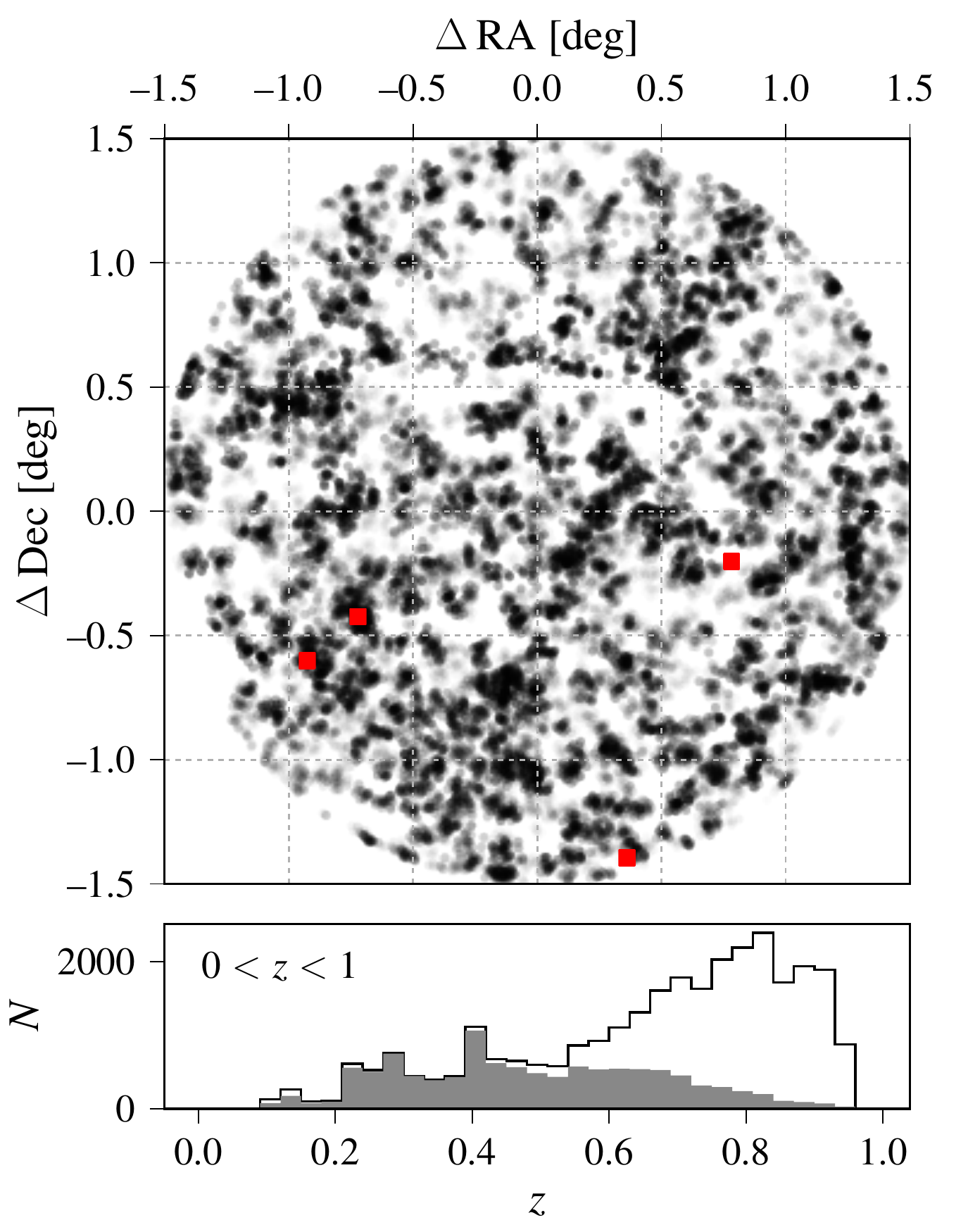}
\includegraphics[width=0.35\linewidth]{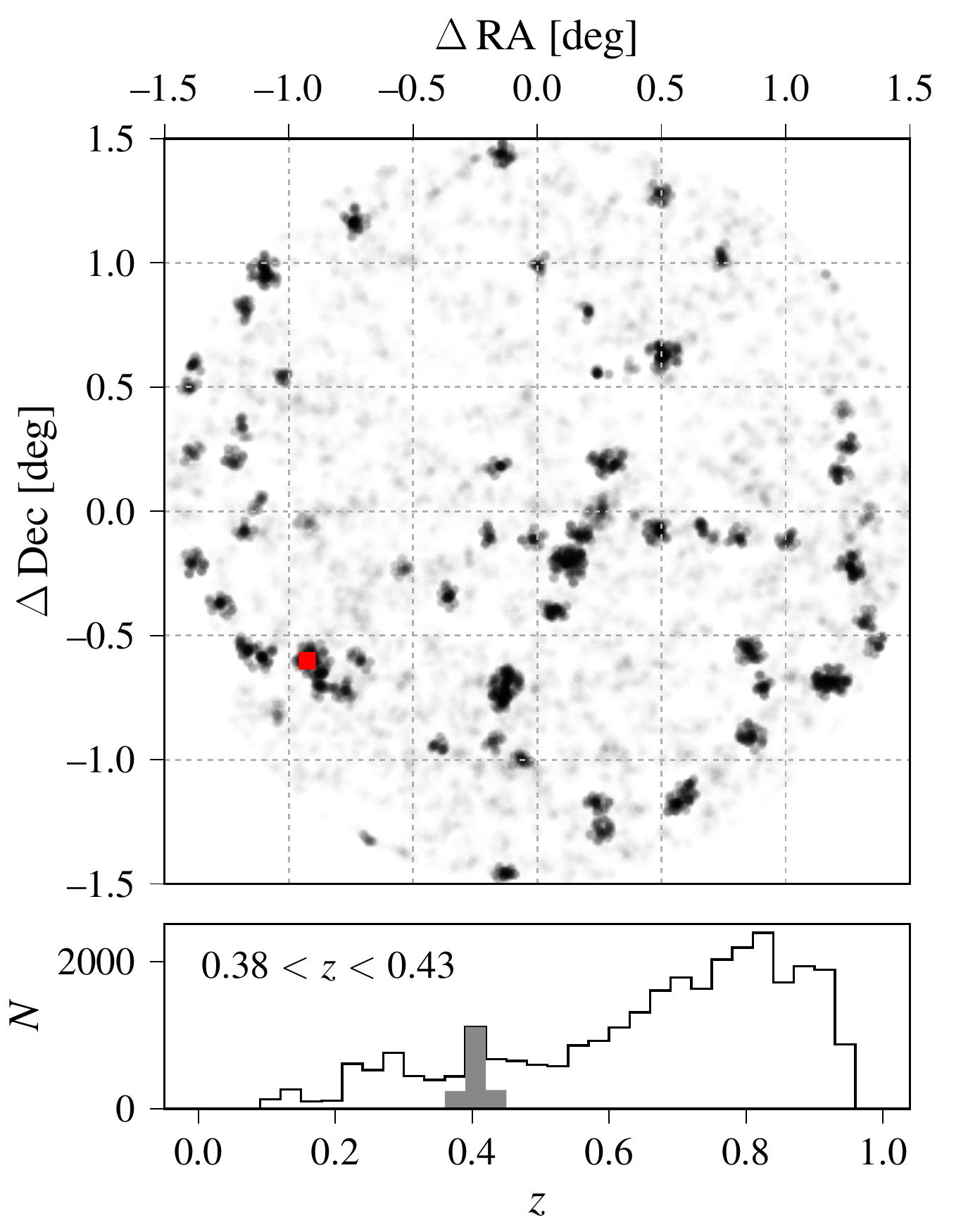}
\caption{This figure shows the cluster galaxy distribution around the biggest overdensity in \Fref{fig:rm_overlay} 
at (RA, Dec) = ($309^\circ$, $-56^\circ$). We show in the left panels all members of 
\redmapper-detected galaxy clusters with $0<z<1$ and $\lambda>5$. The top panel is the spatial distribution, 
projected onto the tangent plane, with shading of each circle indicating the lensing weights. The red dots indicate 
SPT-detected galaxy clusters from \citet{Bleem2015} in the same redshift range. The bottom panel is the redshift 
distribution of the cluster member galaxies, before (light) and after (dark) applying the lensing weights. The right 
panels show the same as the left panels, but for a narrower redshift range of $z=0.405 \pm 0.025$, isolating the 
galaxies associated with a noticeable peak in the redshift distribution. The spatial distribution of the cluster members 
in this narrow redshift range exhibits some hint of a filamentary structure.}
\label{fig:filament}
\end{figure*}

Next, we measure the cross-correlation of the same $\kappa$ maps with foreground \redmagic samples. 
We construct the samples so that the mean and spread of the $n(z)$ distribution is similar to that of the flux-limited 
sample. This corresponds to a redshift selection of $0.15<z<0.45$ ($0.25<z<0.6$) for the flux-limited sample at 
$0.2<z<0.4$ ($0.4<z<0.6$). By doing this, the ratio of the cross-correlation amplitude for the \redmagic sample 
and the flux-limited sample would scale directly as the ratio of the galaxy bias for the two samples. The cross-correlation 
between the $\kappa$ maps and the \redmagic sample is also shown in \Fref{fig:mass_light_corr}.
The error bars for the \redmagic sample are larger due to the lower number density, but overall the shape  
of the cross-correlation as a function of scale is very similar, with an overall multiplicative factor that is nearly 
constant over scales. The value of the multiplicative factor (within the 15--100 arcminute range) is $\sim1.38$ for the 
lower redshift bin $0.2<z<0.4$ and $\sim1.27$ for the higher redshift bin $0.4<z<0.6$. 
\citet{Crocce2016} measured the galaxy bias for a flux-limited galaxy sample in the DES 
SV data using angular clustering measurements; their results give a bias of $\sim1.16$ ($\sim1.29$) if we 
interpolate onto the redshift and magnitude range of our sample at $0.2<z<0.4$ ($0.4<z<0.6$). 
\citet{Kwan2017} measured the galaxy bias for the \redmagic sample to be $\sim$1.6 using joint constraints from 
galaxy clustering and galaxy-galaxy lensing in approximately the redshift range of our data (the redshift evolution 
between the two lens bins considered is much less than the statistical uncertainties at $\sim0.3$). The ratio of the 
galaxy bias between the two samples is thus $\sim$1.38 and 1.24 for the two redshift bins, which is broadly 
consistent with our measurements. 

We defer a more quantitative analysis of galaxy bias to future work, but 
this initial test demonstrates one example of cross-correlation of the mass maps with other maps. The results 
also serve as a test for potential systematics in the mass maps -- by comparing the measurements with simulations, 
we have shown that there is no outstanding systematic issues in using the maps for cross-correlation applications. 

\subsection{Peaks and voids}
\label{sec:superstructure}

Another strength of map-level products is that one can visualise and 
detect pronounced local over- or underdense regions that would otherwise be averaged over in global 
summary statistics. The abundance of the massive peaks is a sensitive cosmological probe, as they occupy the 
highest end of the halo mass function \citep{Bahcall1998,Haiman2001,Holder2001}. Some of the extreme 
structures can also help to constrain a certain class of modified gravity theories \citep{Knox2006, Jain2010}. 
On the other hand, the abundance of large voids has been used as a powerful test of $\Lambda$CDM 
cosmology \citep{Plionis2002,Higuchi2017}.
In this section we seek to briefly characterise the physical nature of peaks and voids 
that are associated with the largest over- or underdense regions in the convergence maps. 

To construct a catalog of peaks and voids, we begin with the 4 tomographic S/N maps presented in \Fref{fig:map}, 
which are smoothed with a $\sigma_{G}=30$ 
arcminutes Gaussian filter. We place a threshold on the pixels value at 2.5$\sigma$ (S/N$>2.5$ for peaks and $S/N<-2.5$ 
for voids), and for all pixels that survive the cut, we use a mean-shift clustering algorithm \citep{Comaniciu2002} to divide them into 
clusters of adjoining pixels. We inspect these clusters visually and place an additional cut requiring there to be more 
than 50 pixels (slightly larger than the smoothing we applied) in order to become a candidate for a peak or a void. 
In this approach, we find 9 (5), 9 (5), 18 (13) and 9 (7) peaks (voids) in the four tomographic maps, respectively.

To study the structures associated with peaks, we make use of the approach presented in \citet{Melchior2015}.
We select all cluster member galaxies of \redmapper clusters with richness $\lambda>5$ within 1.5 degrees of 
the peak centre. We show their distribution around the largest peak in the map of \Fref{fig:rm_overlay} in the 
left panel of \Fref{fig:filament} at (RA, Dec)=(309$^\circ$, -56$^\circ$). 
While some correlated structure appears present in the 2D distribution, the redshift 
distributions of the cluster galaxies in this region appears to be very broad, even after taking the lensing kernel into 
account. This suggests that the large peak cannot be accounted for by one large structure localised in redshift space. 
In the right panel of \Fref{fig:filament} we isolate a particular peak at $z\approx 0.4$ in the redshift distribution and 
select only \redmapper clusters with $z=0.4\pm0.025$, a range that corresponds to about two standard deviations 
of their typical redshift accuracy \citep{Rykoff2016}. A central, possibly filamentary structure becomes more pronounced, 
but there is no evidence of a particularly massive galaxy cluster or even a super-cluster in that region. In fact, of the four 
SPT-detected clusters from \citet{Bleem2015} within the search radius, only one, with $z\approx 0.4$, falls in this 
redshift range.

Performing an analogous analysis at different redshifts or on other peaks yields similar results, namely that overdensities 
smoothed on such a large scale generally do not correspond to massive structures in physical contact. 
Instead, the broad redshift kernel is prone to accumulating multiple layers of mildly overdense structures along the line of 
sight. This outcome demonstrates the difficulty of detecting clusters in weak-lensing mass maps or shear catalogs, 
especially when the number density of source galaxies is low and one cannot go to a smaller smoothing scale. This 
generally needs the construction of optimal matched filters in configuration and redshift space \citep{Maturi2005, 
Simon2009, VanderPlas2011}, which is outside of the scope of this work. 

\begin{figure}
\centering
\includegraphics[width=0.95\linewidth]{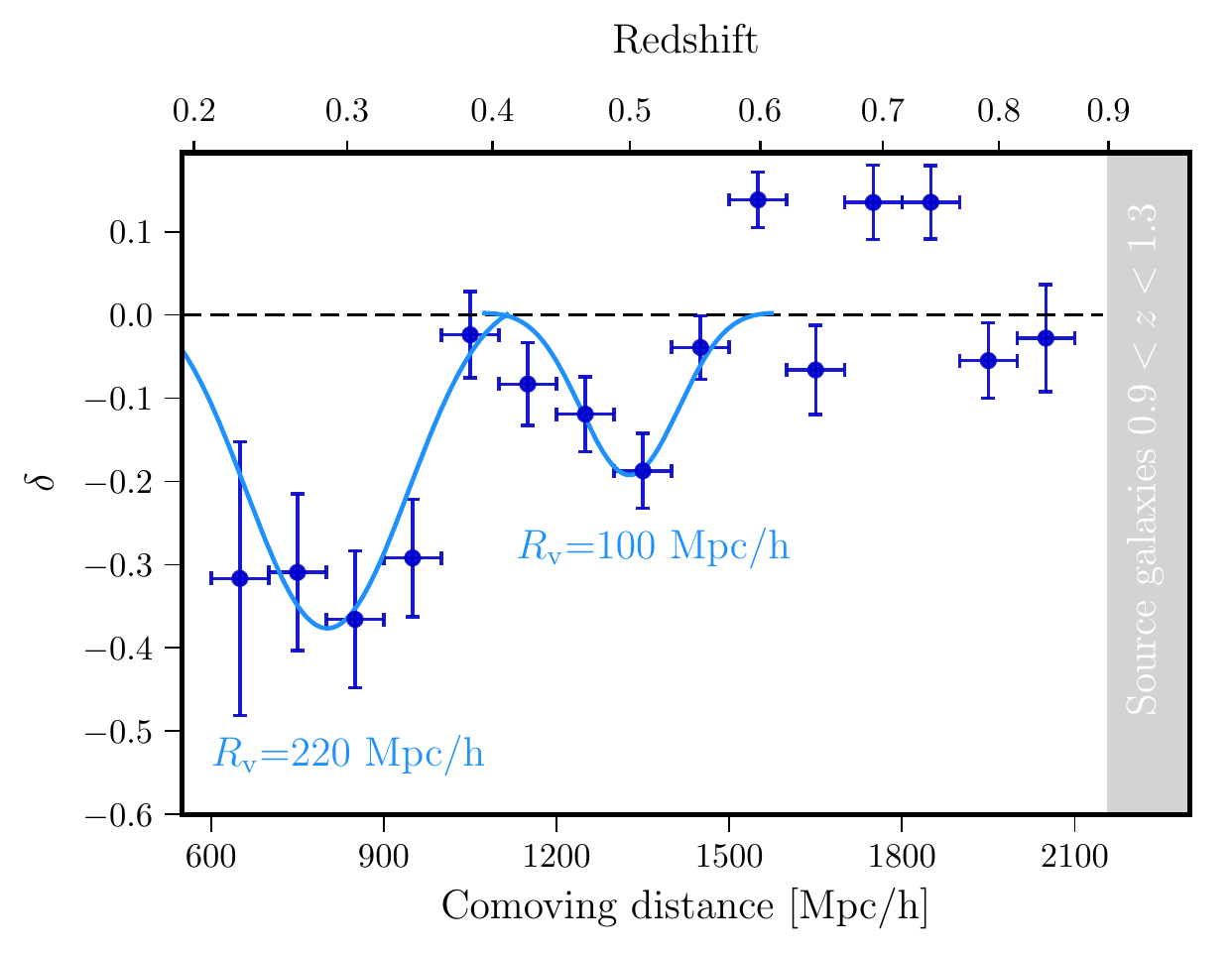}
\caption{The data points show the \redmagic galaxy density contrast $\delta$ along the foreground line-of-sight 
of the largest void identified in the mass map in the highest redshift bin. The profile fits very well with a model 
consisting two supervoids with a size of 220 and 100 $h^{-1}$Mpc, as shown with the cyan line.}
\label{fig:void}
\end{figure}

For voids the situation is more promising. We use \redmagic galaxies with relatively good photo-$z$'s 
(same as that used in \Sref{sec:crosscorr}) 
as tracers of the foreground matter density and 
study their radial distribution. We project the data into 2D slices of 50 $h^{-1}$Mpc along the line-of-sight. We then 
measure the density contrast of the \redmagic galaxies in these 2D slices where the large voids in the maps 
(significant negative convergence values) are measured, compared to the mean \redmagic density at that redshift. 
The density contrast measurements at different redshifts are then used to reconstruct the radial density profile of 
voids.
As an example, we look at the largest void detected in the furthest $0.9<z<1.3$ bin at 
(RA, Dec)=(62$^{\circ}$, -43$^{\circ}$), and count  
galaxies within 2.0 degrees of the void centre, which approximately corresponds to the full angular size of the void in 
the map. We show the resulting line-of-sight density profile measurements of the \redmagic galaxies in \Fref{fig:void}. 
We find two extended underdensities that are consistent with supervoids of radii $R_{\rm v}$=100 $h^{-1}$Mpc and 
$R_{\rm v}$=220 $h^{-1}$Mpc assuming simple Gaussian void profiles \citep{Finelli2016,Kovcs2016,Snchez2017}. 
These supervoids are quite shallow 
even in their centres but their size is comparable to the largest known supervoids. Most probably, these supervoids have 
substructure at smaller scales but that information is not accessible even using high quality photo-$z$ data like 
\redmagic.

We repeated the above analysis for less significant and less extended voids, finding that voids identified in the mass 
maps that extend beyond 
$\sim$0.64 deg$^{2}$ of size can typically be associated with at least one 
$R_{\rm v} \gtrsim$100 $h^{-1}$Mpc supervoid in the \redmagic catalogue. These are of greatest interest in cosmology 
and their integrated Sachs-Wolfe imprint was also studied using DES Y1 data \citep[see e.g.][]{Kovcs2017}.

\section{Conclusion}
\label{sec:conclusion}

Weak lensing allows us to probe the total mass distribution in the Universe. One of the most intuitive ways to visualise 
and comprehend this information is through weak lensing convergence maps, or mass maps. These maps contain the 
Gaussian and non-Gaussian information for the matter field, which could then either be extracted via various statistical 
tools, or analyzed locally for regions of special interest.
  
In this paper, we construct weak lensing mass maps for the first year of Dark Energy Survey data (DES Y1) using two 
independent shear catalogs, \metacal and \imshape, in the redshift range $0.2<z<1.3$ and in the region overlapping with 
the South Pole Telescope footprint. This yields maps covering $\sim1,500$ deg$^{2}$, corresponding to a total 
volume of $\approx$10 Gpc$^{3}$. With the unprecedented large sky coverage, a spherical reconstruction 
approach was used based on decomposing the shear field into spin-2 spherical harmonics, followed by an E/B mode separation. 
The curl-free E-mode and the divergence-free B-mode form the E- and B-mode lensing convergence maps, $\kappa_{E}$ 
and $\kappa_{B}$. The lensing potential $\psi$ and deflection angles $\bm{\eta}$ can also be reconstructed using these 
decomposed spin harmonics. 

We test the mass map reconstruction with simulations, starting with an idealised setup and gradually degrading the 
simulations to match the data. By doing so, we can isolate the effect of individual sources of systematics and noise. 
We use the $F_{1}$ and $F_{2}$ statistics (\Eref{eq:f1f2}) to quantify the performance of the reconstruction in terms of 
the amplitude and the phase information: for perfect reconstruction $F_{1}=F_{2}=1$. Based on these statistics, we 
find that (1) we can reconstruct very well the convergence field in a fully-sampled, full-sky Gaussian simulation for 
scales larger than the pixel scale, as expected; (2) the DES Y1 mask biases the reconstructed maps at the few percent 
level, but the bias mainly comes from the pixels around the edges; (3) finite sampling from galaxies at the DES Y1 density 
does not degrade the reconstruction significantly for our maps at a resolution of 3.44 arcminutes; (4) adding shape noise 
increases the variance of the map and perturbs the phase information, but at the DES Y1 noise level, the 
signal-to-noise is still significant and the resulting $F_{1}$ and $F_{2}$ are consistent with 1; (5) 
we can reconstruct within measurement uncertainty the second moment of the maps on all scales and third moment of 
the maps for scales $>5$ arcminutes, where shape noise is subdominant.

One new application that comes with the large sky coverage is the reconstruction of other lensing maps such as the 
 lensing potential $\psi$ and deflection angles $\bm{\eta}$ maps. We explore briefly in \Aref{sec:deflection_test_sim} 
 this application, finding a $\sim70\%$ ($\sim50\%$) lower amplitude in the reconstruction for 
$\psi$ ($\bm{\eta}$). The reconstruction of $\psi$ and $\bm{\eta}$ is relatively poor compared to that of the $\kappa$ 
maps because information in $\bm{\eta}$ is dominated by scales larger than $\kappa$, 
and the information in $\psi$ is dominated by even larger scales. This suggests that from $\kappa$ to $\bm{\eta}$ to 
$\psi$, the importance of the mask increases while the importance of shot noise and shape noise decreases.    

After rigorous testing with simulations, we generate weak lensing mass maps from the DES Y1 data with a spatial 
resolution of $\sim3.44$ arcminutes. We construct one map that covers the entire redshift range of $0.2<z<1.3$, which 
carries the highest S/N, and also four tomographic bins at the redshift intervals $0.2<z<0.43$, 
$0.43<z<0.63$, $0.63<z<0.9$, and $0.9<z<1.3$. The tomographic maps are relatively noisy, but allow us to explore the 
redshift-dependencies of the maps and can be used for tomographic cross-correlation with other tracers of mass. 
In the highest signal-to-noise map (\metacal, $0.2<z<1.3$), the ratio between the mean S/N in the E-mode 
and the B-mode map is $\sim$1.5 ($\sim$2) when smoothed with a Gaussian filter of $\sigma_{G}=30$ (80) arcminutes.
We examine the PDF of the maps, together with the second and third moments of the PDF as a function of smoothing 
scale and find them to be consistent with realistic 
simulations that incorporate similar noise and mask properties as the data. We further test for systematic effects by 
cross-correlating the maps with various environment and PSF quantities at the one-point and two-point level. We 
find no significant systematic contamination of the maps beyond what is expected from statistical fluctuations. 

Finally, we demonstrate two applications of these mass maps. First, we cross-correlate the mass maps with two sets 
of foreground mass tracers constructed to have similar redshift distributions: a flux-limited galaxy sample and an LRG 
sample. The cross-correlation is done in two redshift bins and shows very good agreement with simulations. The 
ratio of the amplitudes of the cross-correlation, which reflects the ratio of the galaxy bias for the two samples, are 
consistent with previous measurements of similar samples in earlier DES data. Second, we examine the extreme 
peaks and voids identified in the maps. We find that most high S/N peaks in the maps correspond to an accumulated 
mass distribution along the line of sight, even though rare filamentary structures could be found occasionally. For the 
high-S/N voids, however, most of them correspond to real void structures with $R_{\rm v} \gtrsim$100 $h^{-1}$Mpc in the 
foreground. 

The DES Y1 mass maps are the largest weak lensing mass maps to date constructed from galaxy surveys, 
about ten times larger than the previous maps from CFHTLenS \citep{VanWaerbeke2013} and DES SV 
\citep{Vikram2015, Chang2015}. Even though the Y1 depth is shallower (and therefore noisier) than the 
previous maps, these very large maps provide a new perspective on weak lensing map making and the 
various topics one can explore with them. 
Moving onto the larger dataset from DES and other surveys, we expect 
many of the explorations in this paper to be carried out and advanced to serve as complementary probes of 
cosmology alongside more traditional two-point statistics.

\section*{Acknowledgement}

We thank Sandrine Pires for useful discussions. 
CC was supported in part by the Kavli Institute for Cosmological
Physics at the University of Chicago through grant NSF PHY-1125897 and an endowment from Kavli Foundation
and its founder Fred Kavli. 
AP acknowledges support from beca FI and 2009-SGR-1398 from 
Generalitat de Catalunya, project AYA2012-39620 and AYA2015-71825 from MICINN, and from a European Research 
Council Starting Grant (LENA-678282).
Support for DG was provided by NASA through Einstein Postdoctoral Fellowship grant number PF5-160138 awarded 
by the Chandra X-ray Center, which is operated by the Smithsonian Astrophysical Observatory for NASA under contract 
NAS8-03060.
BL was supported by NASA through EinsteinPostdoctoral Fellowship Award Number PF6-170154. 
BJ and MJ are partially supported by the US Department of Energy grant DE-SC0007901 and funds from the University 
of Pennsylvania. 
ES is supported by DOE grant DE-AC02-98CH10886.

Funding for the DES Projects has been provided by the U.S. Department of Energy, the U.S. National Science Foundation, the Ministry of Science and Education of Spain, 
the Science and Technology Facilities Council of the United Kingdom, the Higher Education Funding Council for England, the National Center for Supercomputing 
Applications at the University of Illinois at Urbana-Champaign, the Kavli Institute of Cosmological Physics at the University of Chicago, 
the Center for Cosmology and Astro-Particle Physics at the Ohio State University,
the Mitchell Institute for Fundamental Physics and Astronomy at Texas A\&M University, Financiadora de Estudos e Projetos, 
Funda{\c c}{\~a}o Carlos Chagas Filho de Amparo {\`a} Pesquisa do Estado do Rio de Janeiro, Conselho Nacional de Desenvolvimento Cient{\'i}fico e Tecnol{\'o}gico and 
the Minist{\'e}rio da Ci{\^e}ncia, Tecnologia e Inova{\c c}{\~a}o, the Deutsche Forschungsgemeinschaft and the Collaborating Institutions in the Dark Energy Survey. 

The Collaborating Institutions are Argonne National Laboratory, the University of California at Santa Cruz, the University of Cambridge, Centro de Investigaciones Energ{\'e}ticas, 
Medioambientales y Tecnol{\'o}gicas-Madrid, the University of Chicago, University College London, the DES-Brazil Consortium, the University of Edinburgh, 
the Eidgen{\"o}ssische Technische Hochschule (ETH) Z{\"u}rich, 
Fermi National Accelerator Laboratory, the University of Illinois at Urbana-Champaign, the Institut de Ci{\`e}ncies de l'Espai (IEEC/CSIC), 
the Institut de F{\'i}sica d'Altes Energies, Lawrence Berkeley National Laboratory, the Ludwig-Maximilians Universit{\"a}t M{\"u}nchen and the associated Excellence Cluster Universe, 
the University of Michigan, the National Optical Astronomy Observatory, the University of Nottingham, The Ohio State University, the University of Pennsylvania, the University of Portsmouth, 
SLAC National Accelerator Laboratory, Stanford University, the University of Sussex, Texas A\&M University, and the OzDES Membership Consortium.

The DES data management system is supported by the National Science Foundation under Grant Numbers AST-1138766 and AST-1536171.
The DES participants from Spanish institutions are partially supported by MINECO under grants AYA2015-71825, ESP2015-88861, FPA2015-68048, SEV-2012-0234, SEV-2016-0597, and MDM-2015-0509, 
some of which include ERDF funds from the European Union. IFAE is partially funded by the CERCA program of the Generalitat de Catalunya.
Research leading to these results has received funding from the European Research
Council under the European Union's Seventh Framework Program (FP7/2007-2013) including ERC grant agreements 240672, 291329, and 306478.
We  acknowledge support from the Australian Research Council Centre of Excellence for All-sky Astrophysics (CAASTRO), through project number CE110001020.

This manuscript has been authored by Fermi Research Alliance, LLC under Contract No. DE-AC02-07CH11359 with the U.S. Department of Energy, Office of Science, Office of High Energy Physics. The United States Government retains and the publisher, by accepting the article for publication, acknowledges that the United States Government retains a non-exclusive, paid-up, irrevocable, world-wide license to publish or reproduce the published form of this manuscript, or allow others to do so, for United States Government purposes.

This paper has gone through internal review by the DES collaboration.

\input{massmapping_y1.bbl}

\appendix

\section{Interpolating empty pixels}
\label{sec:inpainting}

In this appendix we test the impact of the empty pixels inside the contiguous footprint and different 
approaches to interpolate over them. We use the same noiseless Buzzard simulations used in 
\Sref{sec:bcc} and test with the redshift bin of $0.2 < z < 1.3$. In this map, the fraction of empty pixels 
inside the footprint occupies $\sim1.67\%$ of the total pixels.

We test the following 4 approaches of assigning values to these empty pixels and calculate the $F_{1}$ 
and $F_{2}$ statistics defined in \Sref{sec:sim_test}:
\begin{itemize}
\item Fiducial: set the empty pixels to 0.
\item Gaussian interpolation: interpolate the values of these empty pixels from a Gaussian kernel with a $\sigma$ corresponding to 3 times the pixel size.
\item Mean interpolation: we assign the empty pixels the mean value of their neighbour pixels.
\item Random interpolation: we assign the empty pixels the value of a random neighbour pixel.
\end{itemize}

\begin{figure}
\centering
\includegraphics[width=0.95\linewidth]{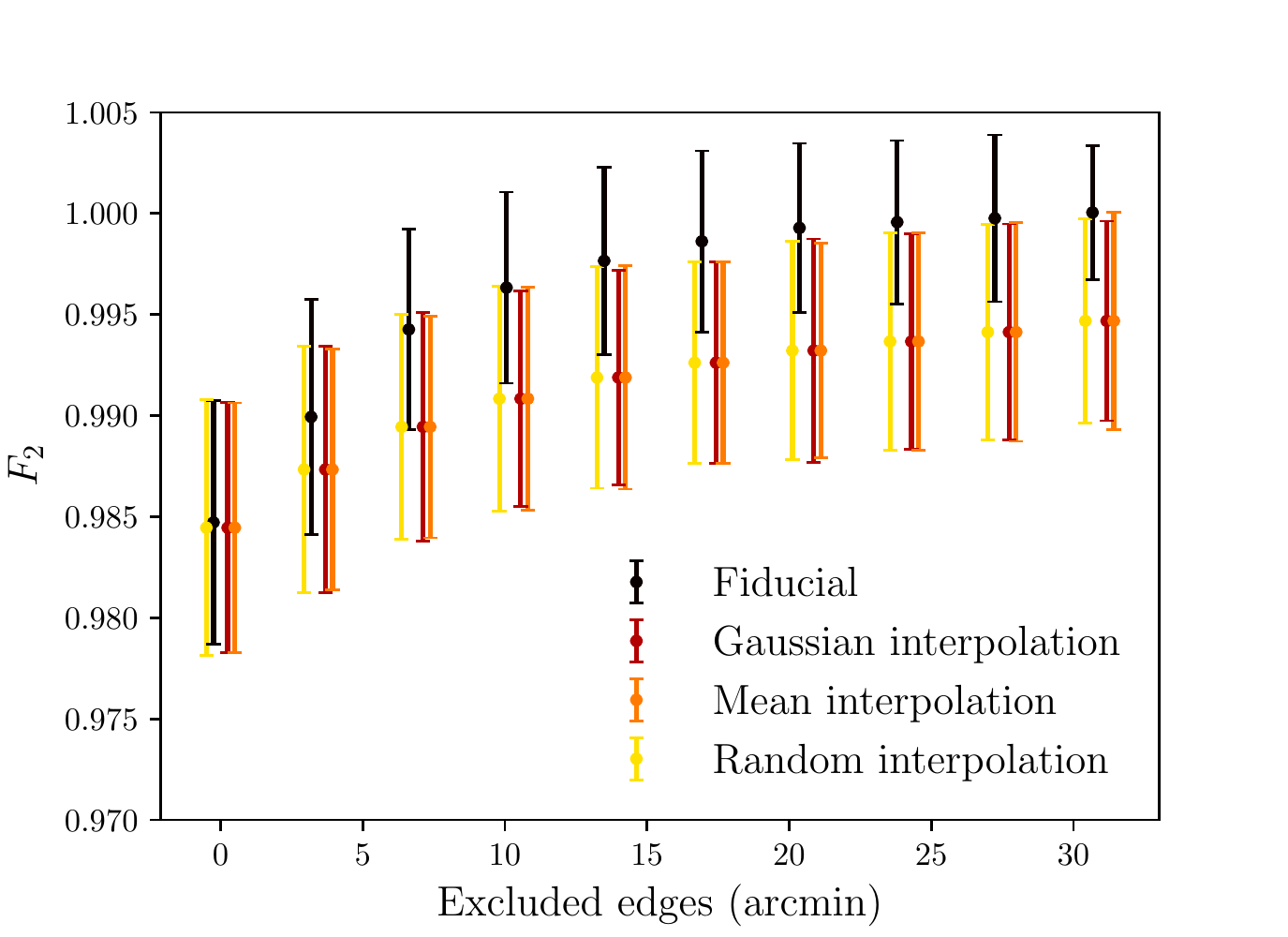}
\caption{$F_2 = \langle \kappa_E \kappa_{\rm sm} \rangle/\langle \kappa_{\rm sm}^2 \rangle$ for different 
interpolation schemes for empty pixels inside the footprint.}
\label{fig:empty_pixels}
\end{figure}

In \Fref{fig:empty_pixels} we show the $F_2$ statistics as a function of the scale excluded from the 
edges for all this cases, similar to \Fref{fig:edges}. The $F_{1}$ statistics looks qualitatively similar to $F_{2}$. 
We see that at our resolution, the different approaches all give very similar results. We therefore adopt the 
fiducial approach for simplicity in our main analysis. 

\section{Reconstructing the lensing potential and deflection map}
\label{sec:deflection_test_sim}

As discussed in \Sref{sec:formalism}, in addition to the convergence maps $\kappa$, we can also construct 
the lensing potential $\psi$ and deflection $\bm{\eta}$ maps with similar formalism. In this appendix we show 
the implementation of the reconstruction for $\psi$ and $\bm{\eta}$. We perform in \Aref{sec:sim_phi_eta} 
similar tests on the reconstruction with simulations as in \Sref{sec:synfast}. Then we apply the method to 
DES Y1 data in \Aref{sec:data_phi_eta}. Although the quality of the reconstruction for $\psi$ and $\bm{\eta}$ 
is not comparable to that of the $\kappa$ maps, they point to an area that we can start to explore as the 
sky coverage of future weak lensing data sets becomes increasingly large.
 
\begin{figure}
\centering
\includegraphics[width=0.99\linewidth]{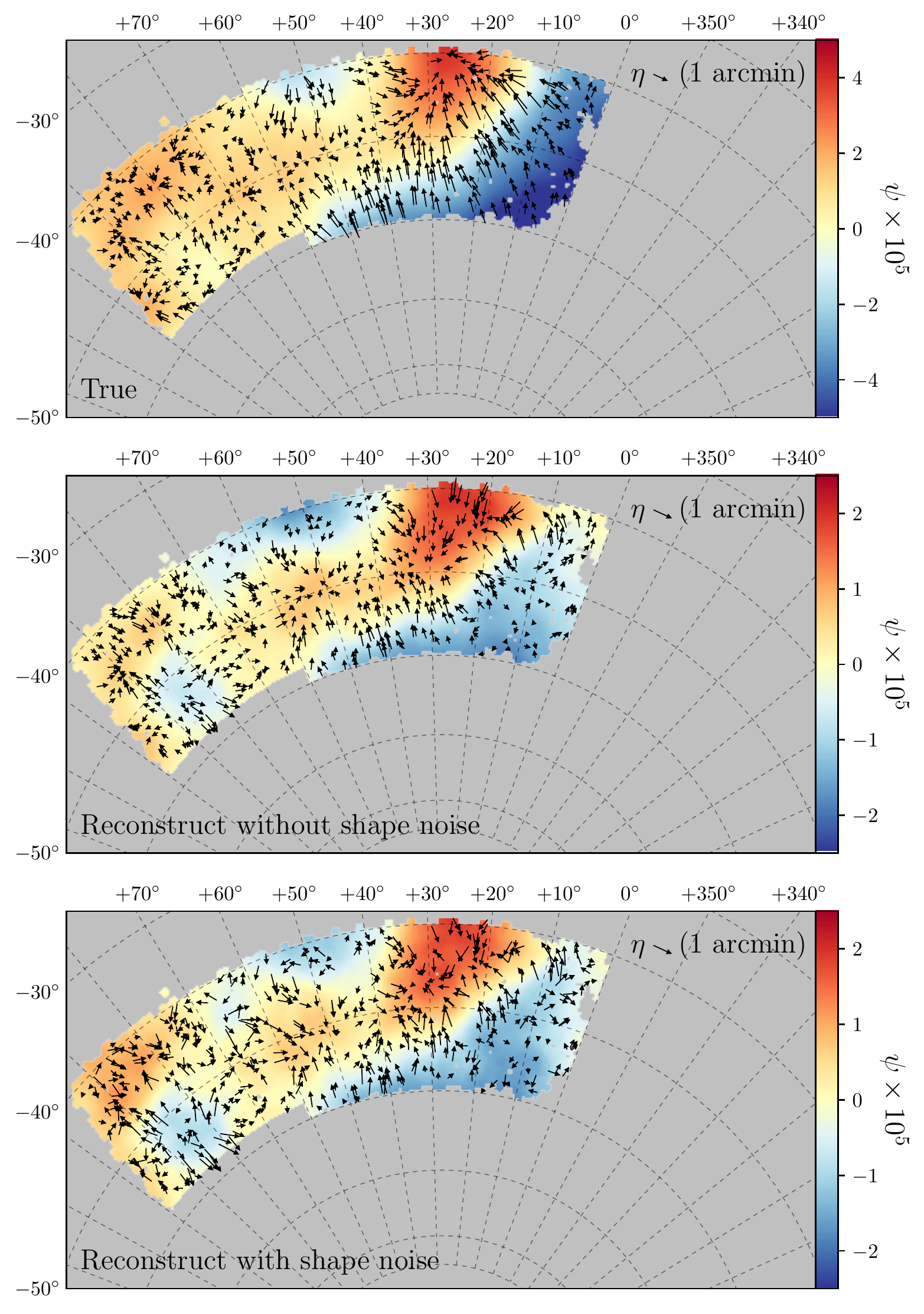}
\caption{The truth (top), noiseless reconstruction (middle) and noisy reconstruction (bottom) of $\psi$ and $\bm{\eta}$ 
field using the Gaussian simulations and a Y1-like mask. The colours indicate the value of the $\psi$ fields while the arrows indicate 
the observed deflection angle caused by lensing. The arrows are not to-scale --- they are enlarged for visualisation purpose. 
The amplitude of the reconstructed field is lower than the true field, therefore the colour bars in the bottom panel span 
over a range 4 times smaller than the top panel; while the arrows in the bottom panel are enlarged 2 times more than 
the top panel, as indicated by the one-arcminute bar on the upper right.}
\label{fig:map_sim_phi_alpha}
\end{figure}

\subsection{Simulation tests}
\label{sec:sim_phi_eta}

Similar to our test in reconstructing the convergence maps in \Sref{sec:sim_test}, we investigate the performance 
of reconstructing the lensing potential field $\psi$ and the deflection field $\bm{\eta}$. The techniques used for 
mapping these quantities are similar to those used for $\kappa$ and utilise the \healpix routines. The 
definition of $\bm{\eta}$ has been introduced in \Eref{eq:alpha_def} and related to $\psi$ in \Eref{eq:alpha}, 
but as $\bm{\eta}$ is a spin-1 field it requires use of the \healpix routine \texttt{alm2map\_spin} function to produce 
the final maps.  
In \Fref{fig:map_sim_phi_alpha} we show an example of a $\psi$ and $\bm{\eta}$ map generated via 
\texttt{synfast}. The top panel displays the true fields; the middle panel shows a reconstructed field with the Y1 
mask imposed and $RA>100^{\circ}$ region excluded; the bottom panel shows the reconstruction with the mask 
and realistic Y1 shape noise. We find that both the $\psi$ and $\bm{\eta}$ fields exhibit significant degradation 
due to the mask, as shown in the difference between the upper two panels, even though some level of 
resemblance remains.  The addition of shape 
noise has a much less significant effect, as can be seen from the bottom panel, which is very similar to the middle 
panel; this is expected as shape noise mainly degrades small-scale information, and is less important for 
the reconstruction of the $\psi$ and $\bm{\eta}$ maps. 

\begin{figure}
\centering
\includegraphics[width=0.95\linewidth]{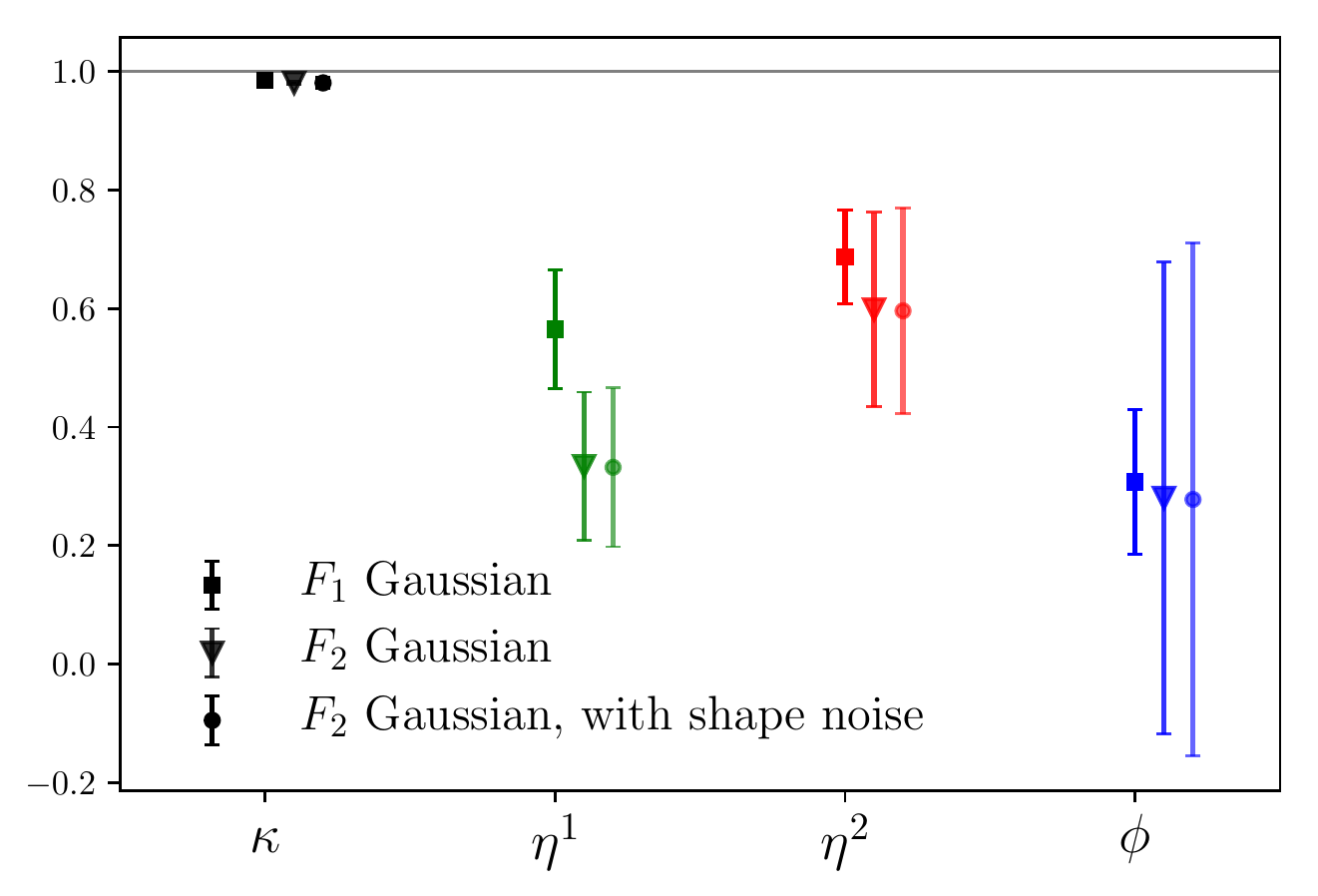}
\caption{$F_1$ (square markers) and $F_2$ (triangle markers) statistics for the reconstruction of the $\kappa$ 
(black), $\bm{\eta}=(\eta^{1}, \eta^{2})$ (green and red) and $\psi$ (blue) fields measured excluding pixels within 
$\sim10$ arcminutes from the edge of the mask. All measurements are done with the Gaussian simulations and 
the Y1-like mask described in \Sref{sec:synfast}. The round markers are the same as the triangle markers except 
for the addition of shape noise.}
\label{fig:edges_F1_F2}
\end{figure}

To quantify the degradation caused by the mask, we calculate the $F_1$ and $F_2$ 
(replacing $\kappa$ by $\psi$ and $\bm{\eta}$ in \Eref{eq:f1f2}) when excluding 10 arcminutes from the edges 
as shown in \Fref{fig:edges_F1_F2}. We generate 500 Gaussian realisations of the sky with the 
same underlying power spectrum to account for the effect of cosmic variance, which is an important factor 
in the reconstruction of $\psi$ and $\bm{\eta}$ since the information is dominated by large scales. 
We show the mean and standard deviation from these 500 simulations in \Fref{fig:edges_F1_F2}. 
As expected, we find that the mask has a stronger effect 
upon these two maps than for $\kappa$, as they use a higher proportion of information from the 
lower $\ell$ modes, which are more poorly constrained. This can be seen as a progressive 
degradation, from $\kappa$ to $\bm{\eta}$, to the most adversely affected $\psi$, but significant 
information is still reconstructed from the maps. The main effect of the mask on $\psi$ can be seen 
from the low value of $F_{1}$, due to the large unobserved sky regions suppressing the power inside the 
masked region by $\sim 70$\%. Similarly, $\bm{\eta}$ also suffers from this but to a 
lesser extent; $\eta^1$ is suppressed by $\sim 60$\% and $\eta^2$ by $\sim40$\%. The 
difference in this amplitude suppression comes from the fact that $\eta^1$ and $\eta^2$ are reconstructing the 
deflection angle in different directions on the sky --- the mask is a non-isotropic and there is more information 
in the $RA$ direction, which contributes mainly to $\eta^{2}$. 

To measure the reliability of the reconstruction of the phase information we use $F_2$. Comparing $F_2$ to $F_1$ 
gives a measurement of the phase reconstruction. These results are also shown in 
\Fref{fig:edges_F1_F2}. We find that for both $\psi$ and $\bm{\eta}$, the mean $F_{2}$ is at a similar level as $F_{1}$, 
but the standard deviation of $F_{2}$ is much larger than that of $F_{1}$, which suggests that the quality of the phase 
reconstruction varies dramatically depending on the specific realisation of the sky. 
Furthermore, we find that the influence of shape noise on $F_2$ is much less compared to the influence from the 
mask, as also suggested by \Fref{fig:map_sim_phi_alpha}.

Taken in combination, 
$F_1$ and $F_2$ suggest that considerable information can be inferred about $\bm{\eta}$ and $\psi$, although 
with much larger uncertainties than for $\kappa$. We do not perform further quantitative analyses on these maps, but 
note that for data sets on areas larger than DES Y1, the reconstruction of these other lensing fields 
becomes interesting. In these scenarios, algorithms that specifically deal with the mask will be particularly 
useful. For example, instead of converting the $\bm{\gamma}$ field to $\bm{\psi}$ and $\bm{\eta}$ directly, 
one can imagine forward-modelling the observed $\bm{\gamma}$ field from some underlying $\psi$ field. 
We defer the study of a forward-fitting mass mapping method to future work. \\

\subsection{Deflection and potential maps for DES Y1}
\label{sec:data_phi_eta}

\begin{figure*}
\centering
\includegraphics[width=0.9\linewidth]{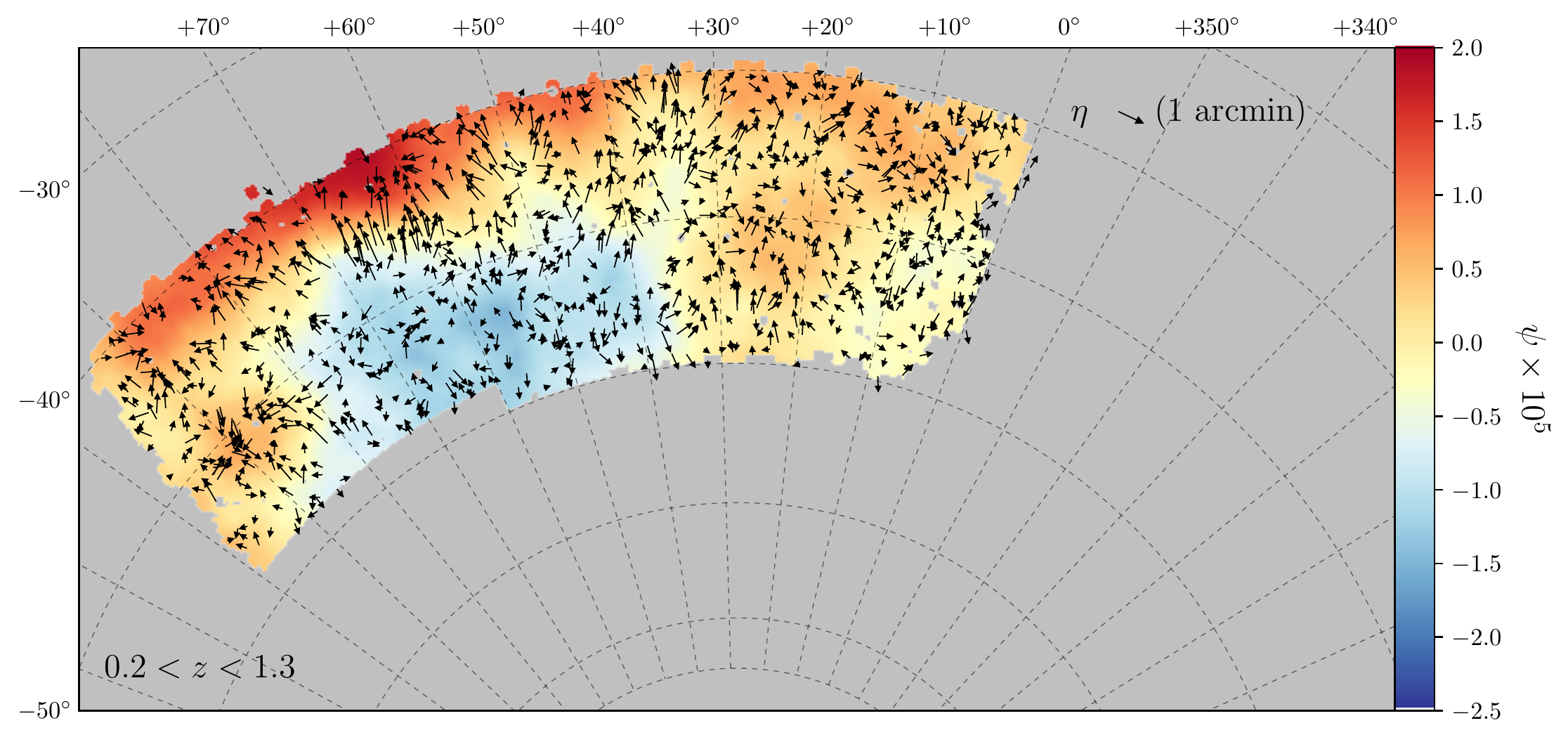}
\caption{The lensing potential (colour map) and deflection field (arrows) reconstructed for the DES Y1 \metacal 
data in the redshift range $0.2<z<1.3$. The arrows are not to-scale, but can be compared to the one-arcminute 
arrow in the upper right corner.}
\label{fig:map_phi_alpha}
\end{figure*}

We now apply the reconstruction method above to DES Y1 data. In \Fref{fig:map_phi_alpha} we show 
these maps constructed using \metacal shear measurements with the full redshift range $0.2<z<1.3$. 
As can be seen from the simulation tests in the previous section, the amplitude of these maps are 
expected to be much lower than the true fields 
due to the effect of masking. However, we can see reasonable correspondence between all the three maps: 
$\kappa$, $\bm{\eta}$ and $\psi$. On large scales, we find the low convergence (underdensed) regions 
are mostly located on the upper half of the footprint in \Fref{fig:map_full}: those correspond to a 
high potential value in \Fref{fig:map_phi_alpha}, and the deflection angle points from low to high potential. 
The characteristic scale of $\psi$ is larger than $\bm{\eta}$, which is larger than $\kappa$. The amplitude of 
the $\psi$ and $\bm{\eta}$ 
map also agrees well with that expected from the simulation tests in the previous section --- the 
potential field is $\sim10^{-4}$ and the deflection angle has a value on the order slightly below an arcminute. 
This map has implications for the mass distribution beyond the footprint. 
For example, the fact that the deflection angle points away from the boundary at the lower boundary of the 
map at RA$\sim30^{\circ}$ could indicate a large-scale overdensity just outside the footprint, which will be 
tested when more data is available.

\section{Catalog consistency}
\label{sec:im3shape}

In this appendix we compare maps from \imshape catalog with the results presented in the main text from \metacal. 
We also compare with the map from DES Science Verification data \citep[SV,][]{Vikram2015, Chang2015} which partly 
overlaps with the Y1 footprint. 

\begin{figure*}
\centering
\includegraphics[width=0.9\linewidth]{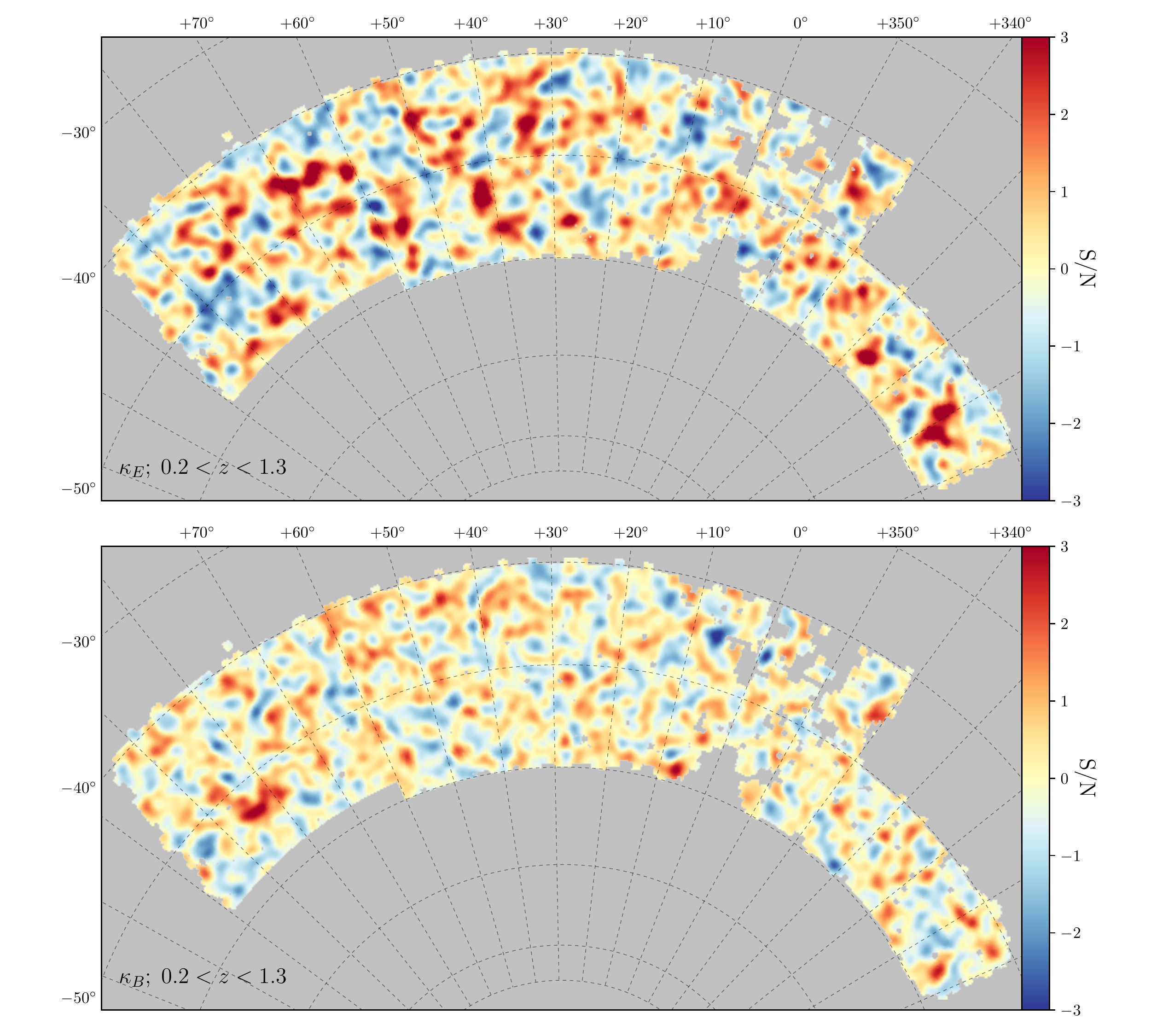}
\caption{Same as \Fref{fig:map_full}, but constructed using the \imshape shear catalog.}
\label{fig:map_im3shape_full}
\end{figure*}

\begin{figure*}
\centering
\includegraphics[width=0.98\linewidth]{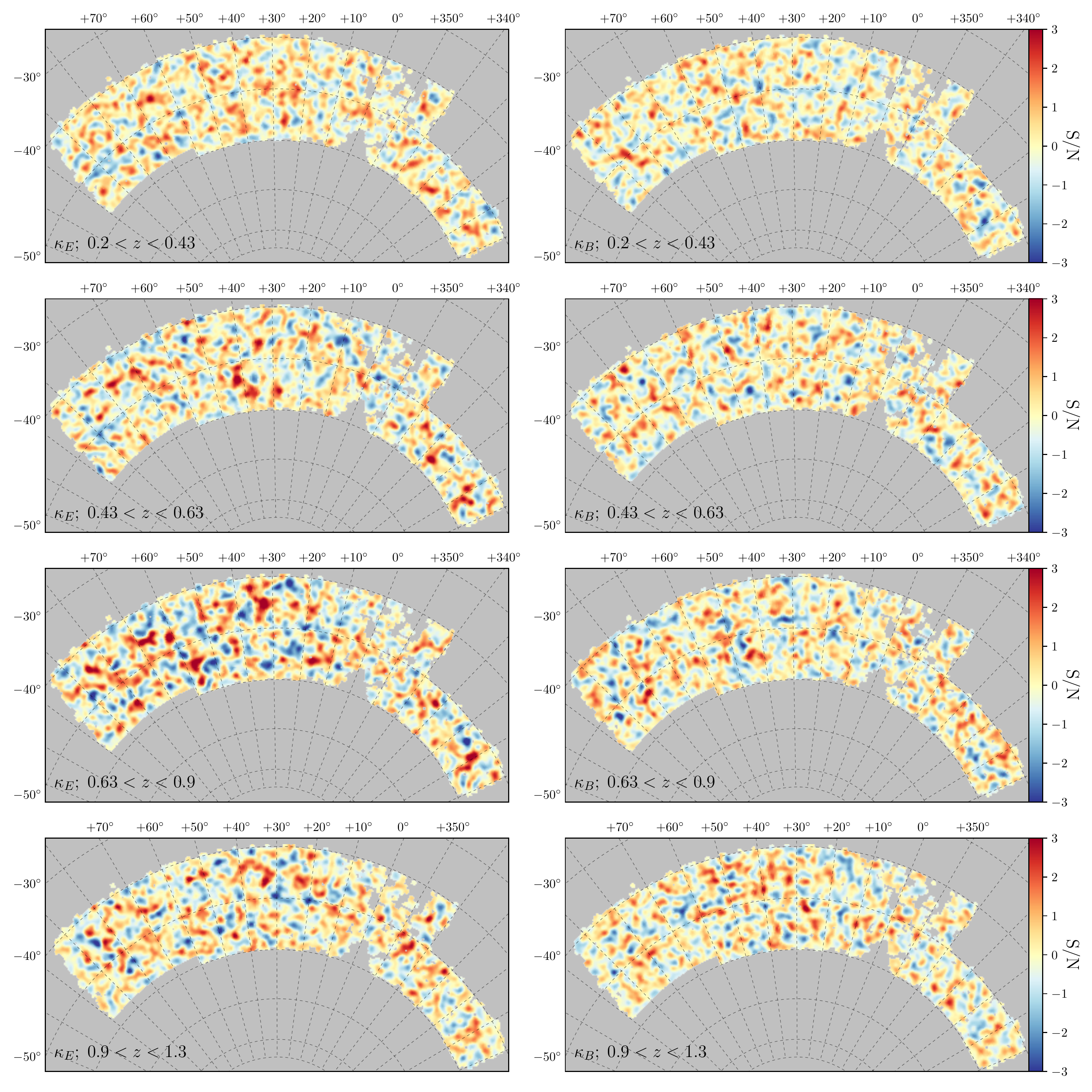}
\caption{Same as \Fref{fig:map}, but constructed using the \imshape shear catalog.}
\label{fig:map_im3shape}
\end{figure*}

In \Fref{fig:map_im3shape_full} and \Fref{fig:map_im3shape} we show the convergence maps generated using the \imshape 
shear catalogs. Comparing to \Fref{fig:map_full} and \Fref{fig:map} ,we can see that the broad structures in the $\kappa_{E}$ 
maps are similar, especially for the high S/N maps. The contrast between the E- and B-mode is less strong compared to 
\metacal due to the overall lower S/N in the \imshape catalog. Nevertheless, the agreement between the two independent 
catalogs provides a check of the shear measurement pipeline. 
 
\begin{figure*}
\centering
\includegraphics[width=0.6\linewidth]{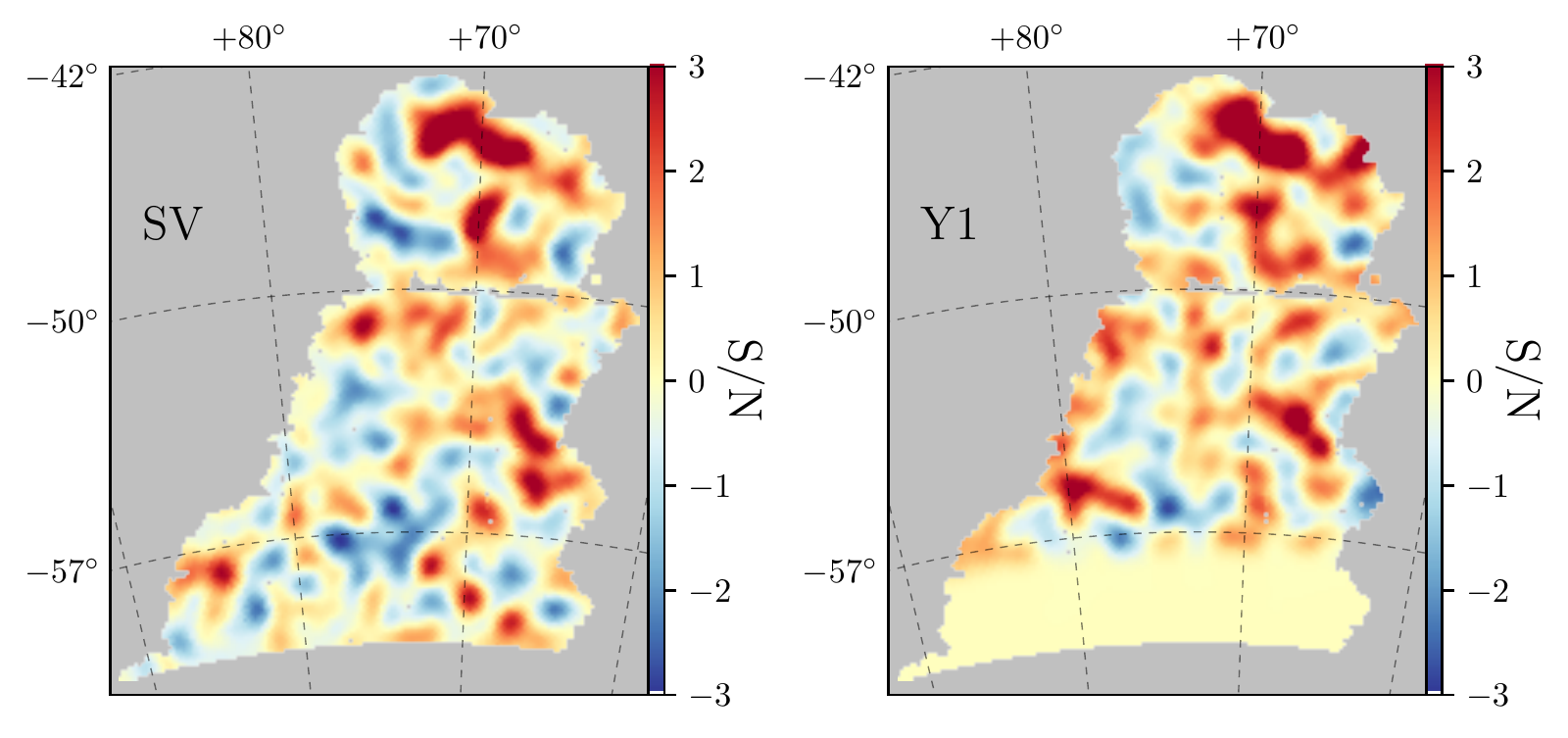}
\caption{Convergence map constructed using the SV \textsc{ngmix} catalog (left) and the Y1 \metacal catalog (right).}
\label{fig:map_sv}
\end{figure*}

In \Fref{fig:map_sv} we show the Y1 map and the SV map in the SV footprint; both maps use galaxies with a mean redshift 
$0.2<z<1.3$, and smoothed with $\sigma_{G}=20$ arcminutes. The SV map was constructed using another independent 
catalog \texttt{ngmix} and a different photo-z code, \textsc{Skynet}. The SV map is also half a magnitude deeper than the Y1 
map. The visual correspondence between the structures in the two maps is very good given the differences in the input data. 
This again serves as a consistency check between the different catalogs. 

\section{Author Affiliations}
\label{sec:affiliations}
{\small
$^{1}$\kicp\\
$^{2}$\ceasaclay\\
$^{3}$\cea\\
$^{4}$\ieec\\
$^{5}$\ports\\
$^{6}$\manchester\\
$^{7}$\princeton\\
$^{8}$\ifae\\
$^{9}$\upenn\\
$^{10}$\nyu\\ 
$^{11}$Einstein Fellow\\
$^{12}$\cambridgekavli\\
$^{13}$\cambridge\\
$^{14}$\lmu\\
$^{15}$\lsst\\
$^{16}$\stanford\\
$^{17}$\kipac\\
$^{18}$\cnrs\\
$^{19}$\ucl\\
$^{20}$\sorbonne\\
$^{21}$\lina\\
$^{22}$\on\\
$^{23}$\ciemat\\
$^{24}$\fermilab\\
$^{25}$\slac\\
$^{26}$\ethz\\
$^{27}$\jpl\\
$^{28}$\ohio\\
$^{29}$\ccap\\
$^{30}$\ua\\
$^{31}$\bnl\\
$^{32}$\maxplanck\\
$^{33}$\anl\\
$^{34}$\edinburgh\\
$^{35}$\ctio\\
$^{36}$\rhodes\\ 
$^{37}$\uiuc\\
$^{38}$\ncsa\\
$^{39}$\iit\\
$^{40}$\cluster\\
$^{41}$\uam\\
$^{42}$\uw\\
$^{43}$\santacruz\\
$^{44}$\aao\\
$^{45}$\brazil\\
$^{46}$\barcelona\\
$^{47}$\sussex\\
$^{48}$\michigan\\
$^{49}$\southhampton\\
$^{50}$\wataghin\\
$^{51}$\oakridge
}
 
\end{document}